\documentclass[twocolumn]{aastex701}

\usepackage{natbib}
\usepackage{rotating}
\usepackage{booktabs}
\usepackage{multirow}
\usepackage{mhchem}
\usepackage{graphicx}

\begin{document}

\title{A Multiwavelength Assessment Disfavoring the X-ray Binary Origin of \ion{He}{3} Regions in Metal-Poor Star-Forming Dwarf Galaxies}

\author{Ivan Altunin}
\affiliation{Department of Physics, University of Nevada, Reno, NV 89557, USA}
\email[show]{ialtunin@unr.edu}

\author{Christopher Ellis}
\affiliation{Department of Physics, University of Nevada, Reno, NV 89557, USA}
\email{harleyellis5@gmail.com}

\author[0000-0002-7092-0326]{Richard M. Plotkin}
\affiliation{Department of Physics, University of Nevada, Reno, NV 89557, USA}
\affiliation{Nevada Center for Astrophysics, Las Vegas, NV 89154, USA}
\email{rplotkin@unr.edu}

\author[0000-0002-4622-796X]{Roberto Soria}
\affiliation{INAF - Osservatorio Astrofisico di Torino, Strada Osservatorio 20, I-10025 Pino Torinese, Italy}
\affiliation{Sydney Institute for Astronomy, School of Physics A28, The University of Sydney, Sydney, NSW 2006, Australia}
\email{roberto.soria@inaf.it}

\author{Ryan Tanner}
\affiliation{Department of Physics, University of Nevada, Reno, NV 89557, USA}
\email{ryantanner@unr.edu}

\author[0000-0002-9165-6245]{Erica Thygesen}
\affiliation{Greenhill Observatory, The School of Natural Sciences, PO Box 807, Sandy Bay, TAS 7006 Australia}
\email{erica.thygesen@utas.edu.au}

\author[0000-0001-5802-6041]{Elena Gallo}
\affiliation{Department of Astronomy, University of Michigan, 1085 South University, Ann Arbor, MI 48109, USA}
\email{egallo@umich.edu}

\author{Manfred W. Pakull}
\affiliation{Universit\'e de Strasbourg, CNRS, Observatoire Astronomique, CNRS, UMR 7550, F-67000 Strasbourg, France}
\email{manfred.pakull@astro.unistra.fr}

\author[0000-0003-3484-0326]{Andrea H. Prestwich}
\affiliation{Institute for Astrophysics and Computational Sciences, Catholic University of America}
\affiliation{Astrophysics Science Division, NASA-Goddard Space Flight Center}
\affiliation{Center for Research and Exploration in Space Science and Technology, NASA-Goddard Space Flight Center, Greenbelt, MD 20771}
\email{andrea.h.prestwich@nasa.gov}

\author[0000-0001-7158-614X]{Amy Reines}
\affiliation{Department of Physics, Montana State University, Bozeman, MT 59717, USA}
\email{amy.reines@montana.edu}

\author[0000-0003-1814-8620]{Ryan Urquhart}
\affiliation{Center for Data Intensive and Time Domain Astronomy, Department of Physics and Astronomy, Michigan State University, East Lansing, MI 48824, USA}
\email{urquha20@msu.edu}

\author[0000-0002-8808-520X]{Aarran W. Shaw}
\affiliation{Department of Physics \& Astronomy, Butler University, 4600 Sunset Avenue, Indianapolis, IN 46208, USA}
\email{awshaw@butler.edu}

\begin{abstract}
Recent observations of metal-poor, star-forming dwarf galaxies reveal He III regions, traced by nebular \ion{He}{2} $\lambda4686$ emission that require a strong source of extreme-ultraviolet (EUV) radiation. The origin of this hard ionizing radiation remains poorly understood, as standard stellar populations fail to account for it, posing key implications for the understanding of early galaxy formation. We present a systematic Chandra X-ray study of 21 nearby (z $<$ 0.1) star-forming galaxies with \ion{He}{2} emission but lacking Wolf–Rayet spectral signatures. Using 7 new and 36 archival Chandra X-ray observations combined with optical stellar population synthesis modelling, we constrain the ionizing continuum required to sustain the observed \ion{He}{2} line, the ionizing continuum available from X-ray objects, and the properties of the host H II regions. We find that the inferred EUV output from accreting X-ray sources in our sample is systematically lower than what is required to produce the observed \ion{He}{2} emission. Our sample is consistent with established empirical scaling relations for X-ray luminosity, indicating that this discrepancy cannot be attributed to an anomalously low number or luminosity of X-ray sources. These results indicate that accreting X-ray sources alone cannot account for the observed He II-ionizing photon budget, pointing to additional or alternative sources of hard EUV radiation in metal-poor star-forming environments. Potential alternative or additional contributors are discussed.

\end{abstract}


\section{Introduction} 
\label{sec:intro}
Low-metallicity dwarf galaxies - characterized by their modest sizes ($< 10^9 M_\odot$) and low-metallicities ($Z < 0.1 Z_\odot$) - situated in the local Universe ($z < 0.1$) serve as galactic laboratories, providing a window into the types of stellar and black hole populations that might have existed in the early Universe \citep[e.g.,][]{Mezcua19, Thuan05}. Not only do some of these galaxies host active galactic nuclei (AGNs), and stellar mass black holes, which can influence how galaxies form and grow over cosmic time \citep[e.g.,][]{Greene20, Fabian12, Manzano-King19}, but their early-universe counterparts are also suspected contributors to cosmic reionization (e.g., \citealt{Kehrig18, Ponnada20, Berg2018}). 

Unexpectedly, many nearby metal-poor dwarf galaxies with high rates of star formation have been observed to host He III regions, i.e. compact zones of doubly ionized helium ({He}$^{++}$) embedded within larger H II regions (e.g., \citealt{Garnett91}; \citealt{Shirazi12}). The existence of such regions requires a substantial flux of photons with energies above the {He}$^{+}$ ionization edge (54.4 eV), corresponding to an extreme-ultraviolet (EUV) radiation field whose origin remains poorly understood. Because He III regions cannot be observed directly, their presence is inferred from narrow nebular emission from the \ion{He}{2} $\lambda4686$ ($\chi_{\rm ion}=54$ eV) recombination line, which arises from the recombination of {He}$^{+}$. Thus, the detection of nebular He II emission provides a direct tracer of an underlying He III zone and, by extension, a hard ionizing continuum extending beyond 54 eV \citep{Pakull86}.

In analogy with direct observations of \ion{He}{3} regions in the Milky Way and other nearby galaxies (e.g., \citealt{Garnett91, Moon11, Pakull02}), we expect the \ion{He}{2} line emission to be associated with single-populations found in low-metallicity star-forming regions. The observed similarity between spatially unresolved star-forming dwarf galaxies and individual \ion{H}{2} regions (e.g., \citealt{Kobulnicky99}) allows their integrated spectra to be interpreted within the same photoionization framework. In a similar manner, this motivates modelling ensembles of unresolved \ion{He}{3} regions as high-ionization zones within large-scale clouds of diffuse gas photoionized by a single embedded population (e.g., \citealt{Charlot2000, Charlot2001}).

In metal-rich environments, strong line-driven winds strip massive stars of their hydrogen envelopes, revealing hot helium-burning cores producing Wolf-Rayet (WR) stars with characteristic spectral features. The ionizing output from such stars is a sufficent source of He$^{+}$-ionizing photons \citep{Schaerer96}, however their formation through single-star stellar winds is expected to become inefficient at low metallicity where weaker winds inhibit envelope removal \citep[e.g.,][]{Guseva00, Shenar2020, Gotberg2023}. Indeed while WR stars and He II emission have been observed to co-exist in some systems \citep[e.g.,][]{Pakull89, deMello98, Roy2025}, an increasing number of low-metallicity galaxies exhibit strong nebular He II emission without detectable WR features \citep[e.g.,][]{Shirazi12, Kehrig15}.

Binary interactions fundamentally alter this picture, with findings that a significant fraction of stars form in binary systems (40-100$\%$ depending on the star type), and that the majority ($\sim$70$\%$) of massive stars exchange mass with a binary companion during their evolution \citep[e.g.,][]{Eldridge17, Eldridge2022}. Mass transfer and stripping in binary systems enable the removal of hydrogen envelopes even in low metallicity environments, producing hot, compact helium stars that extend from classical WR stars with high luminosities to lower-mass stripped stars with weaker, absent, or obscured WR spectral features \citep[e.g.,][]{Shenar2020}. Recent observations confirm such helium stars further supporting their existence as potentially common, and sometimes hidden, ionizing sources \citep[e.g.,][]{Drout2023, Gotberg2023}.

An alternative or additional channel of ionization may be provided by accreting compact objects, including X-ray binaries (XRBs) and AGN, which naturally produce hard ionizing radiation extending from the X-ray into the EUV. The relevance of XRBs is strengthened by the same binary evolution processes found to produce stripped helium stars \citep[e.g.,][]{VanBever1999, Eldridge2022}. Furthermore, XRBs are suitable candidates for established photoionization models given the observationally-derived preference of luminous XRBs towards areas with higher star formation and lower metallicity (e.g., \citealt{Prestwich13, Basu-Zych13, Brorby14, Douna15, Lehmer21}).

The ability for X-ray sources to power \ion{He}{2} line emission is already established for some individual nearby galaxies and nebulae \citep{Pakull86, Pakull02, Kaaret04, Moon11}. However, conflicting results are reported in the literature on whether XRBs can be the primary source of ionization across multiple hosts, with some supporting XRBs (see, e.g., \citealt{Schaerer19, Simmonds21}, and reference therein) and others reaching the opposite conclusion (see, e.g., \citealt{Thuan05, Senchyna20, Thygesen23}, and references therein). Hence, despite the ability of numerous physical sources to produce the required hard ionizing radiation, no primary originator common to all host environments has been identified \citep{Schaerer19}. Here we provide constraints on the properties of the ionizing radiation responsible for the observed \ion{He}{2} emission, and possible accretion-powered sources within the framework of the embedded single-population model.

First, we aim to determine the EUV continuum required to produce the observed \ion{He}{2} line luminosity, as this directly constrains the minimum ionizing power needed to sustain the nebular line emission. Quantifying this requirement allows us to test whether populations of compact accreting objects are energetically capable of powering the nebular \ion{He}{2} emission. Direct measurement of the EUV ionizing continuum is not possible because photons at these energies are efficiently absorbed by the intervening ISM through photoelectric absorption, dominated by neutral hydrogen and helium (\citealt{Wilms2000}). Consequently, the production of EUV photons must be inferred indirectly from X-ray observations. To this end, we perform a \textit{Chandra} X-ray survey of 21 nearby ($z<0.1$) star-forming galaxies that show nebular \ion{He}{2} emission but lack obvious spectral signatures of WRs. This sample was originally identified by \citet{Shirazi12} from the Sloan Digital Sky Survey \citep[SDSS;][]{York00}.

Secondly, we aim to constrain the effective parameters of the host H II regions and predict the EUV continuum present that is responsible for the observed He II emission. The assumption that the ionizing radiation arises from XRB populations formed during recent star formation episodes suggests that the emergent ionizing luminosity will scale with both the star formation rate and metallicity \citep[e.g.,][]{Lehmer21, Brorby14}. Traditionally, these parameters have been determined from emission-line diagnostics. Alternatively, we combine stellar population synthesis (SPS) with photoionization modelling to enable a more comprehensive approach through the modelling of both line and continuum emission (e.g., \citealt{Johnson2021, Garofali2024}). Through the additional modelling of star formation history, stellar mass, age, and dust content we obtain simultaneous constraints on star formation rate and metallicity, while also investigating how the ionization budget changes with time and the possibility of attenuated accreting sources. 

We present our sample and our observations and data processing/modeling in Section \ref{sec:observations}, our results are presented in Section \ref{sec:results}, their implications are discussed in Section \ref{sec:discussion}, and we summarize our findings in Section \ref{sec:conclusion}. Throughout, we adopt a cosmology with $H_0=70$ km s$^{-1}$ Mpc$^{-1}$, $\Omega_m=0.3$, and $\Omega_\Lambda=0.7$. All uncertainties are presented at the 68\% (1$\sigma$) confidence level, and all upper limits on non-detections are presented at the 99\% confidence level.

\section{Observations and Data Reduction} \label{sec:observations}
\subsection{Sample Selection} \label{sec:samples}
\citet{Shirazi12} studied the spectra of nearly 3000 emission-line galaxies from the SDSS Data Release 7 \citep[DR7;][]{sdssdr7} that showed strong nebular \ion{He}{2} emission lines (they defined nebular as emission lines with comparable widths to forbidden lines in each galaxy spectrum). Of these 3000 galaxies, they identified a subset of 83 star-forming galaxies (classified through narrow emission line ratios) that do not show spectroscopic signatures from WRs (specifically, they lacked broad emission around 4650 and 5808~\AA, from \ion{He}{2} blended with other metals and from \ion{C}{4}, respectively; see, e.g., \citealt{Crowther07}). These 83 galaxies are mostly (although not exclusively) in metal-poor galaxies with low Oxygen abundances, $12 + \log\left(O/H\right) < 8.2$ (see Figure 14 in \citealt{Shirazi12})\footnote{ Throughout this paper we refer to metallicity using the gas-phase oxygen abundance, 12 + log(O/H) as a standard nebular tracer of chemical enrichment. However, O/H does not directly trace the iron abundance ([Fe/H]) that more strongly governs massive-star evolution \citep{Eldridge2022}. As a result, and because $\alpha$-enhancement typically increases toward lower metallicities \citep{Byrne25}, the galaxies in our sample may in fact be more metal-poor in [Fe/H] relative to solar than their O/H values reported herein would suggest.}. The lack of WR signatures in these galaxies is unlikely systematically driven by low signal-to-noise in their SDSS spectra (see Section 6.1 of \citealt{Shirazi12}). Given the narrow width of the \ion{He}{2} emission, \citet{Shirazi12} argue that star-forming galaxies in their sample suffer from low contamination from Type 1 AGNs (they expect $<$10\% of the observed \ion{He}{2} line emission to be due to AGN; see their Section 2.2).

We observed a subset of 7 of these 83 galaxies during \textit{Chandra} Cycle 19 (proposal 19620343; PI Plotkin). These galaxies were selected as being amongst the closest ($d<30$ Mpc), as the most metal-poor ($12 + \log\left(O/H\right) < 8.0$), and as having the hardest inferred EUV spectra (based on the line ratio $\log$\, \ion{He}{2}/H$\beta > -1.8$). We also cross-matched the remaining 76 galaxies from \citet{Shirazi12} against the \textit{Chandra} archive, using a 30$\arcsec$ search radius and restricting only to observations taken with the Advanced CCD Imaging Spectrometer (ACIS; \citealt{Garmire03}), without gratings. We found an additional 14 galaxies with archival X-ray coverage (spread over 36 observations).\footnote{Our search also returned \textit{Chandra} observations of M77. We exclude M77 from our analysis because it known to be a Seyfert 2 galaxy, its X-ray population is already well studied in the literature \citep[e.g.,][]{Brinkman02, Liu11}, and it is a distinctly more massive spiral galaxy compared to the other objects in our sample.} 

\begin{deluxetable}{ccccc}
\tablecaption{Galaxy Properties \label{galaxytable}}
\tablehead{
\colhead{Target} &
\colhead{Distance} &
\colhead{Metallicity} &
\colhead{$\log{F_{{4686}}}$} &
\colhead{Galaxy Name} \\
\colhead{(SDSS)} &
\colhead{(Mpc)} &
\colhead{(12+log O/H)} &
\colhead{($\rm{erg\;s^{-1}cm^{-2}}$)} &
\colhead{}
}
\colnumbers
\startdata
$\rm{J000953.09+154404.8}$ &  $\rm{12.8}$ & $7.92\pm0.09$ & $-15.60\pm0.16$ & LEDA 697 \\ 
$\rm{J021513.98-084624.3}$ &  $\rm{21.1}$ & $7.74\pm0.09$ & $-15.77\pm0.18$ & SHOC 111 \\ 
$\rm{J083743.48+513830.2}$ &  $\rm{10.5}$ & $7.95\pm0.08$ & $-15.45\pm0.07$ & MRK 0094 \\ 
$\rm{J101624.51+375445.9}$ &  $\rm{16.7}$ & $7.80\pm0.05$ & $-15.04\pm0.03$ & LEDA 29998 \\ 
$\rm{J103410.15+580349.1}$ &  $\rm{32.2}$ & $7.78\pm0.06$ & $-15.09\pm0.05$ & MRK 1434 \\ 
$\rm{J104653.98+134645.7}$ &  $\rm{45.8}$ & $7.92\pm0.09$ & $-15.82\pm0.13$ & \nodata \\ 
$\rm{J105310.82+501653.2}$ &  $\rm{18.7}$ & $7.94\pm0.08$ & $-15.33\pm0.06$ & MRK 156 \\ 
$\rm{J110458.30+290816.5}$ &   $\rm{9.0}$ & $7.79\pm0.05$ & $-15.30\pm0.21$ & MRK 36 \\ 
$\rm{J111746.30+174424.6}$ &  $\rm{21.1}$ & $7.90\pm0.09$ & $-15.61\pm0.13$ & \nodata \\ 
$\rm{J114107.48+322537.2}$ &  $\rm{25.8}$ & $7.79\pm0.05$ & $-15.27\pm0.05$ & LEDA 36252 \\ 
$\rm{J115237.67-022806.3}$ &  $\rm{15.2}$ & $7.90\pm0.09$ & $-15.54\pm0.13$ & \nodata \\ 
$\rm{J115441.22+463636.3}$ &  $\rm{15.0}$ & $7.80\pm0.05$ & $-15.61\pm0.10$ & LEDA 2284119 \\ 
$\rm{J121749.30+375155.5}$ &   $\rm{2.6}$ & $7.92\pm0.09$ & $-15.62\pm0.11$ & WR 415 \\ 
$\rm{J122225.79+043404.7}$ &  $\rm{18.4}$ & $8.10\pm0.11$ & $-15.57\pm0.20$ & SDSSCGB 20312.2 \\ 
$\rm{J122615.70+482938.4}$ &   $\rm{5.8}$ & $7.80\pm0.04$ & $-14.98\pm0.05$ & UGCA 281 \\ 
$\rm{J124134.25+442639.2}$ & $\rm{185.2}$ & $8.80\pm0.08$ & $-15.35\pm0.09$ & LEDA 2244532 \\ 
$\rm{J140411.23+542518.7}$ &   $\rm{6.9}$ & $7.99\pm0.06$ & $-15.64\pm0.19$ & \nodata* \\ 
$\rm{J140428.62+542352.8}$ &   $\rm{6.9}$ & $8.33\pm0.32$ & $-15.60\pm0.25$ & NGC 5471* \\ 
$\rm{J142628.16+382258.6}$ &  $\rm{97.2}$ & $7.77\pm0.07$ & $-15.77\pm0.19$ & LEDA 51563 \\ 
$\rm{J144852.02+344242.9}$ &  $\rm{9.3}$ & $7.91\pm0.09$ & $-15.73\pm0.17$ & LEDA 200320 \\ 
$\rm{J222510.13-001152.8}$ & $\rm{300.0}$ & $7.90\pm0.04$ & $-15.73\pm0.32$ & \nodata \\ 
\enddata
\tablecomments{Target Galaxy Properties. Column 1: SDSS galaxy name. Column 2: Distance to target galaxy. Column 3: Metallicity of target galaxy. Column 4: Flux of \ion{He}{2} $\lambda$4686 (log). Column 5: Common name of target galaxy, if available.}
\tablenotetext{*}{Satellite dwarfs of M101.}
\end{deluxetable}

Unlike our own observations, we do not explicitly impose a restriction on host-galaxy metallicity for the archival sample. Still, because of the  requirement for galaxies in our archival search to not display WR features, our archival subset tends to be weighted to low Oxygen abundances (only 2 of 14 galaxies in the archival subsample have $\log\left(O/H\right)+12 > 8.2$), (LEDA 2244532, and NGC 5471). Note that in many cases the galaxy of interest was not the main target of the archival \textit{Chandra} observation. In these cases, the galaxy was usually located several arcminutes from the \textit{Chandra} aimpoint, therefore resulting in lower X-ray sensitivity and poorer spatial resolution. 

\begin{figure*}
    \centering
    \includegraphics[width=0.85\linewidth]{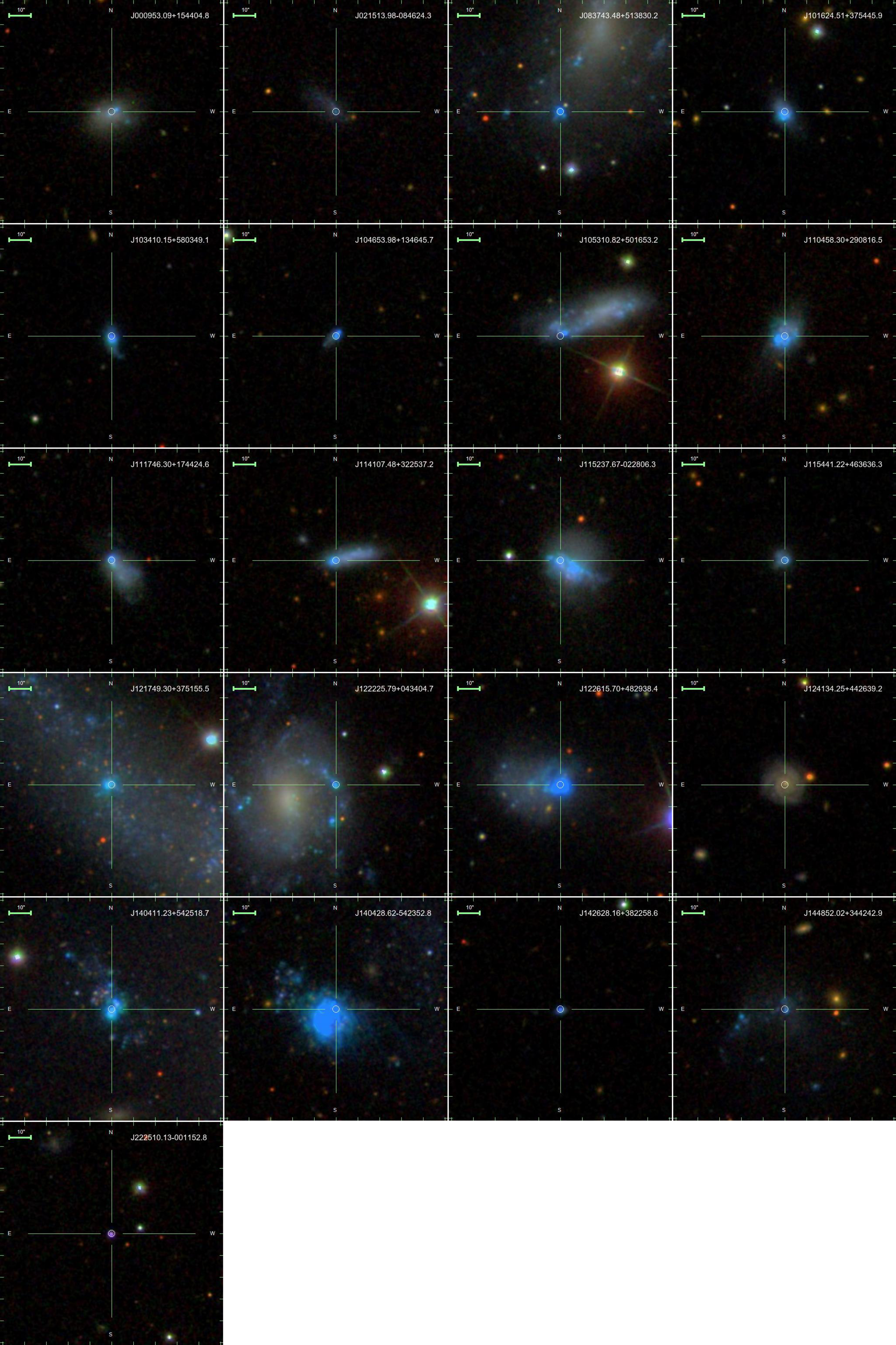}
    \caption{SDSS finding charts for our 21 target galaxies, ordered by increasing right ascension. Each panel shows the SDSS image of a target galaxy \citep{sdssdr7}, with the position of the SDSS 3$\arcsec$ spectroscopic fiber indicated by the centered white circle. The green scale bar in the top left of each plot represents the length of 10$\arcsec$.}
    \label{fig:finding_chart}
\end{figure*}

In summary, our final sample includes 7 new and 36 archival X-ray observations over 21 galaxies (See Table \ref{galaxytable}). Although the SDSS spectra for most sources correspond to individual star-forming H II regions rather than entire galaxies (See Figure \ref{fig:finding_chart}), we refer to them colloquially as galaxies for simplicity, while keeping in mind that our analysis pertains to more localized regions of their larger hosts. Also for simplicity, we identify X-ray observations associated with these regions by their host galaxy name (or SDSS Name in lieu of) and refer to individual Chandra Observation ID's only when discussion of specific observations are warranted.

The properties of the galaxies in our sample are listed in Table \ref{galaxytable}. Distances listed in Table \ref{galaxytable}, given in Mpc, were obtained from the \citealt{NED}\footnote{The NASA/IPAC Extragalactic Database (NED)  is operated by the Jet Propulsion Laboratory, California Institute of Technology, under contract with the National Aeronautics and Space Administration}. When no redshift-independent distance was available, distances were calculated from SDSS redshifts using the cosmology described in Section \ref{sec:intro}. For MRK 36, the median Tully–Fisher distance was adopted, while for UGCA 281 the median tip of the red giant branch distance was used. The distance to LEDA 200320 was taken from its reported Tully–Fisher measurement. As J140411.23+542518.7 and J140428.62+542352.8 (NGC 5471) are satellite dwarf galaxies of M101, the median value of Cepheid distance measurements to M101 was adopted.

\begin{figure*}
    \centering
    \includegraphics[width=0.9\linewidth]{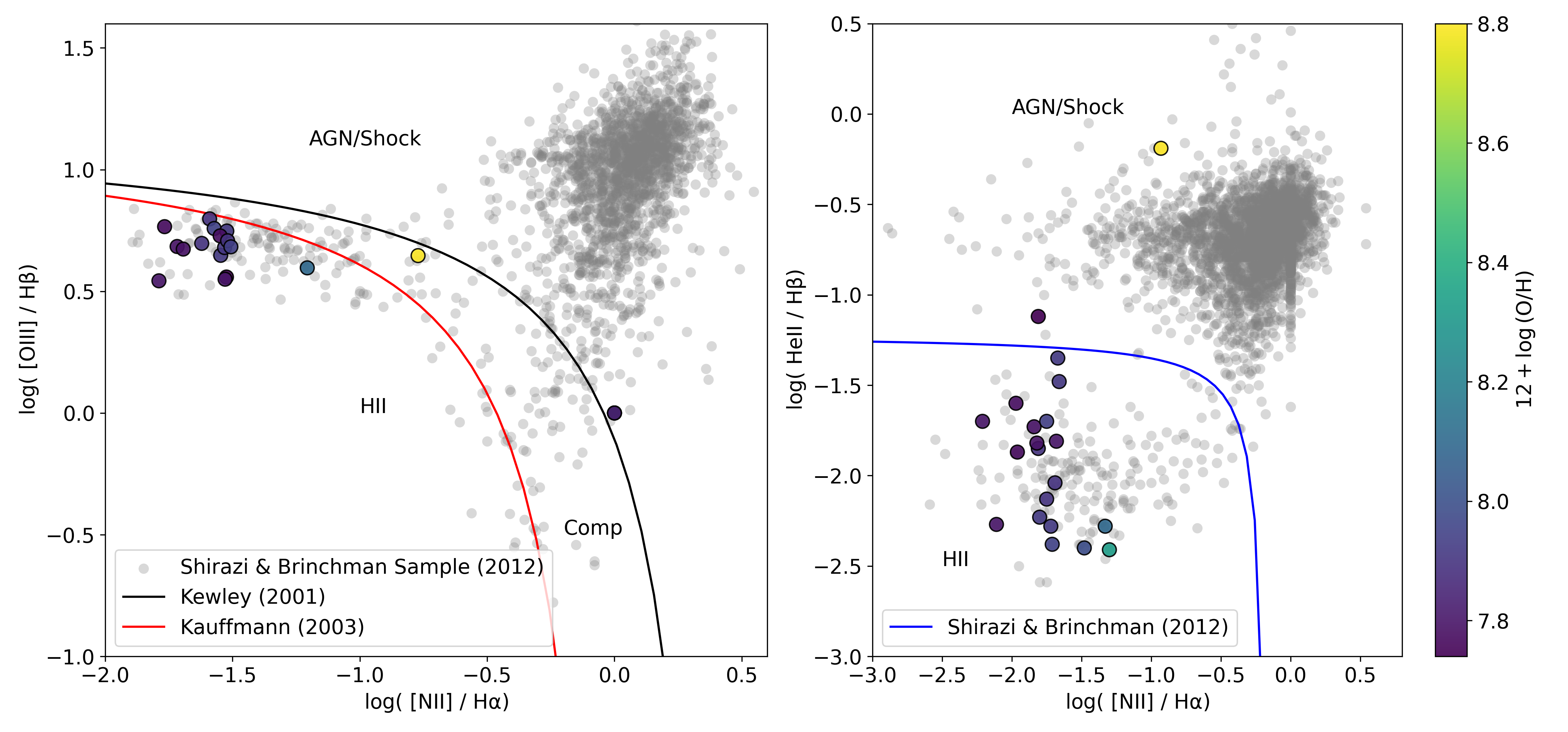}
    \caption{Optical emission-line diagnostic diagrams for our galaxy sample. Left: The standard BPT diagram \citep{Baldwin1981} showing $\log$([O III] $\lambda5007$/H$\beta$) versus $\log$([N II] $\lambda6584$/H$\alpha$). The solid black and red curves denote the \citet{Kewley2001} and \citet{Kauffmann2003} demarcations, respectively, separating star-forming (H II), composite (Comp), and AGN/shock-dominated regions. Right: The $\log$(He II $\lambda4686$/H$\beta$) versus $\log$([N II] $\lambda6584$/H$\alpha$) diagram following \citet{Shirazi12}, which provides enhanced sensitivity to harder ionizing spectra. Both: Gray points represent the parent sample from \citet{Shirazi12}, while colored points correspond to our targets, color-coded by gas-phase metallicity (12 + log(O/H)). The majority of our galaxies lie within the star-forming region in both diagnostics, with onle 1-2 sources exhibiting line ratios consistent with AGN or shock excitation.}
    \label{fig:BPT}
\end{figure*}

Figure \ref{fig:finding_chart} visualizes the SDSS finding chart \citep{sdssdr7} for each of the galaxies in our sample with the position of the SDSS 3-arcsecond spectroscopic fiber showen. In addition to the standard BPT diagram \citep{Baldwin1981}, we employ the \ion{He}{2} $\lambda4686$/H${\beta}$ versus [N II] $\lambda6584$/H${\alpha}$ diagnostic to refine the classification of our sources in Figure \ref{fig:BPT}. This diagram, following the prescription of \citet{Shirazi12}, provides a more sensitive measure of the ionizing spectrum hardness. The majority of our sample falls within the star-forming region in both diagnostics with only a few sources showing harder ionizing spectra and the potential for AGN or shock-dominated emission line ratios.

\subsection{\textit{Chandra} X-ray Observations} \label{sec:data}

The properties of all X-ray observations are listed in Table \ref{xraytable}. For our 7 new \textit{Chandra} observations, each target was placed at the aimpoint of the ACIS-S3 chip, and the data were telemetered in \texttt{VFAINT} mode. Observations were scheduled for 8-12 ks exposure times. All new and archival X-ray observations were reduced using the \textit{Chandra Interactive Analysis of Observations} (CIAO) software v4.14 \citep{Fruscione06}, utilizing \texttt{CALDB} v4.9.4. We ran the reprocessing script {\tt chandra\_repro} to apply calibration files and to create new bad pixel maps (we set the parameter \texttt{check\_vf\_pha=yes} for data telemetered in \texttt{VFAINT} mode). We next checked for background flares, and we did not find any major flaring events. 

For the subset of new observations (i.e., where our target was placed at the ACIS-S3 aimpoint and therefore observed under high spatial resolution), we applied an astrometric correction when possible. We found all X-ray sources on the S3 chip by running {\tt wavdetect} on broadband images (0.5-7.0 keV) created by the tool {\tt fluximage}. When running {\tt wavdetect}, we set parameter {\tt scale}=``1 2 4 6 8 12 16 24 32'', and we set the {\tt sigthresh} parameter to 10$^{-6}$ (so that we expect an average of one false X-ray detection per observation). In preparing the point spread function (psf) map for {\tt wavdetect}, using the tool {\tt mkpsfmap}, we set the encircled counts fraction parameter {\tt ecf=0.9} (which helps to minimize the number of fainter sources with larger positional uncertainties recovered by {\tt wavdetect}\footnote{https://cxc.cfa.harvard.edu/ciao/threads/reproject\_aspect/}; the psf map was created at an energy of 2.3 keV). We compared the locations of all X-ray sources returned by {\tt wavdetect} to the positions of optical (point) sources in SDSS \citep{sdssdr12}
using the task {\tt wcs\_match}, after excluding X-ray regions within the primary optical galaxy (so as to avoid any X-ray sources within our target galaxies biasing our final astrometry). In three observations (obsID 20352, 20355, and 26289) we found 2-5 common X-ray/optical sources (each of which was verified by manual inspection), allowing us to perform an astrometric correction to the event files using {\tt wcs\_match} (typical corrections were less than 1.2 arcsec).
We used {\tt method=trans} to perform translational corrections, except for obsID 26289 we also performed a rotation and scale factor correction ({\tt method=rst}), since we had 5 common X-ray/optical matches for this observation.  If no common X-ray/optical sources were found, then we treat the 1$\sigma$ uncertainty on the positional accuracy as 0.8. arcsec\footnote{https://cxc.harvard.edu/cal/ASPECT/celmon/}

\begin{figure*}
    \centering
    \includegraphics[width=1\linewidth]{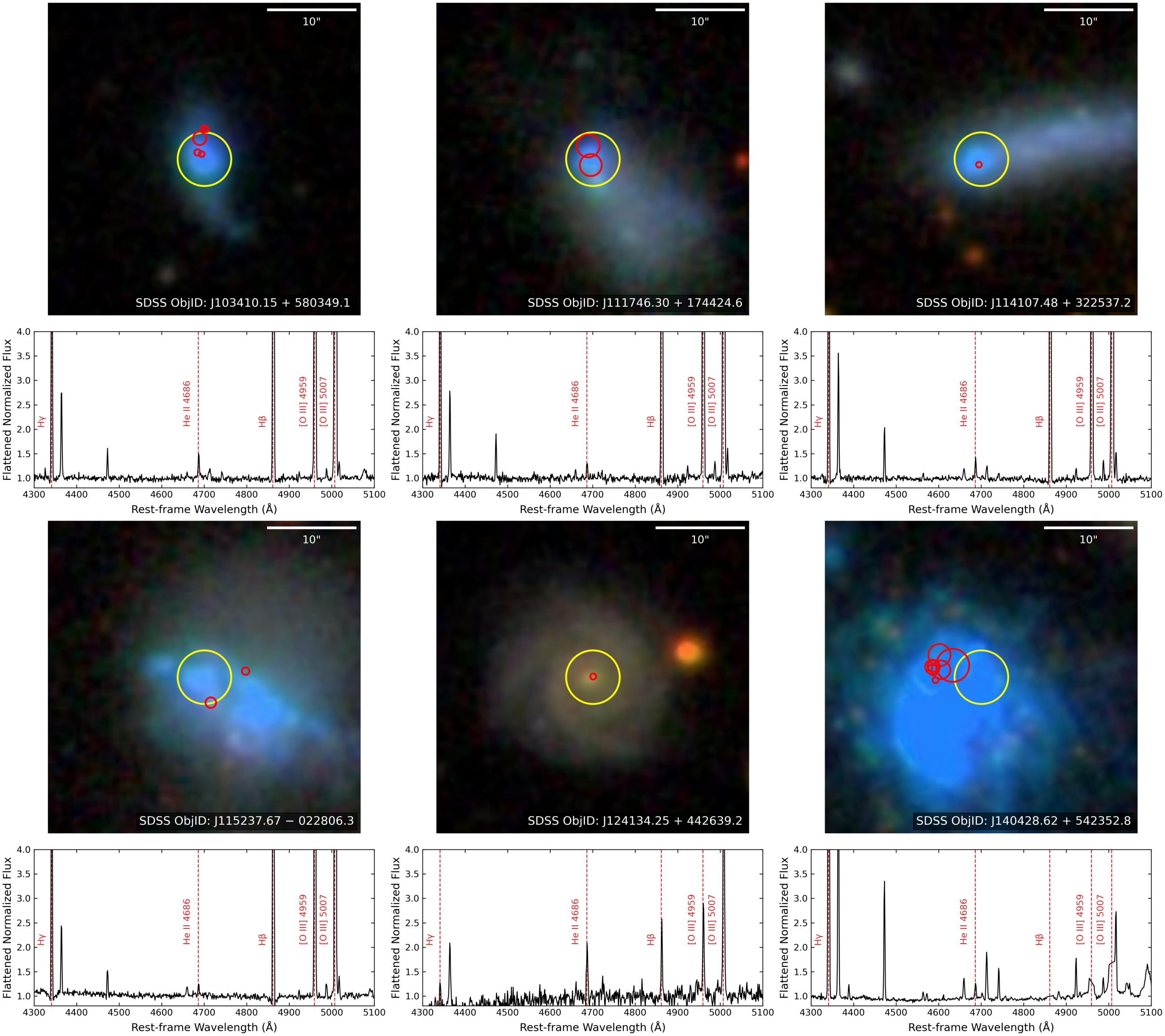}
    \caption{SDSS finding charts for our 7 targets with X-ray detections, ordered by increasing right ascension. Each top panel shows the SDSS image of a target galaxy \citep{sdssdr7}, with the position of the SDSS 3$\arcsec$ spectroscopic fiber indicated by the centered yellow circle. Red circles indicate the relative positions and positional uncertainty of Chandra X-ray detections within each galaxy (See Table \ref{xraytable}). The white scale bar in the top right of each plot represents the length of 10$\arcsec$. The bottom panels shows the normalized and flattened SDSS spectra with relevant emission lines labelled.}
    \label{fig:x-ray_chart}
\end{figure*}

For all observations, we next ran {\tt wavdetect} using the same parameters as above to search for X-ray sources within each galaxy, except we set {\tt ecf=0.393} when creating the psf map to allow sensitivity to fainter targets. We found 6 galaxies to host at least one X-ray source within a 20 arcsec radius of the optical center, which are summarized in Table \ref{xraytable} and pictured in Figure \ref{fig:x-ray_chart}. ObsIDs 20352 and 5836(S) were also initially identified as X-ray source detections, but were excluded after visual inspection revealed inconsistent spatial association with their hosts. Figure \ref{fig:x-ray_chart} also shows which X-ray sources lie within the 3\arcsec\ SDSS spectroscopic fiber and the characteristically narrow He II line widths in their spectra. X-ray detections falling outside the fiber aperture may provide an indication of whether the X-ray emitting sources are physically associated with the regions producing the He II emission. 

We ran {\tt srcflux} on all observations to measure the number of net counts for each X-ray detection, using a source aperture with a radius set to encompass 90\% of the psf (the psf fraction was calculated using the {\tt psfmethod=arfcorr} method). Background regions are defined by an annulus 1.7-5 times the radius of the source aperture. For galaxies with one X-ray source, we estimated the local sky background near each source using annuli centered on each source with outer radii 5 times larger than the source aperture. However, for galaxies with two sources, we manually created 30$\arcsec$ circular regions on a nearby, source-free area of the sky to estimate the local background (to avoid the background region from including counts from the other source). Spectra were extracted for all X-ray detections with $>$15 counts, using the tool {\tt specextract} and adopting the above apertures.

X-ray fluxes were measured via spectral fitting for the observations with $>$15 counts (16 spectra), as described in the next subsection. For objects with too few counts for spectral fitting (4 spectra), and for X-ray non-detections (23 spectra), we estimated unabsorbed model fluxes (or upper limits) using {\tt srcflux}. Model fluxes were estimated for both a {\tt powerlaw} model (with $\Gamma=1.7$ for all but NGC 5471 which used $\Gamma=4.2$) and {\tt diskbb} model (with inner disk temperature $kT_{\rm in}=1.2$ keV; NGC 5471 used $kT_{\rm in}=0.2$), modified by absorption that was set to the Galactic column density along the line of sight. Typical flux limits range from $F_{\rm0.5-10\,keV}$ of 10$^{-13}-10^{-15}$ erg s$^{-1}$ cm$^{-2}$. See Section \ref{sec:srcflux} for details on our choice of spectral parameters.

\begin{deluxetable*}{ccccccccc}
\tabletypesize{\footnotesize}
\tablecaption{X-ray Observation Properties \label{xraytable}}
\tablewidth{0pt}
\tablehead{
\colhead{ObsID} &
\colhead{Target} &
\colhead{Exposure} &
\colhead{Date} &
\colhead{RA} &
\colhead{DEC} &
\colhead{Chip} &
\colhead{$\rm P_{error}$} &
\colhead{Nuclear}\\
\colhead{} &
\colhead{(SDSS)} &
\colhead{(ks)} &
\colhead{(MJD)} &
\colhead{(HH:MM:SS)} &
\colhead{(DD:MM:SS)} &
\colhead{} &
\colhead{(arcsec)} &
\colhead{}
}
\colnumbers
\startdata
20350 & $\rm J000953.09+154404.8$ & 8.0 & 58265.26238 & \nodata & \nodata & S3 & \nodata & \nodata \\
20349 & $\rm J021513.98-084624.3$ & 11.8 & 58396.52322 & \nodata & \nodata & S3 & \nodata & \nodata \\
16985 & $\rm J083743.48+513830.2$ & 2.4 & 57027.57378 & \nodata & \nodata & S3 & \nodata & \nodata \\
11289 & $\rm J101624.51+375445.9$ & 9.4 & 55220.53179 & \nodata & \nodata & S3 & \nodata & \nodata\\
3347  & $\rm J103410.15+580349.1$ & 38.5 & 52396.59476 & 10:34:10.22 & 58:03:46.74 & I1 & 0.74 & N \\
18059 (N) & $\rm J103410.15+580349.1$ & 5.0 & 57413.74896 & 10:34:10.25 & 58:03:48.32 & S3 & 0.35 & N \\
18059 (S) & $\rm J103410.15+580349.1$ & 5.0 & 57413.74896 & 10:34:10.16 & 58:03:45.66 & S3 & 0.36 & N \\
26289 (N) & $\rm J103410.15+580349.1$ & 49.4 & 59935.72262 & 10:34:10.19 & 58:03:48.50 & S3 & 0.31 & Y \\
26289 (S) & $\rm J103410.15+580349.1$ & 49.4 & 59935.72262 & 10:34:10.13 & 58:03:45.72 & S3 & 0.31 & N \\
12808 & $\rm J104653.98+134645.7$ & 44.5 & 55590.64375 & \nodata & \nodata & S2 & \nodata & \nodata \\
20352 & $\rm J105310.82+501653.2$ & 8.0 & 58180.46737 & \nodata & \nodata & S3 & \nodata & \nodata \\ 
20353 & $\rm J110458.30+290816.5$ & 8.0 & 58228.74075 & \nodata & \nodata & S3 & \nodata & \nodata \\
4933  & $\rm J111746.30+174424.6$ & 18.8 & 53174.61866 & 11:17:46.34 & 17:44:23.20 & I3 & 1.28 & Y \\
5836 (N) & $\rm J111746.30+174424.6$ & 45.2 & 53416.97654 & 11:17:46.32 & 17:44:25.35 & I1 & 1.23 & Y \\
5836 (S) & $\rm J111746.30+174424.6$ & 45.2 & 53416.97654 & \nodata & \nodata & I1 & \nodata & \nodata \\ 
20355 & $\rm J114107.48+322537.2$ & 8.0 & 58180.46737 & 11:41:07.51 & 32:25:37.83 & S3 & 0.31 & Y \\
7135 (N) & $\rm J115237.67-022806.3$ & 5.1 & 53777.56385 & 11:52:37.63 & $-02$:28:03.54 & S3 & 0.59 & N \\
7135 (S) & $\rm J115237.67-022806.3$ & 5.1 & 53777.56385 & 11:52:37.37 & $-02$:28:07.07 & S3 & 0.41 & N \\
20354 & $\rm J115441.22+463636.3$ & 8.0 & 58237.91103 & \nodata & \nodata & S3 & \nodata & \nodata \\
942 & $\rm J121749.30+375155.5$ & 49.2 & 51684.53344 & \nodata & \nodata & S4 & \nodata & \nodata \\
17550 & $\rm J122225.79+043404.7$ & 19.8 & 56977.63071 & \nodata & \nodata & S2 & \nodata & \nodata \\
10560 & $\rm J122615.70+482938.4$ & 3.1 & 55028.74413 & \nodata & \nodata & S3 & \nodata & \nodata \\
10729 & $\rm J124134.25+442639.2$ & 9.5 & 55022.08569 & 12:41:34.25 & 44:26:39.17 & S3 & 0.33 & Y \\
934 & $\rm J140411.23+542518.7$ & 98.4 & 51629.01454 & \nodata & \nodata & I3 & \nodata & \nodata \\
4732 & $\rm J140411.23+542518.7$ & 69.8 & 53083.34405 & \nodata & \nodata & I3 & \nodata & \nodata \\
5309 & $\rm J140411.23+542518.7$ & 70.8 & 53078.07040 & \nodata & \nodata & I3 & \nodata & \nodata \\
2065 & $\rm J140428.62+542352.8$ & 9.6 & 51846.33722 & \nodata & \nodata & S2 & \nodata & \nodata \\
2779 & $\rm J140428.62+542352.8$ & 14.3 & 52578.72816 & 14:04:29.22 & 54:23:53.11 & S3 & 0.33 & N \\
4731 & $\rm J140428.62+542352.8$ & 56.2 & 53023.24799 & 14:04:29.17 & 54:23:50.31 & I2 & 1.24 & N\\
4733 & $\rm J140428.62+542352.8$ & 24.8 & 53132.56225 & \nodata & \nodata & S4 & \nodata & \nodata \\
4736 & $\rm J140428.62+542352.8$ & 77.4 & 53310.77997 & 14:04:29.26 & 54:23:51.63 & S2 & 0.81 & N \\
4737 & $\rm J140428.62+542352.8$ & 21.9 & 53371.60480 & \nodata & \nodata & S2 & \nodata & \nodata \\
5296 & $\rm J140428.62+542352.8$ & 3.1 & 53025.45652 & \nodata & \nodata & I2 & \nodata & \nodata \\
5297 & $\rm J140428.62+542352.8$ & 21.7 & 53028.06531 & 14:04:29.00 & 54:23:51.46 & I2 & 1.85 & Y \\
5322 & $\rm J140428.62+542352.8$ & 64.7 & 53128.32263 & \nodata & \nodata & S4 & \nodata & \nodata \\
5323 & $\rm J140428.62+542352.8$ & 42.6 & 53134.13301 & \nodata & \nodata & S4 & \nodata & \nodata \\
6152 & $\rm J140428.62+542352.8$ & 44.1 & 53316.29438 & 14:04:29.15 & 54:23:51.96 & S2 & 1.04 & N \\
6169 & $\rm J140428.62+542352.8$ & 29.4 & 53369.09087 & \nodata & \nodata & S2 & \nodata & \nodata \\
6170 & $\rm J140428.62+542352.8$ & 47.9 & 53361.05414 & 14:04:29.25 & 54:23:51.94 & S2 & 0.61 & N \\
6175 & $\rm J140428.62+542352.8$ & 40.7 & 53363.69327 & 14:04:29.25 & 54:23:51.54 & S2 & 0.63 & N \\
12894 & $\rm J142628.16+382258.6$ & 5.0 & 55890.98244 & \nodata & \nodata & S2 & \nodata & \nodata \\
20351 & $\rm J144852.02+344242.9$ & 9.9 & 58283.00823 & \nodata & \nodata & S3 & \nodata & \nodata \\
3962 & $\rm J222510.13-001152.8$ & 3.5 & 52892.97279 & \nodata & \nodata & S3 & \nodata & \nodata \\
\enddata
\tablecomments{Chandra X-ray Observation Properties. Column 1: Chandra Observation ID. Column 2: SDSS galaxy name. Column 3: Chandra exposure time in kiloseconds. Column 4: Chandra observation date (Modified Julian Date). Column 5: Right ascension of the X-ray source. Column 6: Declination of the X-ray source. Column 7: ACIS Chip. Column 8: Positional error of the X-ray source in arcseconds. Column 9: An X-ray source is labelled as nuclear if the the target galaxy's coordinates lie within three times the positional error of the X-ray detections. 
}
\end{deluxetable*}

\subsection{Inferred EUV Continuum from X-ray Spectral Fitting and Model Fluxes} \label{sec:srcflux}
For X-ray sources with $>$15 counts we performed X-ray spectral fitting with the Interactive Spectral Interpretation System ({\tt ISIS}) v1.6.2.\citep{Houck-DeNicola2000}.  Given the relatively small number of counts from our X-ray detected sources ($\approx$16--500), spectral fitting was performed using Cash statistics \citep{Cash79} after grouping each spectrum to a minimum of one count per bin. We applied two models to each spectrum, an absorbed powerlaw ({\tt tbabs*powerlaw}) and an absorbed multi-temperature accretion disk ({\tt tbabs*diskbb}), allowing the column density $N_{\rm H}$ to vary as a free parameter in each model. Visual examination of the fit residuals suggested that both models provided acceptable fits to all but seven of our spectra. The seven spectra with unacceptable fits (obsIDs 2779, 4731, 4736, 5297, 6152, 6170, 6175) are all of the same source, within SDSS J140428.62+542352.8, and they are discussed later in this section. 

Excluding the seven spectra of SDSS J140428.62+542352.8, and considering that both models contain the same number of free parameters, we consulted the difference between the Cash statistic for the two fits, $\Delta C = C_{\rm powerlaw} - C_{\rm diskbb}$, to assess if there is a statistical preference for one model. In all but three cases $\Delta C < 11.3$, implying similar quality fits within 99\% confidence for three free parameters. For these three spectra (obsIDs 3347, 4736, and 10279) $\Delta C \approx 15$, so the preference for one model is not exceptionally strong. We therefore report results from both the powerlaw and accretion disk models in Table \ref{tab:detectionpropertiestable}.

Regarding the seven spectra of SDSS J140428.62+542352.8, we found that two spectra in particular (obsIDs 6152 and 6170) showed a hard X-ray excess ($\gtrsim$2 keV), especially when comparing the data to the accretion disk model. For these two spectra, we obtained reasonable fits by applying a disk plus powerlaw model, {\tt tbabs*(diskbb+powerlaw)}, and keeping the column density frozen to the Galactic value, $N_{\rm H}=1.1\times10^{20}$ cm$^{-2}$. For the other five spectra of this source, the best-fit residuals do not require the addition of an extra powerlaw component (nor do they suggest a large column density), so we adopt the {\tt tbabs*diskbb} results for those five spectra (obsIDs 2779, 4731, 4736, 5297, and 6175; allowing $N_{\rm H}$ to vary as a free parameter for these five spectra). It is unclear if the lack of a required powerlaw component for these five spectra is reflecting intrinsic variability, or if it is simply an artifact of fewer counts. Regardless, the best-fit inner disk temperature of these five spectra is similar to the inner disk temperature for the two requiring an extra powerlaw component ($KT_{\rm in}\sim0.2$ keV).

On the other hand, when applying the {\tt tbabs*powerlaw} model to the seven spectra of SDSS J140428.62+542352.8, all fits converged to relatively large photon indices ($\Gamma\gtrsim 4$) and non-zero values of $N_{\rm H}$. Since the steep $\Gamma$ is phenomenological, and possibly influenced by a degeneracy between $\Gamma$ and $N_{\rm H}$ during spectral fitting, we opted to freeze the column density to the Galactic value $N_{\rm H}=1.1\times10^{20}$ cm$^{-2}$ when fitting a powerlaw model to this source. For completeness, in Table \ref{tab:detectionpropertiestable} we report results from the powerlaw model for all seven spectra of SDSS J140428.62+542352.8 with $N_{\rm H}$ frozen, since it provides a statistically reasonable fit. However, in light of the two spectra of this source that require a more complex model, we stress that even when $N_{\rm H}$ is frozen, the powerlaw model is likely less physically relevant compared to the accretion disk model for this particular source. Excluding SDSS J140428.62+542352.8, we find an average $\Gamma=1.7$ and $kT_{\rm in}=1.2$ keV, hence our choice of those model parameters when estimating fluxes/upper limits using {\tt srcflux} for lower-count observations in Section~\ref{sec:data}. For SDSS J140428.62+542352.8, the seven fit values have an average $\Gamma= 4.10$ and $kT_{\rm in}=0.19$ keV.

Finally, we measured unabsorbed model fluxes (0.5-10 keV) for all spectra using the {\tt cflux} convolution model in {\tt ISIS}. Unabsorbed fluxes are reported for both the {\tt powerlaw} and {\tt diskbb}/{\tt diskbb+powerlaw} variants of each model in Table \ref{detectionresultstable}. 

\setlength{\tabcolsep}{3pt}
\begin{deluxetable*}{ccccc|ccc|ccc}
\tabletypesize{\footnotesize}
\tablecaption{X-Ray Detection Properties and Parameters \label{tab:detectionpropertiestable}}
\tablehead{
\colhead{} &
\colhead{} &
\colhead{} &
\colhead{} &
\colhead{} \vline &
\multicolumn{3}{c|}{Powerlaw} &
\multicolumn{3}{c}{Diskbb} \\[2pt]
\colhead{ObsID} &
\colhead{Target} &
\colhead{C$_{\rm total}$} &
\colhead{C$_{\rm bkg}$} &
\colhead{Net Rate} \vline &
\colhead{Cash/d.o.f.} &
\colhead{$\Gamma$} &
\colhead{nH} \vline &
\colhead{Cash/d.o.f.} &
\colhead{kT} &
\colhead{nH} \\[2pt]
\colhead{} &
\colhead{(SDSS)} &
\colhead{} &
\colhead{} &
\colhead{($\rm ks^{-1}$)} \vline &
\colhead{} &
\colhead{} &
\colhead{($10^{22}\,\mathrm{cm^{-2}}$)} \vline &
\colhead{} &
\colhead{(eV)} &
\colhead{($10^{22}\,\mathrm{cm^{-2}}$)} \\
\colhead{(1)} &
\colhead{(2)} &
\colhead{(3)} &
\colhead{(4)} &
\colhead{(5)} \vline &
\colhead{(6)} &
\colhead{(7)} &
\colhead{(8)} \vline &
\colhead{(9)} &
\colhead{(10)} &
\colhead{(11)}
}
\startdata
3347 & $\rm J103410.15+580349.1$ & 442 & 6.5 & $13.0^{+1.0}_{-1.0}$ & $124.8/143$ & $1.62^{+0.07}_{-0.06}$ & $<0.15$ & $140.7/143$ & $1.29^{+0.11}_{-0.09}$ & $<0.01$ \\
18059 (N) & $\rm J103410.15+580349.1$ & 38 & 0.1 & $8.2^{+2.4}_{-2.0}$ & $18.8/15$ & $1.17^{+0.23}_{-0.22}$ & $<0.60$ & $18.3/15$ & $1.60^{+1.32}_{-0.51}$ & $<0.34$ \\
18059 (S) & $\rm J103410.15+580349.1$ & 33 & 0.1 & $7.1^{+2.3}_{-1.9}$ & $9.0/13$ & $1.64^{+0.25}_{-0.23}$ & $<0.47$ & $9.8/13$ & $1.18^{+0.48}_{-0.27}$ & $<0.16$ \\
26289 (N) & $\rm J103410.15+580349.1$ & 340 & 0.3 & $7.5^{+0.7}_{-0.7}$ & $127.5/117$ & $2.07^{+0.06}_{-0.06}$ & $0.21^{+0.14}_{-0.13}$ & $128.8/117$ & $1.24^{+0.11}_{-0.04}$ & $<0.09$ \\
26289 (S) & $\rm J103410.15+580349.1$ & 222 & 0.3 & $4.9^{+0.5}_{-0.5}$ & $96.7/82$ & $1.82^{+0.08}_{-0.08}$ & $<0.52$ & $100.6/82$ & $1.33^{+0.15}_{-0.13}$ & $<0.08$ \\
4933 & $\rm J111746.30+174424.6$ & 16 & 1.1 & $0.9^{+0.5}_{-0.4}$ & $4.2/5$ & $1.72^{+0.45}_{-0.38}$ & $<0.40$ & $5.6/5$ & $1.23^{+0.85}_{-0.44}$ & $<0.20$ \\
05836 (N) & $\rm J111746.30+174424.6$ & 15 & 2.6 & $0.3^{+0.2}_{-0.1}$ & \nodata & 1.7* & 0.016* & \nodata & 1.2* & 0.016* \\
07135 (N) & $\rm J115237.67-022806.3$ & 8 & 0.1 & $29.2^{+3.4}_{-3.4}$ & \nodata & 1.7* & 0.023* & \nodata & 1.2* & 0.023* \\
07135 (S) & $\rm J115237.67-022806.3$ & 34 & 0.1 & $1.6^{+1.2}_{-0.8}$ & $9.2/13$ & $1.50^{+0.25}_{-0.23}$ & $<1.45$ & $9.4/13$ & $1.82^{+1.30}_{-0.65}$ & $0.15^{+0.23}_{-0.14}$ \\
20355 & $\rm J114107.48+322537.2$ & 201 & 0.5 & $7.1^{+2.2}_{-1.8}$ & $81.8/85$ & $2.15^{+0.17}_{-0.15}$ & $0.17^{+0.11}_{-0.10}$ & $89.0/85$ & $1.14^{+0.12}_{-0.05}$ & $<0.06$ \\
10729 & $\rm J124134.25+442639.2$ & 86 & 0.1 & $10.5^{+1.9}_{-1.9}$ & $55.1/33$ & $3.17^{+0.31}_{-0.27}$ & $<0.35$ & $69.4/33$ & $0.35^{+0.05}_{-0.04}$ & $<0.04$ \\
2779 & $\rm J140428.62+542352.8$ & 87 & 0.8 & $7.0^{+1.3}_{-1.3}$ & $29.6/26$ & $3.78^{+0.28}_{-0.29}$ & 0.011* & $26.7/25$ & $0.19^{+0.03}_{-0.03}$ & $<0.29$ \\
4731 & $\rm J140428.62+542352.8$ & 170 & 11.6 & $3.2^{+0.4}_{-0.4}$ & $60.5/45$ & $4.14^{+0.23}_{-0.23}$ & 0.011* & $52.3/44$ & $0.19^{+0.03}_{-0.02}$ & $<0.31$ \\
4736 & $\rm J140428.62+542352.8$ & 279 & 15.4 & $3.9^{+0.4}_{-0.4}$ & $68.0/58$ & $4.22^{+0.17}_{-0.18}$ & 0.011* & $53.3/57$ & $0.19^{+0.02}_{-0.02}$ & $<0.23$ \\
5297 & $\rm J140428.62+542352.8$ & 69 & 6.8 & $3.3^{+0.7}_{-0.7}$ & $33.9/22$ & $4.32^{+0.37}_{-0.39}$ & 0.011* & $27.9/21$ & $0.17^{+0.02}_{-0.00}$ & $<0.33$ \\
6152 & $\rm J140428.62+542352.8$ & 129 & 12.3 & $3.9^{+0.6}_{-0.6}$ & $55.3/42$ & $1.23^{+1.09}_{-1.35}$** & 0.011* & $46.4/40$ & $0.20^{+0.03}_{-0.02}$ & 0.011* \\
6170 & $\rm J140428.62+542352.8$ & 130 & 3.4 & $3.0^{+0.5}_{-0.5}$ & $44.0/53$ & $2.77^{+1.09}_{-2.35}$** & 0.011* & $38.8/51$ & $0.17^{+0.02}_{-0.02}$ & 0.011* \\
6175 & $\rm J140428.62+542352.8$ & 112 & 2.7 & $3.1^{+0.5}_{-0.5}$ & $37.2/40$ & $4.39^{+0.26}_{-0.26}$ & 0.011* & $34.7/39$ & $0.20^{+0.02}_{-0.02}$ & $<0.11$ \\
\enddata
\tablecomments{X-Ray Detection Properties and Parameters. Column 1: Chandra Observation ID. Column 2: SDSS galaxy name. Column 3: Total X-ray counts. Column 4: X-ray background counts. Column 5: Background-subtracted X-ray net count rate. Column 6: Cash statistic/Degrees of Freedom for the powerlaw model. Column 7: Powerlaw photon index. Column 8: Powerlaw model galactic neutral hydrogen column density. Column 9: Cash statistic/Degrees of Freedom for Disk blackbody (diskbb) model. Column 10: Diskbb thermal temperature. Column 11: Diskbb galactic neutral hydrogen column density.}
\tablenotetext{*}{Parameter fixed during model fits.}
\tablenotetext{**}{ Fit with a composite diskbb plus powerlaw model, tbabs*(diskbb+powerlaw), keeping the galactic neutral hydrogen column density frozen.}
\end{deluxetable*}

\begin{deluxetable*}{cc|cc|cc}
\tabletypesize{\footnotesize}
\tablecaption{X-Ray Detection Flux and Luminosity \label{detectionresultstable}}
\tablehead{
\colhead{} &
\colhead{} \vline &
\multicolumn{2}{c|}{Powerlaw} &
\multicolumn{2}{c}{Disk BB} \\
\colhead{ObsID} &
\colhead{Target} \vline &
\colhead{$\log F_{0.5-10\ \mathrm{keV}}$} &
\colhead{$\log L_{\rm X,PowerLaw}$} \vline &
\colhead{$\log F_{0.5-10\ \mathrm{keV}}$} &
\colhead{$\log L_{\rm X,Diskbb}$} \\
\colhead{} &
\colhead{(SDSS)} \vline &
\colhead{($\rm erg\ s^{-1}cm^{-2}$)} &
\colhead{($\rm erg\ s^{-1}$)} \vline &
\colhead{($\rm erg\ s^{-1}cm^{-2}$)} &
\colhead{($\rm erg\ s^{-1}$)} \\
\colhead{(1)} &
\colhead{(2)} \vline &
\colhead{(3)} &
\colhead{(4)} \vline &
\colhead{(5)} &
\colhead{(6)} 
}
\startdata
3347 & $\rm J103410.15+580349.1$ & $-12.72^{+0.03}_{-0.03}$ & $40.37^{+0.04}_{-0.04}$ & $-12.84^{+0.04}_{-0.02}$ & $40.25^{+0.04}_{-0.02}$ \\
18059 (N) & $\rm J103410.15+580349.1$ & $-12.83^{+0.14}_{-0.14}$ & $40.26^{+0.14}_{-0.13}$ & $-13.03^{+0.20}_{-0.14}$ & $40.06^{+0.20}_{-0.13}$ \\
18059 (S) & $\rm J103410.15+580349.1$ & $-13.05^{+0.12}_{-0.11}$ & $40.04^{+0.11}_{-0.09}$ & $-13.19^{+0.12}_{-0.10}$ & $39.90^{+0.12}_{-0.10}$ \\
26289 (N) & $\rm J103410.15+580349.1$ & $-12.81^{+0.06}_{-0.04}$ & $40.28^{+0.06}_{-0.05}$ & $-12.98^{+0.02}_{-0.03}$ & $40.11^{+0.02}_{-0.03}$ \\
26289 (S) & $\rm J103410.15+580349.1$ & $-13.03^{+0.05}_{-0.04}$ & $40.06^{+0.04}_{-0.04}$ & $-13.16^{+0.04}_{-0.03}$ & $39.93^{+0.04}_{-0.03}$ \\
4933 & $\rm J111746.30+174424.6$ & $-13.84^{+0.19}_{-0.18}$ & $38.89^{+0.19}_{-0.17}$ & $-13.96^{+0.26}_{-0.17}$ & $38.77^{+0.26}_{-0.17}$ \\
05836 (N) & $\rm J111746.30+174424.6$ & $-14.38^{+0.58}_{-0.31}$ & $38.34^{+0.59}_{-0.32}$ & $-14.35^{+0.58}_{-0.31}$ & $38.38^{+0.58}_{-0.32}$ \\
07135 (N) & $\rm J115237.67-022806.3$ & $-13.83^{+0.61}_{-0.29}$ & $39.07^{+0.62}_{-0.30}$ & $-13.75^{+0.61}_{-0.30}$ & $39.15^{+0.61}_{-0.27}$ \\
07135 (S) & $\rm J115237.67-022806.3$ & $-12.93^{+0.15}_{-0.12}$ & $39.51^{+0.15}_{-0.12}$ & $-13.08^{+0.26}_{-0.15}$ & $39.36^{+0.26}_{-0.15}$ \\
20355 & $\rm J114107.48+322537.2$ & $-12.36^{+0.05}_{-0.05}$ & $40.08^{+0.06}_{-0.04}$ & $-12.53^{+0.04}_{-0.03}$ & $39.91^{+0.04}_{-0.03}$ \\
10729 & $\rm J124134.25+442639.2$ & $-13.26^{+0.20}_{-0.08}$ & $41.35^{+0.20}_{-0.09}$ & $-13.38^{+0.04}_{-0.05}$ & $41.23^{+0.04}_{-0.05}$ \\
2779 & $\rm J140428.62+542352.8$ & $-13.51^{+0.05}_{-0.05}$ & $37.83^{+0.05}_{-0.05}$ & $-13.57^{+0.42}_{-0.08}$ & $37.77^{+0.42}_{-0.08}$ \\
4731 & $\rm J140428.62+542352.8$ & $-13.36^{+0.04}_{-0.03}$ & $37.98^{+0.04}_{-0.03}$ & $-13.38^{+0.15}_{-0.08}$ & $37.96^{+0.15}_{-0.08}$ \\
4736 & $\rm J140428.62+542352.8$ & $-13.38^{+0.03}_{-0.03}$ & $37.96^{+0.03}_{-0.03}$ & $-13.44^{+0.18}_{-0.05}$ & $37.90^{+0.18}_{-0.05}$ \\
5297 & $\rm J140428.62+542352.8$ & $-13.38^{+0.06}_{-0.06}$ & $37.96^{+0.06}_{-0.06}$ & $-12.54^{+1.00}_{-0.69}$ & $38.80^{+1.00}_{-0.71}$ \\
6152 & $\rm J140428.62+542352.8$ & $-13.39^{+0.04}_{-0.04}$ & $37.95^{+0.04}_{-0.04}$ & $-13.33^{+0.07}_{-0.06}$ & $38.01^{+0.07}_{-0.06}$ \\
6170 & $\rm J140428.62+542352.8$ & $-13.38^{+0.04}_{-0.04}$ & $37.96^{+0.03}_{-0.04}$ & $-13.40^{+0.04}_{-0.04}$ & $37.94^{+0.04}_{-0.04}$ \\
6175 & $\rm J140428.62+542352.8$ & $-13.47^{+0.05}_{-0.04}$ & $37.87^{+0.05}_{-0.04}$ & $-13.56^{+0.10}_{-0.05}$ & $37.78^{+0.10}_{-0.05}$ \\
\enddata
\tablecomments{
X-Ray Detection Flux and Luminosity.
Column 1: Chandra Observation ID.
Column 2: SDSS galaxy name.
Column 3: 0.5--10 keV powerlaw model X-ray flux (log).
Column 4: 0.5--10 keV powerlaw model X-ray luminosity (log).
Column 5: 0.5--10 keV diskbb model X-ray flux (log).
Column 6: 0.5--10 keV diskbb model X-ray luminosity (log).
}
\end{deluxetable*}

We infer the continuum of hard ionizing EUV photons associated with this X-ray emission by extrapolating the best fit model into the extreme ultraviolet, and we measure fluxes and photon fluxes from 0.054--0.3 keV using the {\tt cflux} and the {\tt cpflux} convolution models, respectively. Here 0.054 keV is the minimum energy required to doubly ionize helium, and the upper limit is given by the ionization cross-section, which decreases approximately as E$^{-3}$ for photons significantly above the ionization edge, meaning that photons much beyond $\approx$0.3 keV contribute progressively less to He II ionization.

In cases of X-ray non-detections, we calculated 0.054--0.3 keV photon flux upper limits by multiplying the (unabsorbed) 0.5-10 keV flux upper limits (see Section \ref{sec:data}) by bandpass corrections ($1.2 \times 10^{9}$ for the {\tt powerlaw} model assuming $\Gamma=1.7$, and $1.9 \times 10^{8}$ for the {\tt diskbb} model, assuming $KT_{\rm in}=1.2$ keV). Upper limits of unabsorbed fluxes for galaxies with no detections are reported for both the {\tt powerlaw} and {\tt diskbb}/{\tt diskbb+powerlaw} variants of each model in Table \ref{nondetectiontable}. 

\begin{deluxetable*}{cc|cccc}
\tabletypesize{\footnotesize}
\tablecaption{Upper Limits for Galaxies with No X-Ray Sources Detected \label{nondetectiontable}}
\tablehead{
\colhead{ObsID} &
\colhead{Target} \vline &
\colhead{Net Rate} &
\colhead{$\log L_{\rm X,PowerLaw}$} &
\colhead{$\log L_{\rm X,Diskbb}$} \\
\colhead{} &
\colhead{(SDSS)} \vline &
\colhead{($ks^{-1}$)} &
\colhead{($\rm erg\ s^{-1}$)} &
\colhead{($\rm erg\ s^{-1}$)} \\
\colhead{(1)} &
\colhead{(2)} \vline &
\colhead{(3)} &
\colhead{(4)} &
\colhead{(5)}
}
\startdata
20350 & $\rm J000953.09+154404.8$ & 0.33 & 37.98 & 37.89 \\
20349 & $\rm J021513.98-084624.3$ & 0.22 & 38.25 & 38.16 \\
16985 & $\rm J083743.48+513830.2$ & 2.09 & 38.47 & 38.43 \\
11289 & $\rm J101624.51+375445.9$ & 0.46 & 38.13 & 38.09 \\
12808 & $\rm J104653.98+134645.7$ & 0.09 & 38.85 & 38.78 \\
20352 & $\rm J105310.82+501653.2$ & 2.20 & 39.10 & 39.00 \\ 
20353 & $\rm J110458.30+290816.5$ & 0.34 & 37.66 & 37.58 \\
5836 (S) & $\rm J111746.30+174424.6$ & 0.90 & 38.99 & 38.92 \\ 
20354 & $\rm J115441.22+463636.3$ & 1.02 & 38.43 & 38.31 \\
942   & $\rm J121749.30+375155.5$ & 0.17 & 36.33 & 36.28 \\
17550 & $\rm J122225.79+043404.7$ & 0.22 & 38.22 & 38.14 \\
10560 & $\rm J122615.70+482938.4$ & 0.12 & 37.71 & 37.73 \\
934   & $\rm J140411.23+542518.7$ & 0.04 & 36.46 & 36.40 \\
4732  & $\rm J140411.23+542518.7$ & 0.07 & 36.41 & 36.34 \\
5309  & $\rm J140411.23+542518.7$ & 0.07 & 36.36 & 36.29 \\
2065  & $\rm J140428.62+542352.8$ & 5.75 & 38.09 & 38.03 \\
4733  & $\rm J140428.62+542352.8$ & 4.07 & 38.19 & 38.13 \\
4737  & $\rm J140428.62+542352.8$ & 1.32 & 37.73 & 37.67 \\
5296  & $\rm J140428.62+542352.8$ & 7.41 & 38.38 & 38.32 \\
5322  & $\rm J140428.62+542352.8$ & 3.89 & 38.18 & 38.12 \\
5323  & $\rm J140428.62+542352.8$ & 3.89 & 38.18 & 38.12 \\
6169  & $\rm J140428.62+542352.8$ & 1.26 & 37.62 & 37.57 \\
12894 & $\rm J142628.16+382258.6$ & 1.12 & 40.42 & 40.36 \\
20351 & $\rm J144852.02+344242.9$ & 0.27 & 37.59 & 37.51 \\
3962  & $\rm J222510.13-001152.8$ & 1.78 & 41.24 & 41.20 \\
\enddata
\tablecomments{
Column 1: Chandra Observation ID. Column 2: SDSS galaxy name. Column 3: X-ray net count rate. Column 4: X-ray power-law model luminosity (log). Column 5: X-ray diskbb model luminosity (log).
}
\end{deluxetable*}

\subsection{Optical Broadband Spectral Analysis for Host Ionization Environment} \label{sec:HeII}
In Section \ref{sec:srcflux} we inferred the hard ionizing EUV continuum being produced by observed accreting compact objects in our sample of galaxies with \textit{Chandra} X-ray observations. To further characterize the properties of the ionization environment we performed line fitting and modelling to predict the EUV continuum responsible for the observed \ion{He}{2} line luminosity, while also characterizing the host galaxy properties. 

\subsubsection{Required EUV continuum from Line fluxes}
\label{sec:spec_fitting}

We performed custom line fitting for each galaxy spectrum obtained from SDSS Data Release 18; \citep{Kollmeier19} to obtain line flux measurements of the \ion{He}{2} $\lambda4686$, [\ion{N}{2}] $\lambda6584$, [\ion{O}{3}] $\lambda4959$, [\ion{O}{3}] $\lambda5007$, [\ion{O}{2}] $\lambda3727$, $H\alpha \lambda6563$, and $H\beta \lambda4863$ lines. First, we used the redshift provided by the SDSS metadeta to place each spectrum into the rest frame, and we then restricted each spectrum to the wavelength range of 250~\AA\ around the desired emission line to maximize signal-to-noise when measuring the continuum. We then fit a second degree polynomial to the continuum using the {\tt SpecUtils} package in {\tt astropy} \citep{astropy:2013, astropy:2018, astropy:2022}.

After subtracting out the continuum, we used the Python package {\tt SciPy} \citep{Virtanen2020} to fit a single Gaussian to the desired line. Errors on the line flux were estimated by propagating uncertainties on the amplitude and standard deviation of the best-fit Gaussian. The emission line flux for the \ion{He}{2} is reported in Table \ref{galaxytable}. As a cross-check of our fitting method we found our derived line ratio $\log$ \ion{He}{2}/H$\beta$ is consistent with the ratio reported by \citet{Shirazi12}. 

We next take advantage of the photon-counting property of the \ion{He}{2} line to infer the unobservable EUV flux required following the logic  developed by \cite{Pakull86}. The total recombination rate of {He}$^{++}$ to {He}$^+$ is given by  $R_{\text{rec}} = n_e n_{\text{He}^{++}} \alpha_{B}$ where $n_e$ is the electron density, $n_{\text{He}^{++}}$ is the number density of {He}$^{++}$ ions, and $\alpha_{B}$ is the recombination coefficient for {He}$^+$ (the probability of an electron recombining with {He}$^{++}$ per unit volume per unit time to form {He}$^{+}$). The fraction of recombinations that lead from the n = 4 to the n = 3 state is then given by $f_{4686} = \alpha_{4686}/\alpha_B$ where $\alpha_{4686}$ is the effective recombination coefficient leading to the He II $\lambda$4686 transition.

Using empirical and theoretical values from the literature for a representative compact H II region at temperature $T_e$=$2 \cdot 10^4$ K with $100$ cm$^{-3}$ free electrons  ($\alpha_{4686}={1.713 \times 10^{-13} \text{ cm}^3\text{ s}^{-1}}$ and $\alpha_B={9.024 \times 10^{-13} \text{ cm}^3\text{ s}^{-1}}$, See \citealt{Storey1995}) we then approximate that $\approx$ 19 $\%$ of recombinations of {He}$^{++}$ produce a $\lambda$4686 photon. This fraction depends only weakly on the assumed  conditions, varying by $\sim 10$--$20\%$ over typical electron temperatures and densities, and does not dominate the uncertainty in our inferred ionizing flux. Hence, we find that every $\lambda$4686 photon emitted ($N_{\lambda4686,obs}$) requires $1/0.19 \approx 5.2$ ionizing EUV photons ($N_{54–300 eV}$) incident on the cloud of doubly ionized helium \citep{Pakull86}. We then predict the ionizing photon flux by multiplying the required He II photon flux, inferred from the optical spectra, by 5.2. The predicted ionizing photon flux that is producing the observed \ion{He}{2} is then compared to the extrapolated unabsorbed EUV photon flux estimated from the X-ray observations (Section \ref{sec:srcflux}) in Table \ref{ratiostable}.

Additional analysis was performed to obtain star formation rates (SFRs)  for the  galaxies in our sample, which we calculated using the $H\alpha$ indicator equation $SFR_{10} = 5.5 \times 10^{-42} \cdot L_{H\alpha} $ \citep{Calzetti2013}, providing the characteristic SFR of the last time-averaged  10 Myr \citep{Kennicutt1998}. Further analysis was completed in order to find the metallicity of the sample galaxies. \ion{N}{2} $\lambda6584$, \ion{O}{3} $\lambda4959$, \ion{O}{3} $\lambda5007$, and $H\beta$ emission fluxes were calculated using the aforementioned spectral fitting methods. The metallicities were then obtained using $12 + \log(O/H) = 0.950 \log\left(R_{23} - 0.08 O_{32}\right) + 6.805$ \citep{Izotov2019}, where $O_{32}$ is defined as the ratio of line fluxes $O_{32} = f_{5007}/f_{3727}$ and $R_{23} = (f_{3727} + f_{4959} + f_{5007})/{f_{H\beta}}$. The [OII] 3727 line flux is extrapolated using the relationship $F_{3727} = 20\cdot f_{6584}$.


\begin{deluxetable*}{c c c c c c}
\tabletypesize{\footnotesize}
\tablecaption{Comparison of Required He II $\lambda4686$ Photon Flux to Observed EUV Photon Flux \label{ratiostable}}

\tablehead{
\colhead{} &
\colhead{} \vline &
\multicolumn{2}{c}{Powerlaw} \vline &
\multicolumn{2}{c}{Diskbb} \\[2pt]
\colhead{ObsID} &
\colhead{Target} \vline &
\colhead{$\log N_{\rm EUV}$} &
\colhead{$\log(N_{\rm EUV}/N_{\lambda4686})$} \vline &
\colhead{$\log N_{\rm EUV}$} &
\colhead{$\log(N_{\rm EUV}/N_{\lambda4686})$} \\[2pt]
\colhead{} &
\colhead{(SDSS)} \vline &
\colhead{(${\rm s^{-1}\,cm^{-2}}$)} &
\colhead{} \vline &
\colhead{(${\rm s^{-1}\,cm^{-2}}$)} &
\colhead{} \\
\colhead{(1)} &
\colhead{(2)} \vline &
\colhead{(3)} &
\colhead{(4)} \vline &
\colhead{(5)} &
\colhead{(6)}
}

\startdata
20350 & $\rm J000953.09+154404.8$ & $<-5.23$ & $<-1.72$ & $<-6.13$ & $<-2.62$ \\
20349 & $\rm J021513.98-084624.3$ & $<-5.40$ & $<-1.72$ & $<-6.29$ & $<-2.61$ \\
16985 & $\rm J083743.48+513830.2$ & $<-4.56$ & $<-1.20$ & $<-5.42$ & $<-2.06$ \\
11289 & $\rm J101624.51+375445.9$ & $<-5.31$ & $<-2.35$ & $<-6.17$ & $<-3.21$ \\
3347  & $\rm J103410.15+580349.1$ & $-3.74^{+1.32}_{-0.13}$ & $-0.74^{+1.32}_{-0.14}$ & $-4.61^{+0.03}_{-0.03}$ & $-1.61^{+0.06}_{-0.06}$ \\
18059 (N) & $\rm J103410.15+580349.1$ & $-4.52^{+2.53}_{-0.41}$ & $-1.52^{+2.53}_{-0.41}$ & $-4.93^{+3.96}_{-0.13}$ & $-1.93^{+3.96}_{-0.14}$ \\
18059 (S) & $\rm J103410.15+580349.1$ & $-4.05^{+2.04}_{-0.40}$ & $-1.05^{+2.04}_{-0.41}$ & $-4.91^{+0.13}_{-0.13}$ & $-1.91^{+0.14}_{-0.14}$ \\
26289 (N) & $\rm J103410.15+580349.1$ & $-3.21^{+1.45}_{-0.32}$ & $-0.21^{+1.45}_{-0.32}$ & $-4.73^{+0.04}_{-0.05}$ & $-1.73^{+0.06}_{-0.07}$ \\
26289 (S) & $\rm J103410.15+580349.1$ & $-3.83^{+1.93}_{-0.23}$ & $-0.83^{+1.93}_{-0.23}$ & $-4.95^{+0.05}_{-0.05}$ & $-1.95^{+0.07}_{-0.07}$ \\
12808 & $\rm J104653.98+134645.7$ & $<-5.46$ & $<-1.74$ & $<-6.35$ & $<-2.63$ \\
20352 & $\rm J105310.82+501653.2$ & $<-4.15$ & $<-0.91$ & $<-5.34$ & $<-2.10$ \\
20353 & $\rm J110458.30+290816.5$ & $<-5.24$ & $<-2.03$ & $<-6.14$ & $<-2.93$ \\
4933  & $\rm J111746.30+174424.6$ & $-4.74^{+4.91}_{-0.64}$ & $-1.22^{+4.91}_{-0.65}$ & $-5.70^{+0.20}_{-0.21}$ & $-2.18^{+0.23}_{-0.24}$ \\
5836 (N) & $\rm J111746.30+174424.6$ & $-5.30^{+0.58}_{-0.31}$ & $-1.78^{+0.60}_{-0.33}$ & $-6.21^{+0.58}_{-0.31}$ & $-2.69^{+0.59}_{-0.33}$ \\
5836 (S) & $\rm J111746.30+174424.6$ & $<-5.45$ & $<-1.93$ & $<-5.93$ & $<-2.41$ \\ 
7135 (N) & $\rm J115237.67-022806.3$ & $-4.75^{+0.61}_{-0.30}$ & $-1.57^{+0.61}_{-0.30}$ & $-5.63^{+0.61}_{-0.30}$ & $-2.45^{+0.61}_{-0.30}$ \\
7135 (S) & $\rm J115237.67-022806.3$ & $-4.13^{+4.22}_{-0.92}$ & $-0.68^{+4.22}_{-0.93}$ & $-5.01^{+0.23}_{-0.22}$ & $-1.56^{+0.26}_{-0.26}$ \\
20355 & $\rm J114107.48+322537.2$ & $-2.70^{+0.29}_{-0.31}$ & $0.75^{+0.32}_{-0.34}$ & $-4.23^{+0.05}_{-0.05}$ & $-0.78^{+0.14}_{-0.14}$ \\
20354 & $\rm J115441.22+463636.3$ & $<-4.92$ & $<-1.39$ & $<-5.85$ & $<-2.33$ \\
942 & $\rm J121749.30+375155.5$ & $<-5.49$ & $<-1.96$ & $<-6.36$ & $<-2.83$ \\
17550 & $\rm J122225.79+043404.7$ & $<-5.30$ & $<-1.83$ & $<-6.20$ & $<-2.72$ \\
10560 & $\rm J122615.70+482938.4$ & $<-4.81$ & $<-1.92$ & $<-5.61$ & $<-2.72$ \\
10729 & $\rm J124134.25+442639.2$ & $-2.52^{+0.96}_{-0.26}$ & $0.74^{+0.97}_{-0.27}$ & $-4.25^{+0.11}_{-0.11}$ & $-0.99^{+0.14}_{-0.14}$ \\
934 & $\rm J140411.23+542518.7$ & $<-6.21$ & $<-2.66$ & $<-7.09$ & $<-3.54$ \\
4732 & $\rm J140411.23+542518.7$ & $<-5.85$ & $<-2.30$ & $<-6.74$ & $<-3.18$ \\
5309 & $\rm J140411.23+542518.7$ & $<-5.90$ & $<-2.35$ & $<-6.78$ & $<-3.23$ \\
2065 & $\rm J140428.62+542352.8$ & $<-4.17$ & $<-0.66$ & $<-5.04$ & $<-1.53$ \\
2779 & $\rm J140428.62+542352.8$ & $-1.91^{+0.39}_{-0.38}$ & $1.60^{+0.46}_{-0.46}$ & $-3.80^{+0.26}_{-0.24}$ & $-0.29^{+0.36}_{-0.35}$ \\
4731 & $\rm J140428.62+542352.8$ & $-1.38^{+0.26}_{-0.27}$ & $2.13^{+0.36}_{-0.37}$ & $-3.66^{+3.25}_{-0.24}$ & $-0.15^{+3.26}_{-0.35}$ \\
4733 & $\rm J140428.62+542352.8$ & $<-4.07$ & $<-0.56$ & $<-4.95$ & $<-1.44$ \\
4736 & $\rm J140428.62+542352.8$ & $-1.33^{+0.20}_{-0.21}$ & $2.18^{+0.32}_{-0.32}$ & $-3.75^{+3.28}_{-0.12}$ & $-0.24^{+3.29}_{-0.28}$ \\
4737 & $\rm J140428.62+542352.8$ & $<-4.53$ & $<-1.02$ & $<-5.41$ & $<-1.90$ \\
5296 & $\rm J140428.62+542352.8$ & $<-3.88$ & $<-0.37$ & $<-4.76$ & $<-1.25$ \\
5297 & $\rm J140428.62+542352.8$ & $-1.23^{+0.45}_{-0.45}$ & $2.28^{+0.51}_{-0.51}$ & $-2.14^{+0.24}_{-0.28}$ & $1.37^{+0.35}_{-0.38}$ \\
5322 & $\rm J140428.62+542352.8$ & $<-4.08$ & $<-0.57$ & $<-4.96$ & $<-1.44$ \\
5323 & $\rm J140428.62+542352.8$ & $<-4.08$ & $<-0.57$ & $<-4.95$ & $<-1.44$ \\
6152 & $\rm J140428.62+542352.8$ & $-1.97^{+0.28}_{-0.27}$ & $1.54^{+0.38}_{-0.37}$ & $-3.83^{+0.18}_{-0.14}$ & $-0.32^{+0.31}_{-0.29}$ \\
6169 & $\rm J140428.62+542352.8$ & $<-4.64$ & $<-1.13$ & $<-5.51$ & $<-2.00$ \\
6170 & $\rm J140428.62+542352.8$ & $-1.34^{+0.25}_{-0.24}$ & $2.17^{+0.35}_{-0.35}$ & $-3.29^{+0.72}_{-0.40}$ & $0.22^{+0.76}_{-0.47}$ \\
6175 & $\rm J140428.62+542352.8$ & $-1.25^{+0.30}_{-0.30}$ & $2.26^{+0.39}_{-0.39}$ & $-3.92^{+0.16}_{-0.12}$ & $-0.41^{+0.30}_{-0.28}$ \\
12894 & $\rm J142628.16+382258.6$ & $<-4.54$ & $<-0.87$ & $<-5.43$ & $<-1.75$ \\
20351 & $\rm J144852.02+344242.9$ & $<-5.34$ & $<-1.70$ & $<-6.24$ & $<-2.60$ \\
3962 & $\rm J222510.13-001152.8$ & $<-4.70$ & $<-1.06$ & $<-5.57$ & $<-1.92$ \\
\enddata
\tablecomments{Comparison of inferred He II $\lambda4686$ Photon Flux to extrapolated EUV Photon Flux. Column 1: Chandra Observation ID. Column 2: SDSS galaxy name. Column 3: Extrapolated X-ray powerlaw model 0.054--0.30 keV EUV photon flux (log). Column 4: Ratio of extrapolated X-ray powerlaw model EUV photon flux to the required EUV ionizing photon flux inferred from the \ion{He}{2} line. Column 5: Extrapolated X-ray diskbb model 0.054--0.30 keV EUV photon flux (log). Column 6: Ratio of X-ray extrapolated diskbb model EUV photon flux to the required EUV ionizing photon flux infered from the \ion{He}{2} line.
}
\end{deluxetable*}

\subsubsection{Stellar Population Synthesis Modelling}
\label{sec:SPS}

To better understand the physical conditions that give way to the ionized regions in our sample, and the potential role of accreting sources, we require updated and self-consistent estimates of key properties of the host H II regions including gas-phase and stellar metallicity, dust content, SFR, and star formation history (SFH). These quantities allow us to assess our sample against empirical prescriptions used to estimate XRB populations and their integrated luminosities (e.g., \citealt{Lehmer2010}, \citealt{Brorby2016})

To this end, we isolate a subsample of galaxies from the parent sample of 21, consisting of the 12 galaxies with SDSS spectra with $\mathrm{S/N} > 20$ per pixel and SDSS photometric data available (see Table~\ref{SFR100Results} for list of targets). We perform spectral energy distribution (SED) modelling of this sub-sample, where the higher S/N data has been shown to improve the accuracy of the fitted parameters in the current epoch (e.g., \citealt{Wan2024}, \citealt{Wang2023}). We use the \texttt{Prospector} inference framework (\citealt{Leja2017}, \citealt{Leja2019}; \citealt{Johnson2021}), which combines broad-band photometry and rest-frame optical spectroscopy within a Bayesian framework to derive physical parameters using the Flexible Stellar Population Synthesis code \citep{Conroy2009, Conroy2010}. 

Specifically, we adopt a modified \texttt{Prospector-alpha} template, a flexible and extensively tested SED fitting framework that employs a non-parametric SFH, physically motivated dust attenuation, and a self-consistent treatment of stellar and nebular emission \citep{Leja2019}, described in Section \ref{sec:model}. 
The non-parametric, piecewise SFH formulation enables a wide range of physically motivated evolutionary histories while maintaining computational efficiency and interpretability. This approach has been demonstrated to successfully reproduce the SFHs of both simulated and observed galaxies across diverse environments and provides a well-characterized baseline for interpreting galaxy SEDs (e.g., \citealt{Estrada2020}, \citealt{Lower2020}, \citealt{Nersesian2025}, \citealt{Leja20192}, \citealt{Wang2023}).

\subsubsection{Choice of Stellar Models}
Particularly challenging for our SPS modelling is choosing appropriate stellar libraries that are representative of our low-metallicity sample \citep{Conroy2009}. In an effort to model galaxies with potentially obscure and uncertain populations we aim to include the broadest possible range of stellar types and evolutionary timescales. We use the MESA Isochrones and Stellar Tracks (MIST) isochrones, providing high-quality coverage across all possible ages, masses (0.1–300 $M_\odot$), metallicities, and evolutionary phases of stars \citep{Dotter2016}. 

These isochrones inform the selection of appropriate spectra for SED modelling. We use the Medium-resolution Isaac Newton Telescope Library of Empirical Spectra (MILES), which contains real spectra for approximately 1000 stars spanning a wide range of atmospheric parameters. The high-quality spectra were obtained using the 2.5m Isaac Newton Telescope, covering 3525–7500 \AA \citep{Sanchez2006} with a spectral resolution of 2.5 \AA\ (FWHM) \citep{Falcon2011}, providing sufficient detail to resolve important stellar and nebular spectral lines, while remaining computationally \citep{Conroy2013} efficient and matching well to the SDSS spectra used. 

Unfortunately, the MILES library suffers the ill-fate of most empirical libraries, mainly that it consists of mostly stars in the Galactic disk \citep{Peletier2013}. As such, the lower main sequence and red giant branch are well covered by the empirical library over a wide range in metallicity, but hotter stars are rare, especially at lower metallicity including the upper main sequence and supergiants \citep{Conroy2013}. For this reason, \texttt{Prospector} adopts a hybrid of theoretical and empirical libraries, to complement the primary MILES libraries, allowing for more comprehensive modelling of stellar populations across various evolutionary phases \citep{Johnson2021}. Namely, the TP-AGB Library from \cite{Lanccon2000} is used to model the thermally pulsating asymptotic giant branch (TP-AGB) phase - where low and intermediate-mass stars contribute to infrared emission and affect chemical enrichment processes through mass loss in late stages of stellar evolution. The Post-AGB Library from \cite{Rauch2003} is used to model stars in the post-AGB phase, after a star has shed its outer layers, leaving behind a hot, compact core. Finally, the WMBASIC Hot Star Library from \cite{Eldridge17} is used to model hot stars, including O-type. 

Complicating our analysis is the absence of binary-star evolutionary tracks in our stellar libraries. Binary evolution is another channel that can produce stripped helium stars and other hot, ionizing sources that are not captured in single-star frameworks \citep{Eldridge2022}. These stars primarily modify the EUV portion of the SED, however they can also propagate indirectly into the optical regime through nebular emission. The implications of the absence of binary-stellar evolution in our models, and its impact on the interpretation of our results, are discussed further in Section~\ref{sec:discussion}.

\subsubsection{Model and Data}
\label{sec:model}
\begin{figure}
  \centering
  \includegraphics[width  = \linewidth]{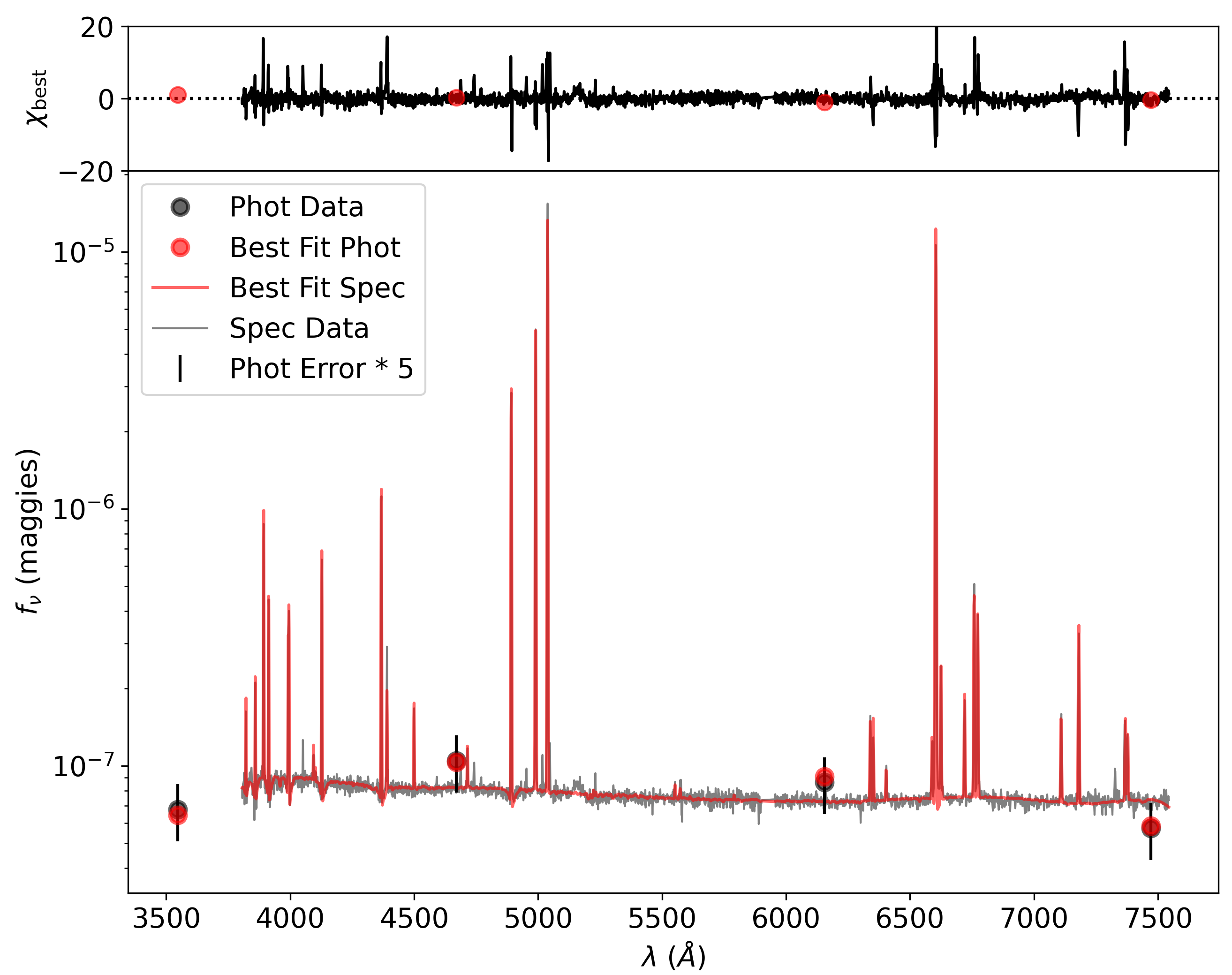}
  \caption{An example of a stellar population synthesis model fit to SDSS photometry and spectroscopy for a single galaxy (Mrk 1434) using \texttt{Prospector} code \citep{Johnson2021}. On the bottom plot the black points with error bars show the observed photometric fluxes (scaled by a factor of five for visibility), while the red circles indicate the corresponding best-fit photometric model values. The gray line represents the observed spectrum after conversion to flux units, sky-line removal, and de-redshifting, and the red line shows the best-fit model spectrum. To match the wavelength range of the MILES spectral library used for fitting, we limit the rest-frame spectrum to $\lambda < 7500$ \AA\ and restrict the photometric data to the SDSS u, g, r, and i filters, assuming a fixed 5\% photometric uncertainty. The top panel shows the residuals ($\chi_{\mathrm{best}}$) between the observed and best-fit spectra.} 
  \label{fig:SPS_model}
\end{figure}

For each galaxy, we fit an SPS model to its SDSS photometry and spectrum (Figure.~\ref{fig:SPS_model}) using the MIST isochrones and MILES spectral libraries. We prepare the spectrum by converting it to flux units, masking sky lines, and de-redshifting. To match the wavelength coverage of the MILES stellar library, we apply a rest-frame cut-off at 7500 \r{A}. For this reason, we also restrict our photometry to the SDSS Fiber magnitudes (flux contained within the aperture of the 3\arcsec spectroscopic fiber) in the u, g, r, and i filters, and we adopt a fixed photometric uncertainty of 5\%. We assume a \cite{Kroupa2001} initial mass function. Our priors are listed in Table \ref{tab:prior} based on a modified \texttt{prospector-alpha} model \citep{Leja2017}. Here we highlight some important components.

The simulated spectra are processed with a dust attenuation model. We use the parametrization of dust attenuation following \cite{Leja2017}. Briefly, the dust attenuation is represented using a two-component model based on \cite{CharlotFall2000}, consisting of a birth cloud and a diffuse interstellar dust screen. The birth cloud component accounts for additional attenuation in dense star-forming H II regions around young stars. This component also influences nebular line emission. Attenuation from the birth clouds follows a wavelength-dependent power law \citep{Calzetti2000}, attenuating stars that have formed in the last 10 Myr. Meanwhile the diffuse dust component attenuates stellar and nebular light. The wavelength dependence follows the framework developed by \cite{Noll2009}. Dust emission is  modelled as the infrared re-emission of any light that is attenuated by dust using \cite{Draine2007}. Nebular emission is modelled using the prescriptions of \cite{Byler2017}, with gas-phase metallicity. Additional components such as spectra noise, velocity dispersion, and gas ionization, are simulated to further improve the realism of the simulated spectra. 

We adopt a non-parametric, piece-wise SFH with seven logarithmically spaced time bins (0, 0.1, 0.3, 1.0, 3.0, 6.0, 13.6 Gyr), spanning the full age of the universe at the redshift of each galaxy. We use the continuity prior \citep{Leja2019}, which imposes a correlated structure on adjacent time bins to prevent unphysical fluctuations in the inferred star formation rates. This prior allows for a wide range of physically motivated SFH shapes - rising, falling, or bursty - while retaining computational tractability and interpretability \citep{Lower2020}.

\setlength{\tabcolsep}{3pt}
\begin{deluxetable*}{lll}
\tabletypesize{\footnotesize}
\tablecaption{Model Parameter Descriptions and Priors \label{tab:prior}}
\tablehead{
\colhead{\textbf{Parameter}} & \colhead{\textbf{Description}} & \colhead{\textbf{Prior}}
}
\startdata
$\log\left(\frac{\mathrm{SFR}(t)}{\mathrm{SFR}(t+\Delta t)}\right)$ & Ratios of SFRs in adjacent time bins \citep{Leja2017} & Student T ($\mu$ = 0, $\sigma$ = 0.3, df = 2) \\[2pt]
SFH$_{\text{age}}$ & Nonparametric SFH age bins & Fixed (0, 0.10, 0.3, 1.0, 3.0, 6.0, 13.6 Gyr) \\[2pt]
IMF & Initial Mass Function & Fixed \citep{Kroupa2001} \\[2pt]
$z$ & Redshift & Uniform $(z_{\text{spec}} - 0.01,\ z_{\text{spec}} + 0.01)$ \\[2pt]
$\sigma_\ast$ (km\,s$^{-1}$) & Stellar velocity dispersion & Uniform (40.0, 400.0) \\[2pt]
$\log(M_\ast / M_\odot)$ & Stellar mass & Uniform (6.0, 10.0) \\[2pt]
$\log(Z / Z_\odot)$ & Gas-phase and stellar metallicity & Uniform ($-2.0$, 0.4) \\[2pt]
$n$ & Attenuation curve power-law index \citep{Calzetti2000} & Uniform ($-2.0$, 2.0) \\[2pt]
$\hat{\tau}_{\text{dust,2}}$ & Diffuse dust optical depth \citep{CharlotFall2000} & Clipped-Normal (min = 0.0, max = 4.0, $\mu$ = 1.0, $\sigma$ = 0.3) \\[2pt]
$\hat{\tau}_{\text{dust,1}} / \hat{\tau}_{\text{dust,2}}$ & Ratio between optical depths of birth cloud & Uniform (min = 0.0, max = 4.0) \\[2pt]
 & and diffuse dust \citep{CharlotFall2000} & \\[2pt]
$U_{\min}$ & Minimum radiation field strength & Uniform (min = 0.1, max = 25.0) \\[2pt]
$\gamma_e$ & Mass fraction of dust heated at $U_{\min}$ & log-Uniform (0.001, 0.150) \\[2pt]
$q_{\text{PAH}}$ & Percent mass fraction of PAHs in dust \citep{Draine2007} & Uniform (0.5, 7.0) \\[2pt]
$\sigma_{\text{gas}}$ (km\,s$^{-1}$) & Velocity dispersion of gas & Uniform (30, 300) \\[2pt]
$\log(U)$ & Gas ionization parameter & Uniform ($-4.0$, $-1.0$) \\[2pt]
$J$ & Spectra multiplicative noise inflation & Uniform (1, 3) \\
\enddata
\tablecomments{Summary of the parameters, their physical interpretations, and adopted priors used in the modelling (Section \ref{sec:SPS}).}
\end{deluxetable*}

In contrast to SED fitting strategies that rely exclusively on photometry, our simultaneous inclusion of spectroscopy enables tighter constraints on parameters that are otherwise highly degenerate \citep{Peletier2013}. In particular, low-metallicity systems pose a challenge for traditional SED modelling due to the weaker metal-line blanketing and more extreme ionizing spectra \citep{Conroy2013}. By fitting both photometric and spectroscopic data within the same framework, we minimize degeneracies between age, dust, and metallicity, and obtain more robust posterior distributions on parameters like the stellar metallicity, mass-weighted age, and SFR (e.g. \citealt{Nersesian2024}, \citealt{Nersesian2025}, \citealt{Pacifici2023})

\section{Results} \label{sec:results}

\subsection{Inferred Embedded EUV Continuum}
\label{inferredZEUV}
\begin{figure*}
    \centering
    \includegraphics[width=1\linewidth]{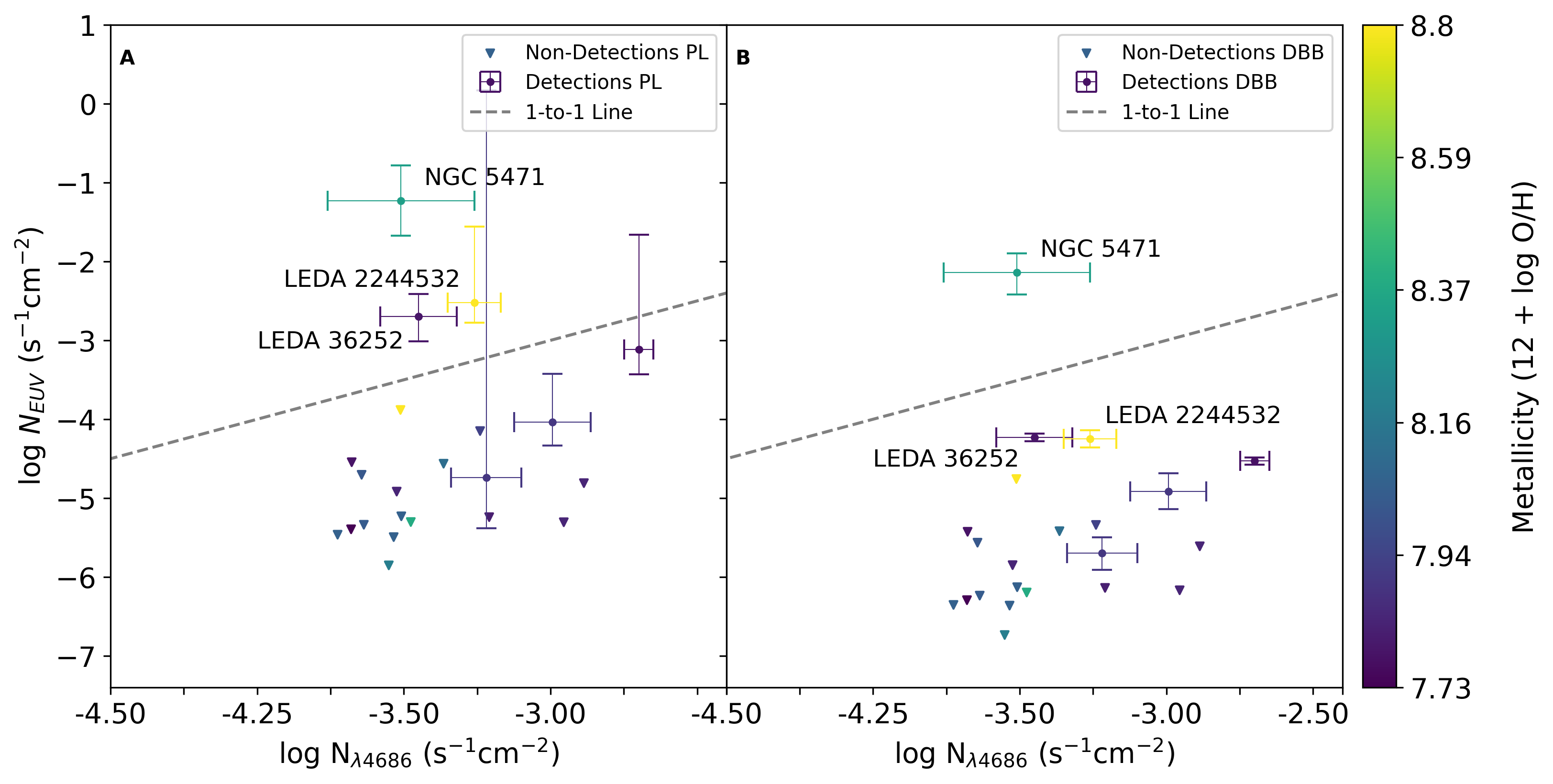}
    \caption{Comparison between the extrapolated unabsorbed EUV photon flux from observed X-ray sources, and the predicted ionizing photon flux required to produce the observed \ion{He}{2} $\lambda4686$ emission for each galaxy in our sample. The predicted EUV flux is obtained by multiplying the required \ion{He}{2} photon flux by 5.2. The dashed gray line marks the 1-to-1 relation, where the number of \ion{He}{2} ionizing photons equals the extrapolated EUV photon count. Panels A and B show results assuming Power-law (PL) and Disk Blackbody ({\tt diskbb}) spectral models, respectively. Galaxies without X-ray detections are shown as downward triangles representing upper limits, while those with detections are shown as squares with error bars corresponding to the uncertainty in the extrapolated EUV photon flux (see Table \ref{ratiostable}). For sources with multiple X-ray observations, only the brightest detection is plotted. Points are color-coded by gas-phase metallicity, with values indicated by the colorbar at right. As expected the {\tt diskbb} model yielded a systematically lower extrapolated EUV photon flux.}
    \label{fig:X-ray resultsgraph}
\end{figure*}

\begin{figure*}
    \centering
    \includegraphics[width=0.9\linewidth]{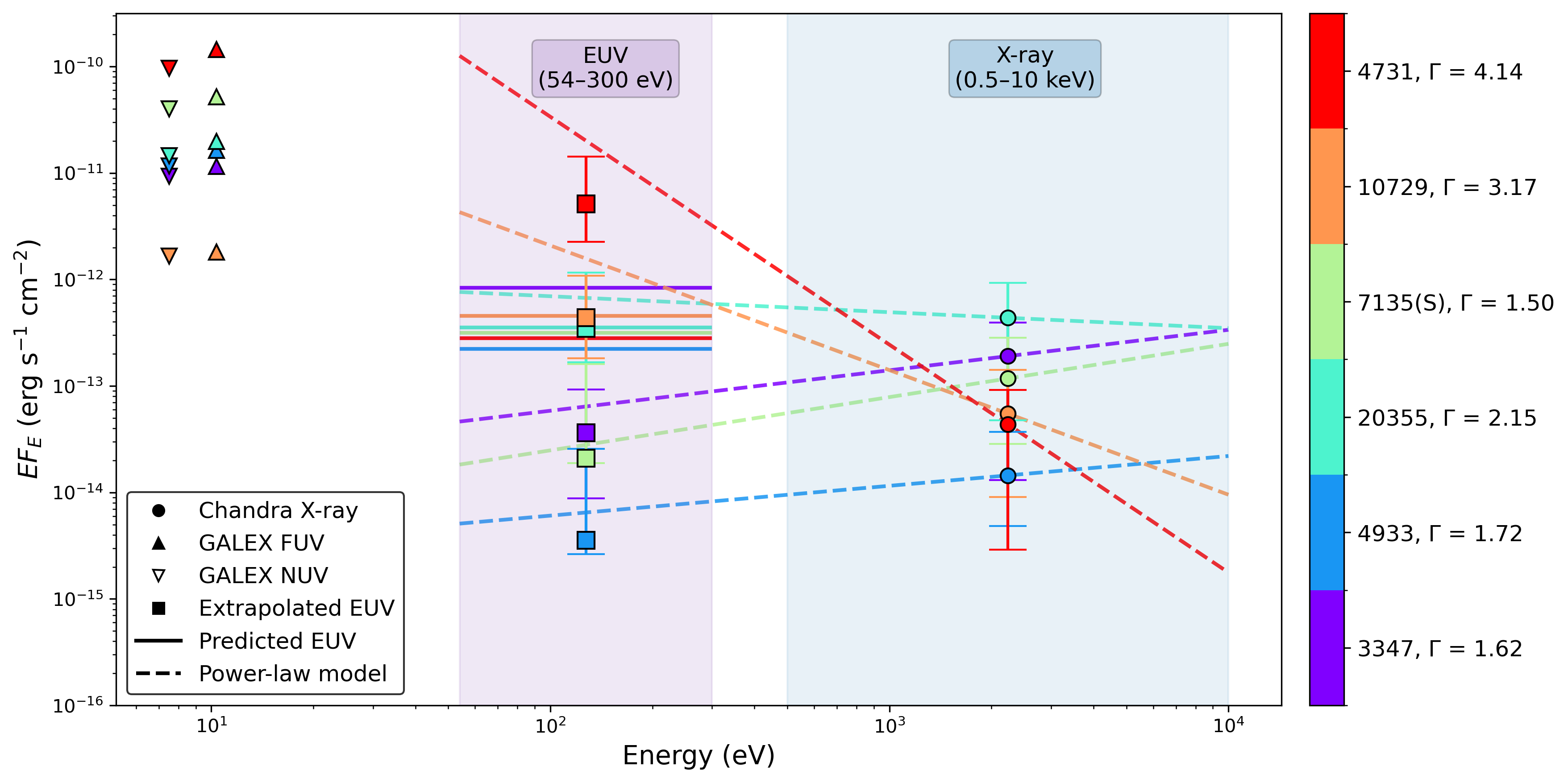}
    \caption{Broadband SEDs for Chandra observations with detections. When multiple detections are available for a given source, only the brightest observation is shown. GALEX \citep{Morrissey2007} FUV (upright triangles) and NUV (inverted triangles) energy fluxes are shown at their effective energies within the 4.4--9.2~eV band. Chandra 0.5--10~keV integrated fluxes (circles) are plotted at their effective energies within the X-ray band (blue shaded region), while square symbols denote EUV fluxes integrated over 54--300~eV (purple shaded region), extrapolated from the X-ray power-law fits. Horizontal solid lines spanning the EUV band indicate the EUV flux predicted from SDSS spectra. Dashed lines show the extrapolated power-law models derived from the Chandra spectra, parameterized by the photon indices $\Gamma$ listed in the color bar. Individual observations are distinguished by color, with the corresponding ObsID and $\Gamma$ values indicated.
}
    \label{fig:SED}
\end{figure*}

From our X-ray analysis in Section \ref{sec:srcflux}, we identified two galaxies amongst our seven new observations that host at least one X-ray source - LEDA 36252 and Mrk 1434. Notably, Mrk 1434 contains two distinct X-ray sources, as outlined in Table \ref{xraytable}. In the archival sample of 14 galaxies, five galaxies displayed significant X-ray detections, with three of these galaxies hosting two X-ray sources each (Mrk 1434, which appeared in both our new and archival samples, SDSS J111746.30+174424.6, and SDSS J115237.67-022806.3). Conversely, the remaining 9 galaxies either exhibited no X-ray sources within their optical galaxy areas or were excluded due to detections occurring at the edge of an ACIS chip, as detailed in Table \ref{xraytable}.

For each X-ray detection and non-detection we calculated the extrapolated EUV photon flux or its upper limit, respectively, following the procedures described in Section \ref{sec:srcflux}. In Figure \ref{fig:X-ray resultsgraph} we compare this extrapolated EUV photon flux from accreting objects to the predicted ionizing EUV photon flux producing the observed He II line emission obtained in Section \ref{sec:HeII}. When multiple X-ray observations are present for an object we plot the highest EUV flux to represent the maximum possible EUV continuum. These results are also tabulated in Table \ref{ratiostable}. Across the sample, the {\tt diskbb} model yielded a systematically lower extrapolated EUV photon flux than the {\tt powerlaw} model. 

Only three observations display enough EUV radiation from X-ray bright objects to adequately explain the observed flux of the He II emission line. First, Galaxy NGC 5471 (obsID 5297) has been shown to host either an energetic explosive hypernova remnant or multiple concentrated core-collapse supernova \citep{Sun2012}. Independent optical spectroscopy has further revealed a substantial population of $\sim$40 WR stars, inferred from the presence of stellar N\,\textsc{v} $\lambda$4602, $\lambda$4620 and C\,\textsc{iv} $\lambda$4658, $\lambda$5808 features \citep{Luridiana2002}. This provides  motivation for continued investigation of WR stars as primary sources of hard ionizing radiation, whose spectral signatures may otherwise be diluted or obscured in integrated observations (see Sections~\ref{sec:intro} and~\ref{sec:discussion}).

Second, Galaxy LEDA 36252 (obsID 20355), is believed to have hosted an unusually high pressure event, which could arise from many dozens of supernova over a period of several Myr \citep{Elmegreen2016}. This follows a recent burst of star formation in a large star-forming region 830 pc in diameter of recently accreted H I. Lastly, Galaxy LEDA 2244532 (ObsID 10729) shows spectral features consistent with the presence of an AGN (\citealt{Gelbord2009}), with the detected X-ray emission spatially consistent with a nuclear origin (Figure~\ref{fig:x-ray_chart}). LEDA 2244532 is identified as AGN-dominated in Figure \ref{fig:BPT} and serves as a valuable control case for validating our diagnostic approach. The astrometric positions of all three X-ray sources lie within three times their positional uncertainties of the Simbad ICRS coordinates \cite{Wenger2000} and are therefore approximated as being nuclear (see Table \ref{galaxytable} and Figure \ref{fig:x-ray_chart}). As shown in Figure~\ref{fig:x-ray_chart}, the X-ray detections in the remaining galaxies are located outside the Sloan fiber aperture, further supporting our findings that these detected X-ray sources are unlikely the sole contributors to the He II line emission measured within the spatially offset optical spectra.

These three sources also stand out in Figure \ref{fig:SED}, which presents the broadband SEDs of the six galaxies with detections using the {\tt powerlaw} model (only the brightest observation for each source is shown). Notably, the power-law spectra for these three galaxies rise toward lower energies, in contrast to the remaining systems. The GALEX FUV and NUV fluxes  \citep{Morrissey2007} were derived from the AB magnitudes using $F_\nu = 10^{-0.4 \,(m_\mathrm{AB} + 48.6)}~\mathrm{erg\,s^{-1}\,cm^{-2}\,Hz^{-1}}$, and then converted to energy flux as $E F_E = h (c/\lambda) F_\nu~\mathrm{erg\,s^{-1}\,cm^{-2}}$, where $\lambda$ is the effective wavelength of the GALEX band (FUV = 1539~\AA, NUV = 2316~\AA). It can be seen that the GALEX fluxes, measured using elliptical Kron apertures, are higher than the extrapolated EUV fluxes from the X-ray power-law fits. This suggests that the UV emission inferred from extrapolating just the X-ray–emitting component does not account for the observed UV luminosities of the entire galaxy and does not dominate the UV output of these systems. 

\begin{figure*}
    \centering
    \includegraphics[width=0.9\linewidth]{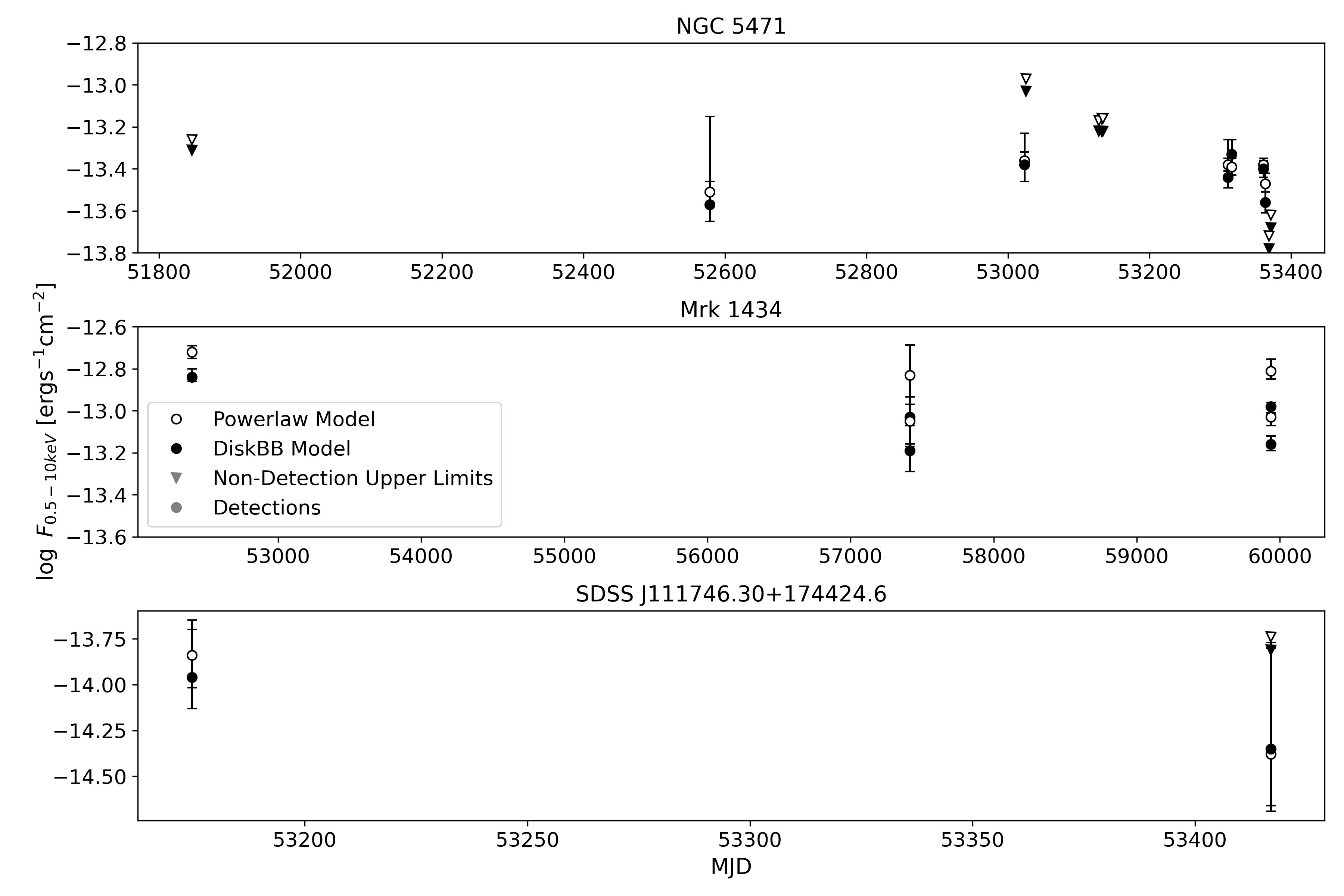}
    \caption{Light-curves of NGC 5471, Mrk 1434, and J111746.30+174424.6, showing the 0.5-10 keV flux in log scale versus the MJD observation. Downward triangles indicate the non-detection upper limits when the galaxy was in view of the Chandra aimpoint but no X-ray emission was detected. Circles indicate a detection with error bars. Solid points represent fluxes obtained from the DiskBB Model and hollow represent fluxes obtained from the powerlaw model. Scatter of less than one order of magnitude is observed}
    \label{fig:LC}
\end{figure*}

A handful of galaxies we sampled were observed over multiple epochs. Here we provide the light curves for NGC 5471, MRK 1434, and J111746.30+174424.6 (Figure \ref{fig:LC}). The largest spread in minimum and maximum of the flux of the light curves, is measured to be $\Delta \log F_{0.5-10keV} = 1.08^{+0.22}_{-0.36}$, which provides a lower bound on the characteristic X-ray variability of our sample. Still, as we only consider the brightest observations when multiple are present, this observed variability is insufficient to move objects that are too X-ray faint to explain He II line to high enough fluxes to explain the He II line via X-ray photoionization.

\subsection{Ionized Region Conditions}
In Table \ref{SFR100Results}, we report the host galaxy properties for the 12  galaxies with high-enough S/N in their SDSS spectra to perform SPS modelling. Overall, our modelled results reflect dwarf-galaxy characteristics including low chemical enrichment and low associated amount of dust, small masses, and very recent star formation. It is emphasized that in this section we address observations for this subset of 12 high-S/N galaxies. 

\begin{figure*}[!t]
    \centering
    \includegraphics[width=0.9\linewidth]{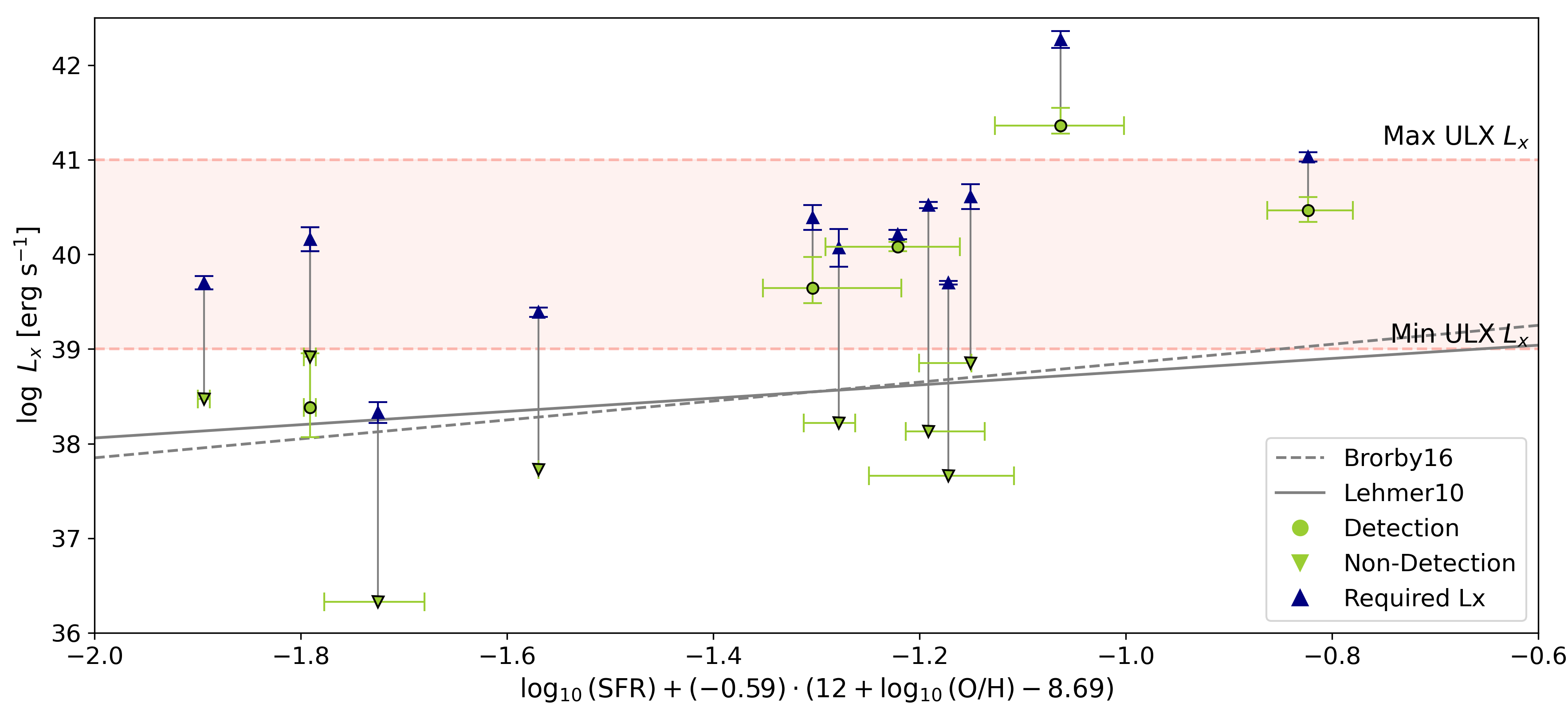}
    \caption{Empirical X-ray luminosity scaling relations compared to our sample galaxies. The solid and dashed lines show the expected XRB luminosity as a function of star formation rate (SFR) and metallicity from \citet{Lehmer2010} and \citet{Brorby2016}, respectively. Green points represent the 12 galaxies in our sample that we performed stellar population synthesis modelling on. SFRs and metallicities are derived from our non-parametric modelling using a 100 Myr time bin in the current epoch. Circles indicate sources in our sample that had X-ray detections and downward triangles denote upper limits for non-detections. Upward-pointing blue triangles mark the minimum required X-ray luminosity of an accreting object to power the observed nebular He II emission. The vertical lines connect each galaxy’s observed X-ray luminosity (or upper limit) to the required luminosity inferred from its He II line. The shaded pink region marks the typical luminosity range of ultraluminous X-ray sources.}
    \label{fig:Lehmer}
\end{figure*}

We find that the galaxies in our sample with X-ray detections follow expectations from XRB empirical scaling relations, given their modelled metallicity and SFRs (Figure \ref{fig:Lehmer}). This is further confirmed by summing the SFRs of the galaxies in our SPS-analysis subsample ($\log(\sum{SFR}_{100}) = -0.53^{0.06}_{-0.06} M_\odot/yr$) and inputting into the scaling relationships with their average metallicity. For these sources we find that the \cite{Brorby2016} treatment for low metallicity galaxies prescribes values of  $\log L_{X,0.5-8keV} = 39.30^{+0.50}_{-0.12}$. These expected values are less than the summed X-ray luminosity of these sources: $\sum\log L_{X,0.5-8keV} = 41.42^{+0.06}_{-0.09}$. Even with a sample biased towards being bright in the X-ray, when the observed X-ray luminosity is compared to the minimum extrapolated X-ray luminosity (between the {\tt diskbb} and {\tt powerlaw} models, see section \ref{sec:srcflux}) required to produce the observed He II line, all galaxies in the SSP-sub-sample are still not powerful enough to power the \ion{He}{3} nebulae (Figure \ref{fig:Lehmer}). 

Given that dust attenuation is wavelength-dependent and strongest in the ultraviolet and optical regimes, its impact on hard X-ray photons ($>$ 1-2 keV) is generally negligible unless column densities are extreme \citep{Wilms2000}. For our sample, the derived extinction parameters from SED fitting (e.g., $A_V$) are uniformly low, reinforcing the notion that dust plays a minor role in shaping the X-ray SED. This conclusion is consistent with the typically low dust-to-gas ratios expected in low-metallicity environments \citep{Fisher2014}, which dominate our galaxy sample. Across our sample we find a median dust attenuation in the $V$-band due to ISM of $A_{V,\mathrm{ISM}} = 0.088^{+0.040}_{-0.042} \mathrm{mag}$ and a value of $A_{V,\mathrm{BC}} = 0.282^{+0.049}_{-0.054} \mathrm{mag}$ due to dust within stellar birth clouds (BC) averaged over the last 100 Myr.  The two-component dust model reveals that the birth cloud consistently contains more dust than surrounding ISM consistent with \cite{CharlotFall2000}. 

To estimate the effect of dust and gas extinction on the observed X-ray luminosities of our sources, we modelled the energy-dependent photoelectric absorption. The total visual extinction, $A_V$ (the sum of BC and ISM), was converted to the neutral hydrogen column density, $N_{\mathrm{H}}$, assuming the empirical relation $N_{\mathrm{H}} \simeq (2.21 \pm 0.09) \times 10^{21} \, A_V \;\; \mathrm{cm^{-2}}$ \citep{Guver2009}. For a photon energy $E$ (in keV), the transmission $T(E)$ along the line of sight is given by $T(E) = \exp \left[ - N_{\mathrm{H}} \, \sigma(E) \right]$. The total photoelectric cross-section per hydrogen atom, $\sigma(E)$, was calculated using the analytic fits provided by \citet{Verner1996}, adopting the interstellar medium elemental abundances from \citet{Wilms2000}. 

\begin{deluxetable*}{cccccc}
\tabletypesize{\footnotesize}
\tablecaption{100 Myr Burst Stellar Population Synthesis Modelling Results \label{SFR100Results}}
\tablehead{
\colhead{Target (SDSS)\tablenotemark{}} & 
\colhead{$12 + \log_{10}{(O/H)}$\tablenotemark{}} & 
\colhead{$\log_{10}{SFR_{100}}$\tablenotemark{}} & 
\colhead{$\log_{10}(M_\ast / M_\odot)$\tablenotemark{}} & 
\colhead{$A_{V,\mathrm{BC}}$\tablenotemark{}} & 
\colhead{$A_{V,\mathrm{ISM}}$\tablenotemark{}} \\
\colhead{} & 
\colhead{} & 
\colhead{($M_\odot~\mathrm{yr^{-1}}$)} & 
\colhead{($M_\odot$)} & 
\colhead{(mag)} & 
\colhead{(mag)}
}
\colnumbers
\startdata
J115237.67$-$022806.3 & $\rm{7.90^{+0.01}_{-0.01}}$ & $\rm{-1.8^{+0.1}_{-0.1}}$ & $\rm{7.2^{+0.1}_{-0.1}}$ & $\rm{0.630^{+0.109}_{-0.098}}$ & $\rm{0.812^{+0.227}_{-0.171}}$ \\
J083743.48+513830.2 & $\rm{8.00^{+0.01}_{-0.01}}$ & $\rm{-2.3^{+0.1}_{-0.1}}$ & $\rm{6.1^{+0.1}_{-0.1}}$ & $\rm{0.011^{+0.011}_{-0.011}}$ & $\rm{0.003^{+0.002}_{-0.001}}$ \\
J103410.15+580349.1 & $\rm{7.88^{+0.01}_{-0.01}}$ & $\rm{-1.3^{+0.1}_{-0.1}}$ & $\rm{7.7^{+0.1}_{-0.1}}$ & $\rm{0.565^{+0.065}_{-0.087}}$ & $\rm{0.447^{+0.114}_{-0.085}}$ \\
J123134.25+442639.2 & $\rm{8.66^{+0.05}_{-0.05}}$ & $\rm{-1.1^{+0.1}_{-0.1}}$ & $\rm{9.6^{+0.1}_{-0.1}}$ & $\rm{0.904^{+0.371}_{-0.246}}$ & $\rm{0.489^{+0.217}_{-0.163}}$ \\
J101624.51+375445.9 & $\rm{7.83^{+0.01}_{-0.01}}$ & $\rm{-1.7^{+0.1}_{-0.1}}$ & $\rm{7.2^{+0.1}_{-0.1}}$ & $\rm{0.869^{+0.109}_{-0.195}}$ & $\rm{0.117^{+0.048}_{-0.043}}$ \\
J122615.70+482938.4 & $\rm{7.96^{+0.01}_{-0.01}}$ & $\rm{-2.0^{+0.1}_{-0.1}}$ & $\rm{6.2^{+0.1}_{-0.1}}$ & $\rm{0.001^{+0.001}_{-0.001}}$ & $\rm{0.000^{+0.001}_{-0.000}}$ \\
J104653.98+134645.7 & $\rm{7.98^{+0.01}_{-0.01}}$ & $\rm{-1.6^{+0.1}_{-0.1}}$ & $\rm{6.6^{+0.1}_{-0.1}}$ & $\rm{0.011^{+0.001}_{-0.001}}$ & $\rm{0.009^{+0.003}_{-0.004}}$ \\
J121749.30+375155.5 & $\rm{8.06^{+0.01}_{-0.01}}$ & $\rm{-2.1^{+0.1}_{-0.1}}$ & $\rm{6.1^{+0.3}_{-0.1}}$ & $\rm{0.076^{+0.022}_{-0.011}}$ & $\rm{0.058^{+0.015}_{-0.040}}$ \\
J114107.48+322537.2 & $\rm{7.88^{+0.01}_{-0.01}}$ & $\rm{-1.7^{+0.1}_{-0.1}}$ & $\rm{6.9^{+0.1}_{-0.1}}$ & $\rm{0.261^{+0.033}_{-0.022}}$ & $\rm{0.050^{+0.032}_{-0.016}}$ \\
J110458.30+290816.5 & $\rm{7.90^{+0.01}_{-0.01}}$ & $\rm{-1.6^{+0.1}_{-0.1}}$ & $\rm{7.4^{+0.1}_{-0.1}}$ & $\rm{2.878^{+0.098}_{-0.109}}$ & $\rm{1.131^{+0.095}_{-0.094}}$ \\
J111746.30+174424.6 & $\rm{7.96^{+0.01}_{-0.01}}$ & $\rm{-2.2^{+0.1}_{-0.1}}$ & $\rm{7.0^{+0.1}_{-0.1}}$ & $\rm{0.303^{+0.125}_{-0.086}}$ & $\rm{0.217^{+0.109}_{-0.043}}$ \\
J122225.79+043404.7 & $\rm{8.17^{+0.01}_{-0.01}}$ & $\rm{-1.6^{+0.1}_{-0.1}}$ & $\rm{6.8^{+0.1}_{-0.1}}$ & $\rm{0.013^{+0.013}_{-0.000}}$ & $\rm{0.000^{+0.011}_{-0.000}}$ 
\enddata
\tablecomments{100 Myr Burst Stellar Population Synthesis Modelling Results. Column 1: SDSS galaxy name. Column 2: Gas-phase Metallicity. Column 3: Star formation rate averaged over the past 100 Myr. Column 4: Stellar mass. Column 5: Dust attenuation toward birth clouds (BCs). Column 6: Dust attenuation toward the diffuse ISM.}
\end{deluxetable*}

For each source, we computed $T(E)$ over the $0.054$--$10~\mathrm{keV}$ band using a dense grid of energies. The resulting attenuation curves were smoothed using cubic splines and are shown in Figure~\ref{fig:attenuation_curves}. The shaded regions represent the uncertainties in $T(E)$ propagated from the reported lower and upper errors in $A_V$. For most sources attenuation between 0.054-3 keV is significant, while non-existent from 3-10 keV. Notably, several target galaxies still exhibit partial transmission in the EUV.

To quantify the total X-ray power absorbed by the dust, we integrated the unabsorbed flux across the $0.5$--$10~\mathrm{keV}$ band:
\begin{equation}
    F_{\mathrm{lost}} = \int_{{0.5 keV}}^{{10 keV}} F_{\mathrm{int}}(E) \,
    \left[ 1 - T(E) \right] \, dE,
    \label{eq:lost_luminosity}
\end{equation}
where $F_{\mathrm{int}}(E)$ is the intrinsic (unabsorbed) X-ray flux across the band measured. This yields the integrated lost flux for each source reported in Table \ref{tab:Dust}. We use the brightest flux between the DiskBB and Powerlaw fits to show that even when the most optimistic dust correction factor is applied to the brightest sources, their resulting change in flux is not significant enough (~$>$1 order of magnitude too small) to change our embedded EUV continuum enough to explain the He II line. 

Analysis of the SFHs of our galaxies reveals a broadly consistent picture: all show recent bursts of star formation within the last 100 Myr (Figure \ref{fig:SFH}). These bursts are temporally compact and intense, consistent with recent findings that extreme emission line galaxies often exhibit elevated specific SFRs and bursty SFHs (e.g., \citealt{Emami2019}). This indicates that the inferred burstiness may be partially limited by the adopted temporal resolution, and motivates the use of more finely binned SFH models in the future to better characterize recent star formation activity.

Figure \ref{fig:metallicity} compares our sample to the parent \citealt{Shirazi12} sample. We obtain values for this figure following methodology in Section \ref{sec:spec_fitting}, notably providing the characteristic SFR averaged of the last 10 Myr. The right panel of Figure \ref{fig:metallicity} shows both a divided population and a possible exponential relationship between HeII/H$\beta$ line ratio and gas-phase metallicity when looking at the \citealt{Shirazi12} sample as a whole. Indeed, current literature points to an existing relationship between these two metrics \citep{Schaerer19}, however, this correlation is suggested to be inversely proportional. This disparity is worth exploring, as sample selection may play a role in this behaviour. Our sample does appear to follow this inverse proportionality, aligning with the metallicity range of the \citealt{Schaerer19} sample (7.0-9.0 12 + log[O/H]). Nevertheless, the low-metallicity of our sample is reflected in this figure, which supports our selected methodology in determining gas-phase metallicities within our sample.

The left panel of Figure \ref{fig:metallicity} recovers an approximately normal distribution among the \citealt{Shirazi12} galaxies when comparing SFR and HeII/H$\beta$ line ratios. Conversely, our subsample appears to be low in HeII/H$\beta$ emission compared to the population while maintaining a diverse range of SFR. This suggests a weak or negligible relation between SFR and HeII/H$\beta$ line ratio between both samples.

\section{Discussion} \label{sec:discussion}

\subsection{Modelling Approaches and Challenges in Uncertainty Connecting Observations to Theory}
\label{Modeling Approaches}

Traditional photoionization models interpret \ion{He}{2} emission observed in star forming dwarf galaxies as compact He III zones ionized by high energy photons originating from embedded populations formed in recent bursts of star formation \citep{Charlot2000}. These zones, nested inside H II regions, are themselves embedded within larger H I envelopes \citep{Charlot2001}. Low-energy non-ionizing photons then propagate through these H I regions into the diffuse interstellar medium, such that the emergent radiation field depends on the properties of the embedded population and the gas-phase conditions. 

For unresolved galaxies, a typical resolution element may encompass a number of H II regions with embedded populations of different ages. It has been shown observationally that the optical-line ratios in the unresolved spectra of galaxies are similar to those in the spectra of individual H II regions \citep{Kobulnicky99}. This allows for the determination of the effective properties of the ionization environment (e.g., metallicity, star formation rate) and the ability to characterize predominant excitation mechanism in extragalactic objects through line-ratios for even unresolved data (e.g., BPT diagrams, see \citealt{Baldwin1981}). 

In our case, the spectra are obtained over the 3$\arcsec$ SDSS fibres, which samples projected diameters of approximately 0.04–4 kpc across the distances spanned by our sample. Although this aperture does not capture the full galaxy (Figure \ref{fig:finding_chart}), the derived effective parameters still trace the ensemble properties of the H II regions and surrounding diffuse gas within the fibre coverage. Ultimately, this has allowed for the the modelling of \ion{He}{2} $\lambda4686$ emission of star forming dwarf-galaxies, as giant H II regions with diffuse gas ionized by a single stellar population (e.g., \citealt{Charlot2000, Charlot2001})

The SPS modelling was unable to reproduce the observed strength of the He II  line for any of the observed galaxies, as modelled by our mix of empirical and theoretical stellar libraries (Section \ref{sec:HeII}). This deficit of ionizing photons above 54.4 eV coming from our modelled stellar population reflects the absence of very hot short-lived stellar populations (including WR and helium stripped stars), other exotic stellar populations, or alternative ionizing sources (such as XRBs) not included in standard models (\citealt{Drout2023}, \citealt{Gotberg2023}). This also marks areas of uncertainties in stellar evolution physics, particularly for binary interactions, mass transfer, and compact object formation. Unfortunately, no single spectral library, whether theoretical or empirical, fully covers the parameter space of all stellar sources and processes highlighting an area where significant advancements can still be made \citep{Conroy2013}. Hence, there is the justification to revisit our modelling approach in the future when more comprehensive stellar libraries are made available.

\setlength{\tabcolsep}{3pt}
\begin{deluxetable*}{cccc}
\tabletypesize{\footnotesize}
\tablecaption{Dust Attenuation From Stellar Population Synthesis Modelling \label{tab:Dust}}
\tablehead{
\colhead{Target (SDSS)\tablenotemark{}} \vline  & 
\colhead{$A_{V}$\tablenotemark{}}  & 
\colhead{$\Delta \log(L_{x,0.5-10~keV})$\tablenotemark{}} & 
\colhead{$\Delta \log F_{0.5-10~keV}$\tablenotemark{}} \\[2pt]
\colhead{(1)} \vline  & 
\colhead{(2)} & 
\colhead{($\rm{erg\;s^{-1}}$)} & 
\colhead{($\rm{erg\;s^{-1}cm^{-2}}$)} 
}
\startdata
J115237.67$-$022806.3 & $\rm{1.442^{+0.168}_{-0.135}}$ & $\rm{-14.303^{+0.051}_{-0.050}}$ & $\rm{-0.126^{+0.031}_{-0.000}}$ \\
J083743.48+513830.2 & $\rm{0.014^{+0.007}_{-0.006}}$ & $\rm{-16.371^{+0.840}_{-0.280}}$ & $\rm{-0.001^{+0.000}_{-0.002}}$ \\
J103410.15+580349.1 & $\rm{1.012^{+0.090}_{-0.086}}$ & $\rm{-13.750^{+0.050}_{-0.042}}$ & $\rm{-0.039^{+0.032}_{-0.045}}$ \\
J123134.25+442639.2 & $\rm{1.393^{+0.294}_{-0.205}}$ & $\rm{-14.316^{+0.088}_{-0.083}}$ & $\rm{-0.037^{+0.019}_{-0.052}}$ \\
J101624.51+375445.9 & $\rm{0.986^{+0.079}_{-0.119}}$ & $\rm{-15.524^{+0.075}_{-0.038}}$ & $\rm{-0.031^{+0.026}_{-0.033}}$ \\
J122615.70+482938.4 & $\rm{0.001^{+0.001}_{-0.001}}$ & $\rm{-17.777^{+0.000}_{-0.476}}$ & $\rm{0.000^{+0.00}_{0.000}}$ \\
J104653.98+134645.7 & $\rm{0.020^{+0.002}_{-0.003}}$ & $\rm{-17.099^{+0.123}_{-0.078}}$ & $\rm{-0.001^{+0.001}_{-0.001}}$ \\
J121749.30+375155.5 & $\rm{0.134^{+0.019}_{-0.026}}$ & $\rm{-16.371^{+0.190}_{-0.093}}$ & $\rm{-0.007^{+0.005}_{-0.009}}$ \\
J114107.48+322537.2 & $\rm{0.311^{+0.033}_{-0.019}}$ & $\rm{-13.832^{+0.045}_{-0.064}}$ & $\rm{-0.014^{+0.012}_{-0.019}}$ \\
J110458.30+290816.5 & $\rm{4.009^{+0.097}_{-0.102}}$ & $\rm{-15.148^{+0.011}_{-0.010}}$ & $\rm{-0.061^{+0.060}_{-0.063}}$ \\
J111746.30+174424.6 & $\rm{0.520^{+0.117}_{-0.065}}$ & $\rm{-14.979^{+0.091}_{-0.109}}$ & $\rm{-0.030^{+0.016}_{0.000}}$ \\
J122225.79+043404.7 & $\rm{0.013^{+0.012}_{-0.000}}$ & $\rm{-17.148^{+0.000}_{-0.445}}$ & $\rm{-0.001^{+0.001}_{0.000}}$ \\
\enddata
\tablecomments{Dust Attenuation From Stellar Population Synthesis Modelling. Column 1: SDSS galaxy name. Column 2: Visual extinction ($A_V$). Column 3: Change in X-ray luminosity for the 100 Myr model due to dust attenuation. Column 4: Change in X-ray flux due to dust attenuation.}
\end{deluxetable*}

One library that explicitly incorporates binary interactions into stellar evolution is Binary Population and Spectral Synthesis (BPASS; \citealt{Eldridge17}, \citealt{Eldridge2022}). Models of this kind, along with related studies, have been shown to enhance the production of ionizing photons (e.g., \citealt{VanBever1999}; \citealt{Lecroq2024}; \citealt{Bray2025}), boosting the EUV SED. While having a modest impact on the optical continuum, this increased EUV continuum can significantly alter optical emission line strengths (\citealt{Stanway2014}; \citealt{Xiao2018}). Because our fitting procedure simultaneously constrains the spectral continuum and emission lines (with the aid of independent photometry) the relative impact of systematic uncertainties associated with EUV ionizing sources altering the optical SED remains difficult to fully quantify (\citealt{Conroy2009}).

For example, \citet{Johnson2021} find that metallicity and dust attenuation are primarily constrained by emission line features, whereas other parameters are driven mainly by broadband photometry and the spectral continuum (with emission lines providing secondary constraints). Although further work is needed to quantify these systematics in detail, the use of single-star models in place of binary evolution models can bias inferred physical parameters. In particular, several studies find that matching the enhanced ionizing photon production of binary populations may require higher inferred star formation rates and, in some cases, lower metallicities (e.g., \citealt{Xiao2018}; \citealt{Lecroq2024}, \citealt{Topping2020}), though the magnitude and direction of these effects depend on the specific data and modeling assumptions. If we are under-estimating SFR and/or over-estimating metallicity than we are likely under-reporting how X-ray luminous our sample is relative to expectations from standard scaling relations. This reinforces our conclusion that the inability of XRB's to power the observed emission lines cannot be explained by a deficit of XRB's and other accreting sources in our sample.

Similar conclusions have been reached in studies that incorporate binary evolution as a natural component of stellar population modeling (e.g., \citealt{Garofali2024}, \citealt{Bray2025}, \citealt{Lecroq2024}, \citealt{Simmonds21}). These efforts often model the ionizing output of XRB populations arising from instantaneous bursts of star formation by combining stellar SED's with those of XRBs to produce composite SEDs. Such SEDs, constructed either from theoretical predictions \citep[e.g.,][]{Garofali2024, Bray2025, Lecroq2024} or empirical spectra \citep[e.g.,][]{Simmonds21}, are then used to generate grids of nebular emission across age and metallicity based on empirical scaling relations or self consistent population modelling. While these models show that including XRBs can enhance the high-energy ionizing budget, in some cases boosting nebular lines such as He II $\lambda4686$ by factors of two or more, they also consistently find that XRBs alone are insufficient to power the observed emission. One possible explanation is that commonly used scaling relations may overestimate the true X-ray output, further contributing to the discrepancy \citep{Lecroq2024}.

\begin{figure*}
    \centering
    \includegraphics[width=1.0\linewidth]{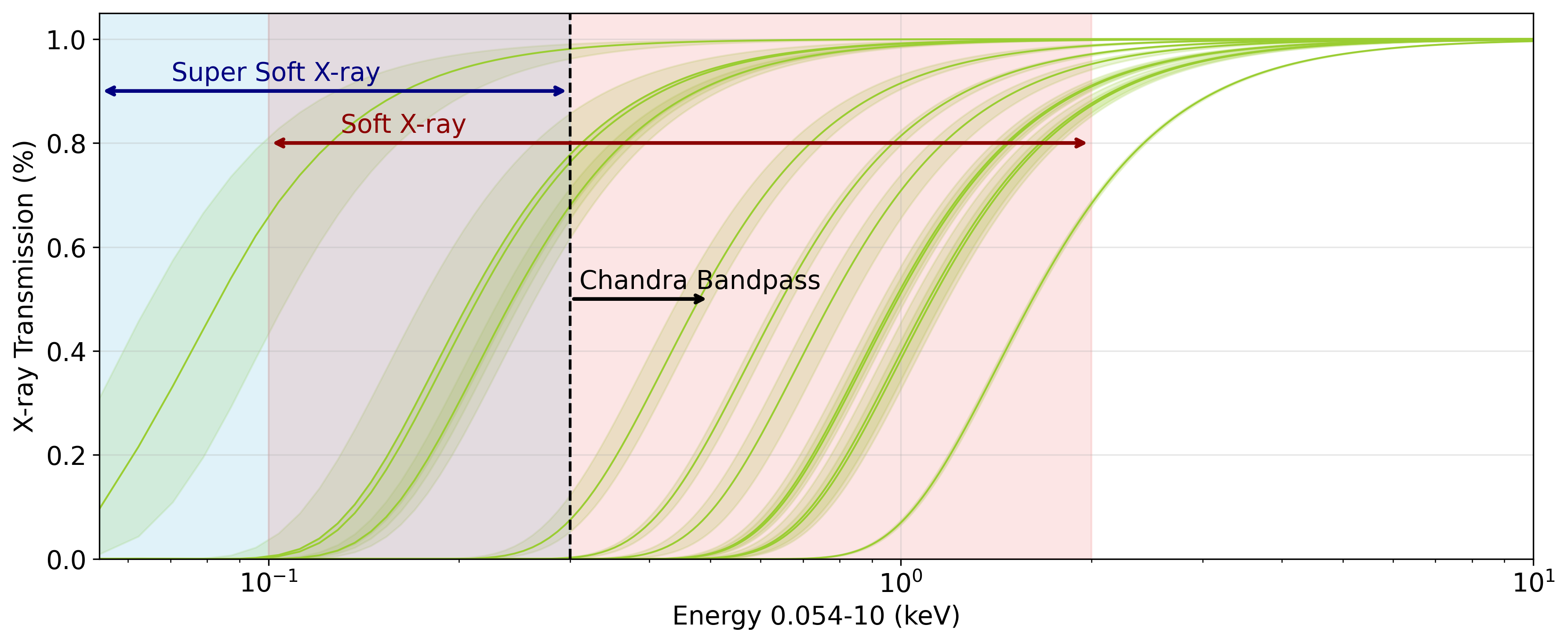}
    \caption{X-ray transmission curves for sources in our subsample that we performed stellar population synthesis modelling on. Transmission is derived from their measured $A_V$ values and shaded regions show the $1\sigma$ uncertainty propagated from the reported $A_V$ errors. The vertical dashed line marks the lower bound of the Chandra bandpass (0.3 keV), while the shaded blue and red regions indicate the super-soft (0.054-0.3 keV)and soft X-ray (0.1-2 keV) regimes, respectively. Low-energy photons ($<$0.5 keV) experience the greatest absorption with few sources having transmission through the dust at these low energies. At higher energies ($>$3 keV), the effects of dust extinction are negligible.}
    \label{fig:attenuation_curves}
\end{figure*}

A central uncertainty in constraining these ionizing contributions is the estimation of SFR, which can differ by orders of magnitude between galaxy types \citep{Kennicutt1998}. Common indirect tracers—including H$\alpha$, UV, and FIR fluxes—are subject to dust attenuation and aperture effects in spatially unresolved data \citep{Brinchmann2004}. Additionally, our modelling shows that many highly ionized glaxies exhibit bursty star formation histories, consistent with recent findings from JWST studies that distant galaxies exhibit unexpectedly high SFRs and mature structures (e.g., \citealt{Xiao24}). The birthrate parameter $b$, which compares current to past-average SFR, is typically four times higher in starbursts than in quiescent galaxies. While indirect indicators average over 0–100 Myr, burst durations are often just $\sim$5–10 Myr \citep{Sparre2017}, leading to systematic underestimation of current SFRs. Since these diagnostics reflect present ionization conditions, they may also fail to capture past star formation episodes or the delayed appearance of ionizing sources \citep{Shirazi12}.  The models used in this study cannot yet resolve such bursts at fine timescales with nebular emission modelling limited to a 5 Myr resolution \citep{Johnson2021}, limiting our ability to identify the most recent bursts that may drive the EUV emission.

Additional uncertainty arises from XRB variability. Measured X-ray fluxes can differ substantially from the flux at the epoch of SFR estimation varying in luminosity by factors of $>$10 on ~150–200 day timescales \citep{Webb2014}. In our sample, out of the targets that had multiple observations, the largest measured luminosity change was on the order of one magnitude (Section \ref{sec:results}). Dust attenuation represents another possible source of uncertainty, but we find that the effect on the integrated X-ray luminosity is minimal.

From an observational standpoint, an important uncertainty is introduced by the differing spatial regions over which the optical emission lines and X-ray emission are measured. The \ion{He}{2} emission is measured within the 3$\arcsec$ SDSS fibre apertures, which generally encompass larger regions than the Chandra extraction regions used to identify individual X-ray point sources. As a result, any extended or unresolved X-ray emission within the optical aperture (such as diffuse hot gas or XRBs below the point-source detection threshold) is not explicitly captured in our analysis. However, the total X-ray luminosity is expected to be dominated by the brightest detected sources, with the cumulative contribution from unresolved populations contributing only a modest fraction. 

Finally, additional observational uncertainties remain, most notably in distance estimates for nearby galaxies where redshift based distances (as used for the majority of our sample, see Section \ref{sec:samples}) are affected by peculiar velocities. Although these uncertainties do not alter our conclusions by more than an order of-magnitude, research of galaxy evolution in the nearby universe could benefit from continued follow-up distance measurements of nearby dwarf galaxies.

\subsection{Other potential Sources of He II} 
Both our population synthesis and observational studies suggest XRB's are unlikely to be a single-population source of He III nebulae in our sample. Although their luminosities do track metallicity and star formation activity, even in galaxies with elevated $L_{\mathrm{X}}$/SFR, the integrated X-ray output typically falls short of producing the ionizing flux needed to power the observed He II emission (Figure \ref{fig:Lehmer}). Reproducing the line strengths in such systems would require assuming unrealistically high luminosity per unit SFR, pointing to a more complex ionizing budget (e.g. \citealt{Schaerer19}, \citealt{Senchyna19}).

\begin{figure}
    \centering
    \includegraphics[width=1\linewidth]{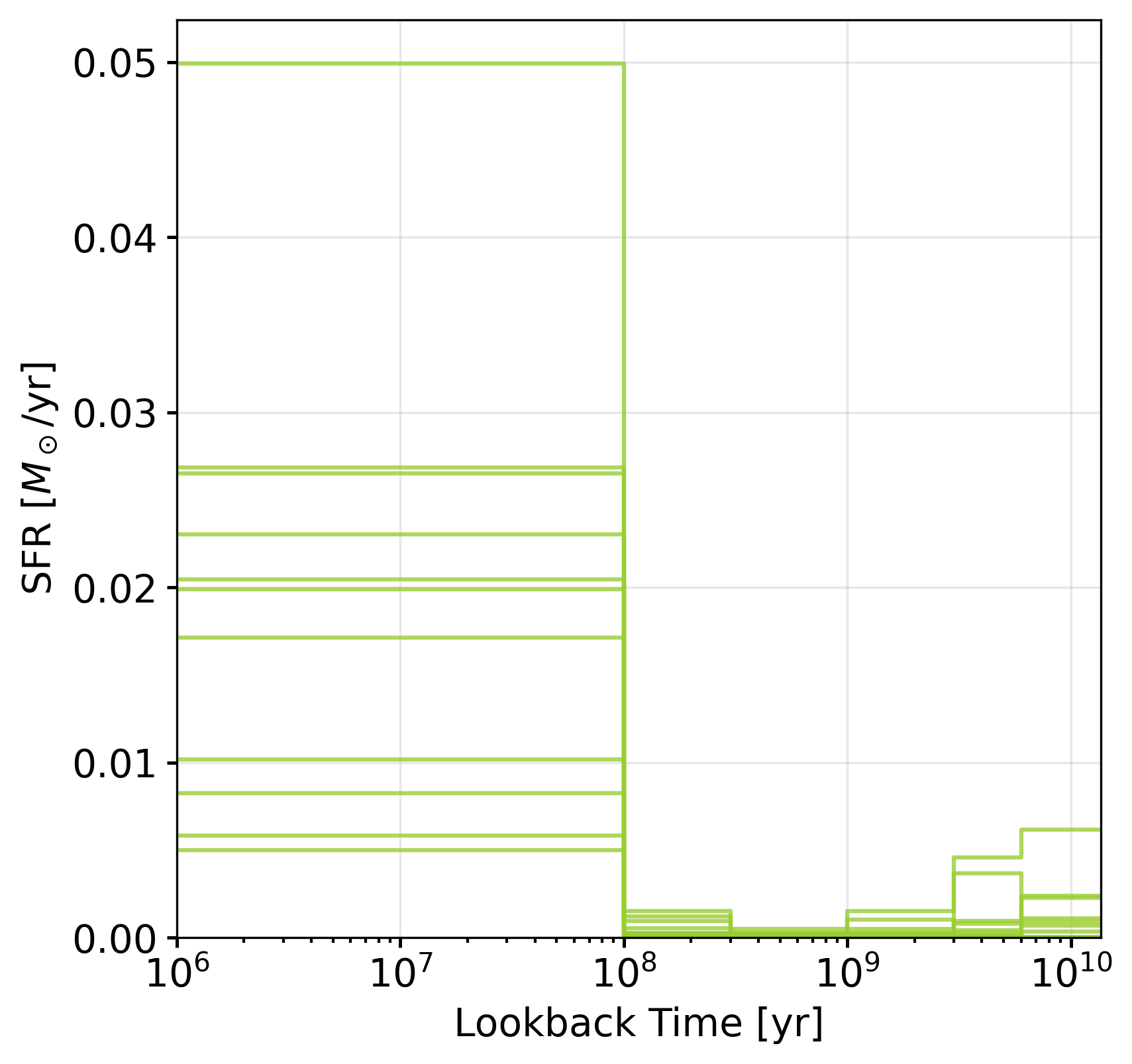}
    \caption{Non-parametric star formation histories (SFHs) derived for our subsample of 12 galaxies that we performed stellar population synthesis modelling on. All galaxies exhibit recent bursts of star formation within the past 100 Myr. This may indicate unresolved, bursty SFHs with rapid and intense star formation episodes at the current epoch.}
    \label{fig:SFH}
\end{figure}

\begin{figure*}
    \centering
    \includegraphics[width=0.9\linewidth]{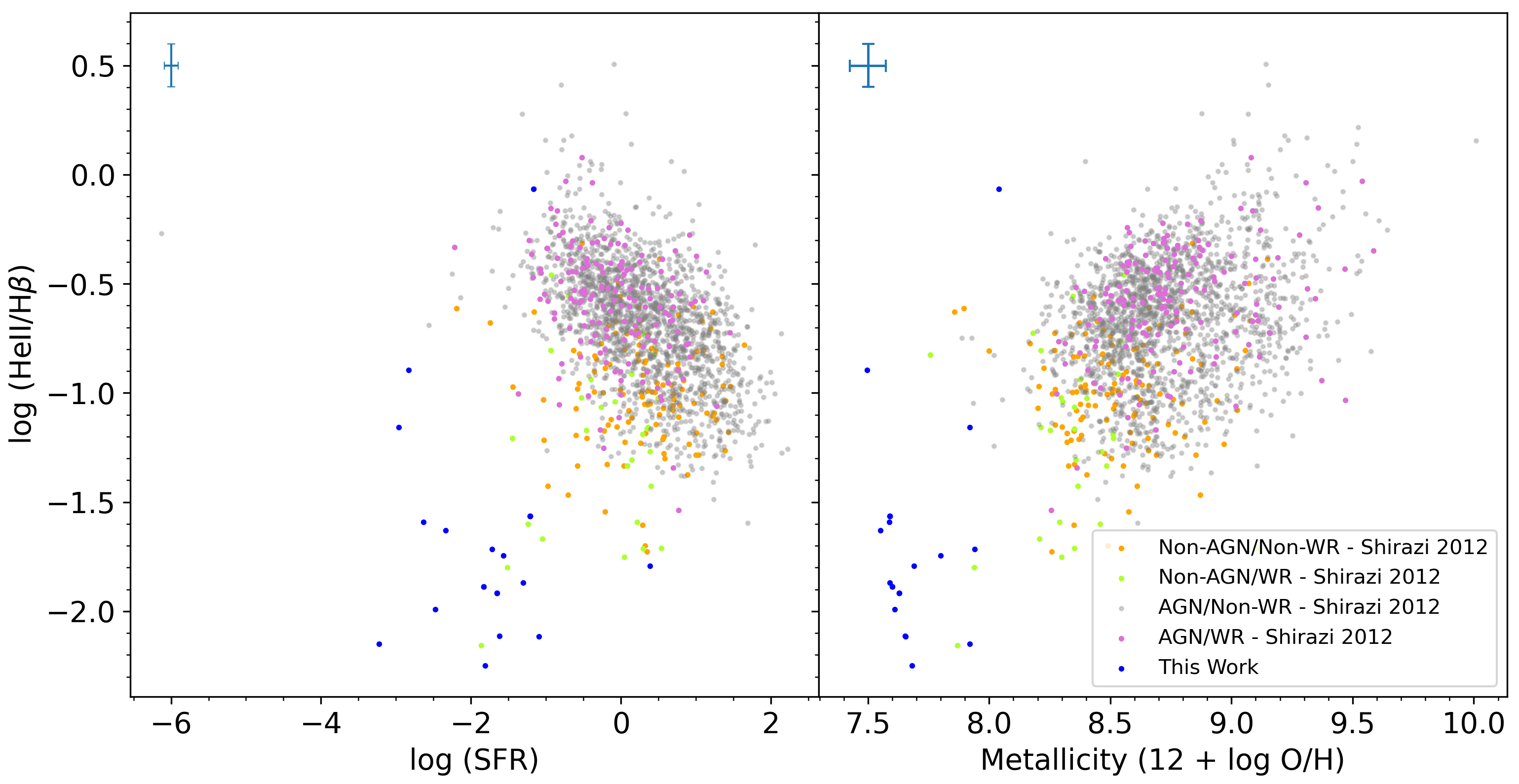}
    \caption{Comparison between our sample and the parent sample from \citet{Shirazi12}. The HeII/H$\beta$ line ratio is plotted against the star formation rate (SFR; left) and gas-phase metallicity (12 + log O/H; right). Values for our sample are derived following the methodology described in Section \ref{sec:spec_fitting}, including the characteristic SFR averaged over the past 10 Myr.}
    \label{fig:metallicity}
\end{figure*}

Binary stellar evolution introduces additional populations of hard ionizing sources that are not captured in single-star frameworks (e.g.\citealt{Eldridge2022}; \citealt{Stanway2014}). In particular, mass transfer and envelope stripping in binary systems have been shown to produce a continuum of hydrogen-deficient helium stars. These span from classical line-driven WR stars in high-metallicity environments down to companion-stripped helium stars in low-metallicity environments that no longer exhibit the strong, optically thick winds required for classical WR spectral classification \citep{Shenar2020}. Such stars can also be observationally challenging to identify because they are often accompanied by a more luminous mass-accreting companion, which can dominate the optical continuum and dilute diagnostic helium features (e.g. \citealt{Quintero2024}).

In this regime, systems may not be classified as WR stars observationally, despite contributing significantly to the EUV photon budget. Recent observational studies have now confirmed the existence of these stripped helium stars in nearby galaxies, directly supporting their predicted role as binary products giving a mechanism for these powerful ionizing sources to form in low-metallicty environments (e.g. \citealt{Drout2023}; \citealt{Gotberg2023}; \citealt{Quintero2024}; \citealt{Dougherty2000})

Another pathway to producing WR-like ionizing sources without classical WR spectral signatures is through quasi-homogeneous evolution in rapidly rotating massive stars (e.g. \citealt{Szcesi15}; \citealt{Xiao2018}). In this scenario, efficient rotational mixing continuously transports helium and nuclear-processed material from the core to the stellar surface. This prevents the formation of a strong chemical gradient between the core and envelope, allowing the star to remain well-mixed throughout much of its main-sequence lifetime, avoiding the red giant phase. This is favored in low-metallicity environments because reduced line opacities lead to weaker stellar winds, which in turn limits angular momentum loss and allows the star to maintain the high rotation rates required for sustained mixing \citep{Eldridge2022}. As a result, these stars can remain compact and hot, producing hard ionizing photons, without developing the dense, optically thick winds that generate prominent WR emission features \citep{Dorottya2025}.

A key implication of these results is that a substantial fraction of WR-like ionizing sources may remain observationally hidden or diluted in integrated spectra with dim, or missing, typical spectral features. As a result, our use of a sample without detection of strong WR features does not rule out a potential substantial contribution from binary-stripped helium stars and other exotic sources. 

Interestingly, these sources are also expected to constitute a significant fraction of Type Ib/c supernova progenitors (e.g. \citealt{Drout2023}; \citealt{Gotberg2023}). The few X-ray candidates capable of driving the He II ionization in our sample  were primarily associated with observations of supernovae and their remnants (Section \ref{inferredZEUV}), suggesting that in some low-metallicity dwarf galaxies, these objects might constitute a population of ionizing sources  (e.g. \citealt{Sun2012}, \citealt{Elmegreen2016}). 

Indeed, in low-metallicity environments, supernova remnants are expected to produce harder radiation owing to reduced cooling efficiencies \citep[e.g.,][]{Dopita2006}. These remnants can power extended soft X-ray emission ($L_{\rm X} \sim 10^{36-38}$ erg s$^{-1}$) on kiloparsec scales, while galactic winds driven by recent starbursts may generate diffuse X-ray halos \citep{Strickland2009}. Such diffuse components, although faint, can overlap spatially with nebular regions and thereby contribute to the ionization of low-density gas \citep{Lehmer2015}.

Another alternative source of ionizing radiation is emission from hot gas associated with stellar feedback and fast radiative shocks (e.g. \citealt{Dopita96}; \citealt{Thuan05}; \citealt{Plat19}). Recent models suggest that cluster winds and superbubbles can generate appreciable flux in low-metallicity environments. In particular, \citet{Oskinova2022} show that superbubble plasmas with characteristic temperatures below $\sim$2.5 MK and $L_{\rm X} \sim 10^{40}$ erg s$^{-1}$ naturally emit sufficient ionizing photons to reproduce the observed nebular He II strengths, without the need to invoke extreme binary populations.

Ubiquitously these types of sources can be defined as soft X-ray sources (0.1–2 keV) which may arise from multiple mechanisms, including hot gas in supernova-driven winds, shock-heated ISM, unresolved low-luminosity XRB populations, and radiative shocks from stellar feedback or outflows.

Another class of sources, supersoft X-ray sources, have recently been revived as plausible contributors. Although their individual luminosities are modest ($L_{\rm X} \sim 10^{36-37}$ erg s$^{-1}$), \cite{Triani2024} show that large populations of blackbody emitters with kT $\sim$ 10–100 eV, such as accreting white dwarfs, neutron stars, and black holes can elevate He II line ratios to observed levels when added to stellar UV spectra. Unfortunately these are fainter in Chandra’s observable bandpass ($>$0.3 keV) and thus largely missed in our survey. Additionally we show that at these energies these components are expected to be strongly obscured by dust for most galaxies although we did recover several galaxies with partial transmission warranting further study in the soft X-ray/UV (Figure \ref{fig:attenuation_curves}). 

Recent JWST NIRSpec observations have significantly advanced our understanding of nebular He II by delivering rest-frame UV spectroscopy of high-redshift star-forming galaxies with unprecedented sensitivity \citep{Schaerer2022}. It has been found that many high-redshift galaxies closely resemble their local-universe analogs, exhibiting characteristically bursty star-formation histories, rapid gas accretion, low stellar masses and metallicities, and an unexplained excess of ultraviolet emission (e.g., \citealt{Katz2024}, \citealt{Schaerer2022},  \citealt{Sandro2023}). In individual cases, such as the $z\approx2.98$ galaxy GNHeII J1236+6215, small pockets of highly ionized interstellar medium are observed, which are thought to be powered by young Population III stars or very low-metallicity massive stars rather than by AGN or WRs \citep{Mondal2025}. Conversely, \cite{Marco2024} show that UV line-intensity ratios in another system (GHZ2/GLASS-z12, z = 12.34) are compatible with both AGN and stellar ionization, predominantly favouring stellar origin for the ionizing photons.

Direct detections of high‑redshift XRB's are also beginning to emerge with JWST. For example, a spectroscopically confirmed galaxy pair at $z=2.544$ shows X‑ray emission dominated by high‑mass XRBs, consistent with the enhanced XRB contribution expected at low metallicity and high star formation rates in the early universe \citep{Cai2025}. Theoretical models of XRB populations also predict that XRBs can significantly contribute to the hard ionizing photon budget at high redshift \citep{Sartorio2023}. 

Although our observational results suggest that XRBs are generally secondary contributors, they may still play a supporting role in galaxies with compact or bursty star formation histories. In these environments, binary stellar evolution and feedback can rapidly produce diverse X-ray emitter populations with evolving spectral hardening (e.g., \citealt{Fragos2013}, \citealt{Prestwich2015}). Multiwavelength observations of dwarf galaxies (e.g., \citealt{Thygesen23}) reveal that off-nuclear X-ray sources associated with He II emission may be luminous XRBs or ultraluminous X-ray sources rather than AGN.

Taken together, this leaves room to consider the possibility where no single class of object is responsible for the full EUV photon budget in low-metallicity He II emitters. Instead, a multi-channel model, where diffuse hard, soft, and super soft X-ray sources, supernova remnants, stellar populations, and shocks act in concert may better account for both the diversity of observed He II emission sources, as well as the wide range of X-ray-to-SFR ratios \citep[e.g.,][]{Senchyna20}. 

If nebular He II emission in low-metallicity galaxies is not dominated by a single population of hard ionizing sources, such as XRBs, but instead originates from a mixture of mechanisms, then extrapolations based solely on local X-ray–SFR relations (e.g., \citealt{Brorby2016}, \citealt{Lehmer2010}) may systematically underestimate the EUV output of early galaxies (Figure \ref{fig:SED}). Several of these channels can contribute substantially to He II-ionizing photon production while generating little detectable emission in the Chandra band, leading to biases in estimates of the ionizing radiation available for escape and propagation during cosmic reionization. Disentangling the individual contributions therefore requires coordinated multiwavelength investigations, including X-ray imaging, optical and ultraviolet spectroscopy, and radio continuum mapping, to fully resolve the complex interplay of ionizing channels across diverse galactic environments.

\section{Conclusion} \label{sec:conclusion}
Through a combined analysis of Chandra X-ray data and stellar population synthesis/photoionization modelling, we have constrained both the environments of He II–emitting dwarf galaxies and the ionizing photon budget required to sustain their observed emission. Our optical SED fits confirm that these galaxies are low-metallicity systems with little dust and recent, bursty star formation. At the same time, extrapolated EUV continua from observed compact accreting sources consistently fall short of the flux required to reproduce the He II $\lambda4686$ line, even though their X-ray luminosities are in line with empirical expectations from metallicity and SFR. This persistent shortfall indicates that standard stellar and accreting compact object populations alone cannot account for the ionization conditions. This implies new physics may be needed, or that alternate sources (supernova remnants, super bubble winds, or previously undetected stellar populations) must be systematically included in future population models. Deep far-UV or soft X-ray surveys, and improved theoretical modelling of bursty star formation and feedback, will be critical for moving forward.

\section{Data Availability}
All the {\it SDSS} and {\it GALEX} data used in this paper can be found in MAST: \dataset[10.17909/16df-fg78]{https://doi.org/10.17909/16df-fg78}

\begin{acknowledgments}
The authors would like to thank the anonymous reviewer for their valuable input. IA acknowledges modelling advice from Joel Leja. IA acknowledges support by NASA NVSGC under award No. 80NSSC20M0043.
RMP acknowledges support from NASA under award No. 80NSSC23M0104.
RS acknowledges support from the INAF grant number 1.05.23.04.04. 
ET has been supported by the Australian Government through the Australian Research Council Discovery Project Grant 240101842.
AER gratefully acknowledges support provided by NSF through CAREER award 2235277.
This work is supported by NASA under award number 80GSFC24M0006.

Funding for the Sloan Digital Sky Survey V has been provided by the Alfred P. Sloan Foundation, the Heising-Simons Foundation, the National Science Foundation, and the Participating Institutions. SDSS acknowledges support and resources from the Center for High-Performance Computing at the University of Utah. The SDSS web site is \url{www.sdss.org}.

SDSS is managed by the Astrophysical Research Consortium for the Participating Institutions of the SDSS Collaboration, including the Carnegie Institution for Science, Chilean National Time Allocation Committee (CNTAC) ratified researchers, the Gotham Participation Group, Harvard University, Heidelberg University, The Johns Hopkins University, L’Ecole polytechnique f\'{e}d\'{e}rale de Lausanne (EPFL), Leibniz-Institut f\"{u}r Astrophysik Potsdam (AIP), Max-Planck-Institut f\"{u}r Astronomie (MPIA Heidelberg), Max-Planck-Institut f\"{u}r Extraterrestrische Physik (MPE), Nanjing University, National Astronomical Observatories of China (NAOC), New Mexico State University, The Ohio State University, Pennsylvania State University, Smithsonian Astrophysical Observatory, Space Telescope Science Institute (STScI), the Stellar Astrophysics Participation Group, Universidad Nacional Aut\'{o}noma de M\'{e}xico, University of Arizona, University of Colorado Boulder, University of Illinois at Urbana-Champaign, University of Toronto, University of Utah, University of Virginia, Yale University, and Yunnan University.

The scientific results reported in this article are based on observations made by the Chandra X-ray Observatory and data obtained from the Chandra Data Archive. This research has made use of software provided by the Chandra X-ray Center (CXC) in the application packages CIAO. 

This research has made use of the SIMBAD database, operated at CDS, Strasbourg, France. 

This research has made use of the NASA/IPAC Extragalactic Database (NED),
which is operated by the Jet Propulsion Laboratory, California Institute of Technology,
under contract with the National Aeronautics and Space Administration.

This work made use of Astropy:\footnote{http://www.astropy.org} a community-developed core Python package and an ecosystem of tools and resources for astronomy \citep{astropy:2013, astropy:2018, astropy:2022}.

\end{acknowledgments}

\vspace{5mm}
\facilities{CXO}

\software{Prospector \citep{Johnson2021, Leja2019}, CIAO 4.14 \citep{Fruscione06}, Astropy \citep{astropy:2013, astropy:2018, astropy:2022}}

\bibliography{References}{}

\begin{thebibliography}{}
\expandafter\ifx\csname natexlab\endcsname\relax\def\natexlab#1{#1}\fi
\providecommand{\url}[1]{\href{#1}{#1}}
\providecommand{\dodoi}[1]{doi:~\href{http://doi.org/#1}{\nolinkurl{#1}}}
\providecommand{\doeprint}[1]{\href{http://ascl.net/#1}{\nolinkurl{http://ascl.net/#1}}}
\providecommand{\doarXiv}[1]{\href{https://arxiv.org/abs/#1}{\nolinkurl{https://arxiv.org/abs/#1}}}

\bibitem[{{Abazajian} {et~al.}(2009){Abazajian}, {Adelman-McCarthy}, {Ag{\"u}eros}, {Allam}, {Allende Prieto}, {An}, {Anderson}, {Anderson}, {Annis}, {Bahcall}, {Bailer-Jones}, {Barentine}, {Bassett}, {Becker}, {Beers}, {Bell}, {Belokurov}, {Berlind}, {Berman}, {Bernardi}, {Bickerton}, {Bizyaev}, {Blakeslee}, {Blanton}, {Bochanski}, {Boroski}, {Brewington}, {Brinchmann}, {Brinkmann}, {Brunner}, {Budav{\'a}ri}, {Carey}, {Carliles}, {Carr}, {Castander}, {Cinabro}, {Connolly}, {Csabai}, {Cunha}, {Czarapata}, {Davenport}, {de Haas}, {Dilday}, {Doi}, {Eisenstein}, {Evans}, {Evans}, {Fan}, {Friedman}, {Frieman}, {Fukugita}, {G{\"a}nsicke}, {Gates}, {Gillespie}, {Gilmore}, {Gonzalez}, {Gonzalez}, {Grebel}, {Gunn}, {Gy{\"o}ry}, {Hall}, {Harding}, {Harris}, {Harvanek}, {Hawley}, {Hayes}, {Heckman}, {Hendry}, {Hennessy}, {Hindsley}, {Hoblitt}, {Hogan}, {Hogg}, {Holtzman}, {Hyde}, {Ichikawa}, {Ichikawa}, {Im}, {Ivezi{\'c}}, {Jester}, {Jiang}, {Johnson}, {Jorgensen}, {Juri{\'c}}, {Kent}, {Kessler}, {Kleinman}, {Knapp},
  {Konishi}, {Kron}, {Krzesinski}, {Kuropatkin}, {Lampeitl}, {Lebedeva}, {Lee}, {Lee}, {French Leger}, {L{\'e}pine}, {Li}, {Lima}, {Lin}, {Long}, {Loomis}, {Loveday}, {Lupton}, {Magnier}, {Malanushenko}, {Malanushenko}, {Mandelbaum}, {Margon}, {Marriner}, {Mart{\'\i}nez-Delgado}, {Matsubara}, {McGehee}, {McKay}, {Meiksin}, {Morrison}, {Mullally}, {Munn}, {Murphy}, {Nash}, {Nebot}, {Neilsen}, {Newberg}, {Newman}, {Nichol}, {Nicinski}, {Nieto-Santisteban}, {Nitta}, {Okamura}, {Oravetz}, {Ostriker}, {Owen}, {Padmanabhan}, {Pan}, {Park}, {Pauls}, {Peoples}, {Percival}, {Pier}, {Pope}, {Pourbaix}, {Price}, {Purger}, {Quinn}, {Raddick}, {Re Fiorentin}, {Richards}, {Richmond}, {Riess}, {Rix}, {Rockosi}, {Sako}, {Schlegel}, {Schneider}, {Scholz}, {Schreiber}, {Schwope}, {Seljak}, {Sesar}, {Sheldon}, {Shimasaku}, {Sibley}, {Simmons}, {Sivarani}, {Allyn Smith}, {Smith}, {Smol{\v{c}}i{\'c}}, {Snedden}, {Stebbins}, {Steinmetz}, {Stoughton}, {Strauss}, {SubbaRao}, {Suto}, {Szalay}, {Szapudi}, {Szkody}, {Tanaka},
  {Tegmark}, {Teodoro}, {Thakar}, {Tremonti}, {Tucker}, {Uomoto}, {Vanden Berk}, {Vandenberg}, {Vidrih}, {Vogeley}, {Voges}, {Vogt}, {Wadadekar}, {Watters}, {Weinberg}, {West}, {White}, {Wilhite}, {Wonders}, {Yanny}, {Yocum}, {York}, {Zehavi}, {Zibetti}, \& {Zucker}}]{sdssdr7}
{Abazajian}, K.~N., {Adelman-McCarthy}, J.~K., {Ag{\"u}eros}, M.~A., {et~al.} 2009, \apjs, 182, 543, \dodoi{10.1088/0067-0049/182/2/543}

\bibitem[{{Alam} {et~al.}(2015){Alam}, {Albareti}, {Allende Prieto}, {Anders}, {Anderson}, {Anderton}, {Andrews}, {Armengaud}, {Aubourg}, {Bailey}, {Basu}, {Bautista}, {Beaton}, {Beers}, {Bender}, {Berlind}, {Beutler}, {Bhardwaj}, {Bird}, {Bizyaev}, {Blake}, {Blanton}, {Blomqvist}, {Bochanski}, {Bolton}, {Bovy}, {Shelden Bradley}, {Brandt}, {Brauer}, {Brinkmann}, {Brown}, {Brownstein}, {Burden}, {Burtin}, {Busca}, {Cai}, {Capozzi}, {Carnero Rosell}, {Carr}, {Carrera}, {Chambers}, {Chaplin}, {Chen}, {Chiappini}, {Chojnowski}, {Chuang}, {Clerc}, {Comparat}, {Covey}, {Croft}, {Cuesta}, {Cunha}, {da Costa}, {Da Rio}, {Davenport}, {Dawson}, {De Lee}, {Delubac}, {Deshpande}, {Dhital}, {Dutra-Ferreira}, {Dwelly}, {Ealet}, {Ebelke}, {Edmondson}, {Eisenstein}, {Ellsworth}, {Elsworth}, {Epstein}, {Eracleous}, {Escoffier}, {Esposito}, {Evans}, {Fan}, {Fern{\'a}ndez-Alvar}, {Feuillet}, {Filiz Ak}, {Finley}, {Finoguenov}, {Flaherty}, {Fleming}, {Font-Ribera}, {Foster}, {Frinchaboy}, {Galbraith-Frew}, {Garc{\'\i}a},
  {Garc{\'\i}a-Hern{\'a}ndez}, {Garc{\'\i}a P{\'e}rez}, {Gaulme}, {Ge}, {G{\'e}nova-Santos}, {Georgakakis}, {Ghezzi}, {Gillespie}, {Girardi}, {Goddard}, {Gontcho}, {Gonz{\'a}lez Hern{\'a}ndez}, {Grebel}, {Green}, {Grieb}, {Grieves}, {Gunn}, {Guo}, {Harding}, {Hasselquist}, {Hawley}, {Hayden}, {Hearty}, {Hekker}, {Ho}, {Hogg}, {Holley-Bockelmann}, {Holtzman}, {Honscheid}, {Huber}, {Huehnerhoff}, {Ivans}, {Jiang}, {Johnson}, {Kinemuchi}, {Kirkby}, {Kitaura}, {Klaene}, {Knapp}, {Kneib}, {Koenig}, {Lam}, {Lan}, {Lang}, {Laurent}, {Le Goff}, {Leauthaud}, {Lee}, {Lee}, {Licquia}, {Liu}, {Long}, {L{\'o}pez-Corredoira}, {Lorenzo-Oliveira}, {Lucatello}, {Lundgren}, {Lupton}, {Mack}, {Mahadevan}, {Maia}, {Majewski}, {Malanushenko}, {Malanushenko}, {Manchado}, {Manera}, {Mao}, {Maraston}, {Marchwinski}, {Margala}, {Martell}, {Martig}, {Masters}, {Mathur}, {McBride}, {McGehee}, {McGreer}, {McMahon}, {M{\'e}nard}, {Menzel}, {Merloni}, {M{\'e}sz{\'a}ros}, {Miller}, {Miralda-Escud{\'e}}, {Miyatake}, {Montero-Dorta}, {More},
  {Morganson}, {Morice-Atkinson}, {Morrison}, {Mosser}, {Muna}, {Myers}, {Nandra}, {Newman}, {Neyrinck}, {Nguyen}, {Nichol}, {Nidever}, {Noterdaeme}, {Nuza}, {O'Connell}, {O'Connell}, {O'Connell}, {Ogando}, {Olmstead}, {Oravetz}, {Oravetz}, {Osumi}, {Owen}, {Padgett}, {Padmanabhan}, {Paegert}, {Palanque-Delabrouille}, \& {Pan}}]{sdssdr12}
{Alam}, S., {Albareti}, F.~D., {Allende Prieto}, C., {et~al.} 2015, \apjs, 219, 12, \dodoi{10.1088/0067-0049/219/1/12}

\bibitem[{{Astropy Collaboration} {et~al.}(2013){Astropy Collaboration}, {Robitaille}, {Tollerud}, {Greenfield}, {Droettboom}, {Bray}, {Aldcroft}, {Davis}, {Ginsburg}, {Price-Whelan}, {Kerzendorf}, {Conley}, {Crighton}, {Barbary}, {Muna}, {Ferguson}, {Grollier}, {Parikh}, {Nair}, {Unther}, {Deil}, {Woillez}, {Conseil}, {Kramer}, {Turner}, {Singer}, {Fox}, {Weaver}, {Zabalza}, {Edwards}, {Azalee Bostroem}, {Burke}, {Casey}, {Crawford}, {Dencheva}, {Ely}, {Jenness}, {Labrie}, {Lim}, {Pierfederici}, {Pontzen}, {Ptak}, {Refsdal}, {Servillat}, \& {Streicher}}]{astropy:2013}
{Astropy Collaboration}, {Robitaille}, T.~P., {Tollerud}, E.~J., {et~al.} 2013, \aap, 558, A33, \dodoi{10.1051/0004-6361/201322068}

\bibitem[{{Astropy Collaboration} {et~al.}(2018){Astropy Collaboration}, {Price-Whelan}, {Sip{\H{o}}cz}, {G{\"u}nther}, {Lim}, {Crawford}, {Conseil}, {Shupe}, {Craig}, {Dencheva}, {Ginsburg}, {Vand erPlas}, {Bradley}, {P{\'e}rez-Su{\'a}rez}, {de Val-Borro}, {Aldcroft}, {Cruz}, {Robitaille}, {Tollerud}, {Ardelean}, {Babej}, {Bach}, {Bachetti}, {Bakanov}, {Bamford}, {Barentsen}, {Barmby}, {Baumbach}, {Berry}, {Biscani}, {Boquien}, {Bostroem}, {Bouma}, {Brammer}, {Bray}, {Breytenbach}, {Buddelmeijer}, {Burke}, {Calderone}, {Cano Rodr{\'\i}guez}, {Cara}, {Cardoso}, {Cheedella}, {Copin}, {Corrales}, {Crichton}, {D'Avella}, {Deil}, {Depagne}, {Dietrich}, {Donath}, {Droettboom}, {Earl}, {Erben}, {Fabbro}, {Ferreira}, {Finethy}, {Fox}, {Garrison}, {Gibbons}, {Goldstein}, {Gommers}, {Greco}, {Greenfield}, {Groener}, {Grollier}, {Hagen}, {Hirst}, {Homeier}, {Horton}, {Hosseinzadeh}, {Hu}, {Hunkeler}, {Ivezi{\'c}}, {Jain}, {Jenness}, {Kanarek}, {Kendrew}, {Kern}, {Kerzendorf}, {Khvalko}, {King}, {Kirkby}, {Kulkarni},
  {Kumar}, {Lee}, {Lenz}, {Littlefair}, {Ma}, {Macleod}, {Mastropietro}, {McCully}, {Montagnac}, {Morris}, {Mueller}, {Mumford}, {Muna}, {Murphy}, {Nelson}, {Nguyen}, {Ninan}, {N{\"o}the}, {Ogaz}, {Oh}, {Parejko}, {Parley}, {Pascual}, {Patil}, {Patil}, {Plunkett}, {Prochaska}, {Rastogi}, {Reddy Janga}, {Sabater}, {Sakurikar}, {Seifert}, {Sherbert}, {Sherwood-Taylor}, {Shih}, {Sick}, {Silbiger}, {Singanamalla}, {Singer}, {Sladen}, {Sooley}, {Sornarajah}, {Streicher}, {Teuben}, {Thomas}, {Tremblay}, {Turner}, {Terr{\'o}n}, {van Kerkwijk}, {de la Vega}, {Watkins}, {Weaver}, {Whitmore}, {Woillez}, {Zabalza}, \& {Astropy Contributors}}]{astropy:2018}
{Astropy Collaboration}, {Price-Whelan}, A.~M., {Sip{\H{o}}cz}, B.~M., {et~al.} 2018, \aj, 156, 123, \dodoi{10.3847/1538-3881/aabc4f}

\bibitem[{{Astropy Collaboration} {et~al.}(2022){Astropy Collaboration}, {Price-Whelan}, {Lim}, {Earl}, {Starkman}, {Bradley}, {Shupe}, {Patil}, {Corrales}, {Brasseur}, {N{"o}the}, {Donath}, {Tollerud}, {Morris}, {Ginsburg}, {Vaher}, {Weaver}, {Tocknell}, {Jamieson}, {van Kerkwijk}, {Robitaille}, {Merry}, {Bachetti}, {G{"u}nther}, {Aldcroft}, {Alvarado-Montes}, {Archibald}, {B{'o}di}, {Bapat}, {Barentsen}, {Baz{'a}n}, {Biswas}, {Boquien}, {Burke}, {Cara}, {Cara}, {Conroy}, {Conseil}, {Craig}, {Cross}, {Cruz}, {D'Eugenio}, {Dencheva}, {Devillepoix}, {Dietrich}, {Eigenbrot}, {Erben}, {Ferreira}, {Foreman-Mackey}, {Fox}, {Freij}, {Garg}, {Geda}, {Glattly}, {Gondhalekar}, {Gordon}, {Grant}, {Greenfield}, {Groener}, {Guest}, {Gurovich}, {Handberg}, {Hart}, {Hatfield-Dodds}, {Homeier}, {Hosseinzadeh}, {Jenness}, {Jones}, {Joseph}, {Kalmbach}, {Karamehmetoglu}, {Ka{l}uszy{'n}ski}, {Kelley}, {Kern}, {Kerzendorf}, {Koch}, {Kulumani}, {Lee}, {Ly}, {Ma}, {MacBride}, {Maljaars}, {Muna}, {Murphy}, {Norman}, {O'Steen},
  {Oman}, {Pacifici}, {Pascual}, {Pascual-Granado}, {Patil}, {Perren}, {Pickering}, {Rastogi}, {Roulston}, {Ryan}, {Rykoff}, {Sabater}, {Sakurikar}, {Salgado}, {Sanghi}, {Saunders}, {Savchenko}, {Schwardt}, {Seifert-Eckert}, {Shih}, {Jain}, {Shukla}, {Sick}, {Simpson}, {Singanamalla}, {Singer}, {Singhal}, {Sinha}, {Sip{H{o}}cz}, {Spitler}, {Stansby}, {Streicher}, {{{S}}umak}, {Swinbank}, {Taranu}, {Tewary}, {Tremblay}, {Val-Borro}, {Van Kooten}, {Vasovi{'c}}, {Verma}, {de Miranda Cardoso}, {Williams}, {Wilson}, {Winkel}, {Wood-Vasey}, {Xue}, {Yoachim}, {Zhang}, {Zonca}, \& {Astropy Project Contributors}}]{astropy:2022}
{Astropy Collaboration}, {Price-Whelan}, A.~M., {Lim}, P.~L., {et~al.} 2022, \apj, 935, 167, \dodoi{10.3847/1538-4357/ac7c74}

\bibitem[{{Baldwin} {et~al.}(1981){Baldwin}, {Phillips}, \& {Terlevich}}]{Baldwin1981}
{Baldwin}, J.~A., {Phillips}, M.~M., \& {Terlevich}, R. 1981, \pasp, 93, 5, \dodoi{10.1086/130766}

\bibitem[{{Basu-Zych} {et~al.}(2013){Basu-Zych}, {Lehmer}, {Hornschemeier}, {Gon{\c{c}}alves}, {Fragos}, {Heckman}, {Overzier}, {Ptak}, \& {Schiminovich}}]{Basu-Zych13}
{Basu-Zych}, A.~R., {Lehmer}, B.~D., {Hornschemeier}, A.~E., {et~al.} 2013, \apj, 774, 152, \dodoi{10.1088/0004-637X/774/2/152}

\bibitem[{{Berg} {et~al.}(2018){Berg}, {Erb}, {Auger}, {Pettini}, \& {Brammer}}]{Berg2018}
{Berg}, D.~A., {Erb}, D.~K., {Auger}, M.~W., {Pettini}, M., \& {Brammer}, G.~B. 2018, \apj, 859, 164, \dodoi{10.3847/1538-4357/aab7fa}

\bibitem[{{Bray} {et~al.}(2025){Bray}, {Stanway}, \& {Eldridge}}]{Bray2025}
{Bray}, J.~C., {Stanway}, E.~R., \& {Eldridge}, J.~J. 2025, \mnras, 542, 2087, \dodoi{10.1093/mnras/staf1348}

\bibitem[{{Brinchmann} {et~al.}(2004){Brinchmann}, {Charlot}, {White}, {Tremonti}, {Kauffmann}, {Heckman}, \& {Brinkmann}}]{Brinchmann2004}
{Brinchmann}, J., {Charlot}, S., {White}, S.~D.~M., {et~al.} 2004, \mnras, 351, 1151, \dodoi{10.1111/j.1365-2966.2004.07881.x}

\bibitem[{{Brinkman} {et~al.}(2002){Brinkman}, {Kaastra}, {van der Meer}, {Kinkhabwala}, {Behar}, {Kahn}, {Paerels}, \& {Sako}}]{Brinkman02}
{Brinkman}, A.~C., {Kaastra}, J.~S., {van der Meer}, R.~L.~J., {et~al.} 2002, \aap, 396, 761, \dodoi{10.1051/0004-6361:20020918}

\bibitem[{{Brorby} {et~al.}(2014){Brorby}, {Kaaret}, \& {Prestwich}}]{Brorby14}
{Brorby}, M., {Kaaret}, P., \& {Prestwich}, A. 2014, \mnras, 441, 2346, \dodoi{10.1093/mnras/stu736}

\bibitem[{{Brorby} {et~al.}(2016){Brorby}, {Kaaret}, {Prestwich}, \& {Mirabel}}]{Brorby2016}
{Brorby}, M., {Kaaret}, P., {Prestwich}, A., \& {Mirabel}, I.~F. 2016, \mnras, 457, 4081, \dodoi{10.1093/mnras/stw284}

\bibitem[{{Byler} {et~al.}(2017){Byler}, {Dalcanton}, {Conroy}, \& {Johnson}}]{Byler2017}
{Byler}, N., {Dalcanton}, J.~J., {Conroy}, C., \& {Johnson}, B.~D. 2017, \apj, 840, 44, \dodoi{10.3847/1538-4357/aa6c66}

\bibitem[{{Byrne} {et~al.}(2025){Byrne}, {Eldridge}, \& {Stanway}}]{Byrne25}
{Byrne}, C.~M., {Eldridge}, J.~J., \& {Stanway}, E.~R. 2025, \mnras, 537, 2433, \dodoi{10.1093/mnras/staf178}

\bibitem[{{Cai} {et~al.}(2025){Cai}, {Cai}, {Lyu}, {Wu}, {Lin}, {Li}, {Mao}, {Chen}, \& {Lu}}]{Cai2025}
{Cai}, S., {Cai}, Z., {Lyu}, J., {et~al.} 2025, \apjl, 989, L35, \dodoi{10.3847/2041-8213/adf5bc}

\bibitem[{Calzetii(2013)}]{Calzetti2013}
Calzetii, D. 2013, Cambridge University Press

\bibitem[{{Calzetti} {et~al.}(2000){Calzetti}, {Armus}, {Bohlin}, {Kinney}, {Koornneef}, \& {Storchi-Bergmann}}]{Calzetti2000}
{Calzetti}, D., {Armus}, L., {Bohlin}, R.~C., {et~al.} 2000, \apj, 533, 682, \dodoi{10.1086/308692}

\bibitem[{Cash(1979)}]{Cash79}
Cash, W. 1979, Astrophysical Journal, 228, 939

\bibitem[{{Castellano} {et~al.}(2024){Castellano}, {Napolitano}, {Fontana}, {Roberts-Borsani}, {Treu}, {Vanzella}, {Zavala}, {Arrabal Haro}, {Calabr{\`o}}, {Llerena}, {Mascia}, {Merlin}, {Paris}, {Pentericci}, {Santini}, {Bakx}, {Bergamini}, {Cupani}, {Dickinson}, {Filippenko}, {Glazebrook}, {Grillo}, {Kelly}, {Malkan}, {Mason}, {Morishita}, {Nanayakkara}, {Rosati}, {Sani}, {Wang}, \& {Yoon}}]{Marco2024}
{Castellano}, M., {Napolitano}, L., {Fontana}, A., {et~al.} 2024, \apj, 972, 143, \dodoi{10.3847/1538-4357/ad5f88}

\bibitem[{{Charlot} \& {Fall}(2000{\natexlab{a}})}]{Charlot2000}
{Charlot}, S., \& {Fall}, S.~M. 2000{\natexlab{a}}, \apj, 539, 718, \dodoi{10.1086/309250}

\bibitem[{{Charlot} \& {Fall}(2000{\natexlab{b}})}]{CharlotFall2000}
---. 2000{\natexlab{b}}, \apj, 539, 718, \dodoi{10.1086/309250}

\bibitem[{{Charlot} \& {Longhetti}(2001)}]{Charlot2001}
{Charlot}, S., \& {Longhetti}, M. 2001, \mnras, 323, 887, \dodoi{10.1046/j.1365-8711.2001.04260.x}

\bibitem[{{Conroy}(2013)}]{Conroy2013}
{Conroy}, C. 2013, \araa, 51, 393, \dodoi{10.1146/annurev-astro-082812-141017}

\bibitem[{{Conroy} \& {Gunn}(2010)}]{Conroy2010}
{Conroy}, C., \& {Gunn}, J.~E. 2010, \apj, 712, 833, \dodoi{10.1088/0004-637X/712/2/833}

\bibitem[{{Conroy} {et~al.}(2009){Conroy}, {Gunn}, \& {White}}]{Conroy2009}
{Conroy}, C., {Gunn}, J.~E., \& {White}, M. 2009, \apj, 699, 486, \dodoi{10.1088/0004-637X/699/1/486}

\bibitem[{{Crowther}(2007)}]{Crowther07}
{Crowther}, P.~A. 2007, \araa, 45, 177, \dodoi{10.1146/annurev.astro.45.051806.110615}

\bibitem[{{de Mello} {et~al.}(1998){de Mello}, {Schaerer}, {Heldmann}, \& {Leitherer}}]{deMello98}
{de Mello}, D.~F., {Schaerer}, D., {Heldmann}, J., \& {Leitherer}, C. 1998, \apj, 507, 199, \dodoi{10.1086/306317}

\bibitem[{{Dopita} \& {Sutherland}(1996)}]{Dopita96}
{Dopita}, M.~A., \& {Sutherland}, R.~S. 1996, \apjs, 102, 161, \dodoi{10.1086/192255}

\bibitem[{{Dopita} {et~al.}(2006){Dopita}, {Fischera}, {Sutherland}, {Kewley}, {Leitherer}, {Tuffs}, {Popescu}, {van Breugel}, \& {Groves}}]{Dopita2006}
{Dopita}, M.~A., {Fischera}, J., {Sutherland}, R.~S., {et~al.} 2006, \apjs, 167, 177, \dodoi{10.1086/508261}

\bibitem[{Dotter(2016)}]{Dotter2016}
Dotter, A. 2016, The Astrophysical Journal Supplement Series, 222, \dodoi{10.3847/0067-0049/222/1/8}

\bibitem[{{Dougherty} {et~al.}(2000){Dougherty}, {Williams}, \& {Pollacco}}]{Dougherty2000}
{Dougherty}, S.~M., {Williams}, P.~M., \& {Pollacco}, D.~L. 2000, \mnras, 316, 143, \dodoi{10.1046/j.1365-8711.2000.03504.x}

\bibitem[{{Douna} {et~al.}(2015){Douna}, {Pellizza}, {Mirabel}, \& {Pedrosa}}]{Douna15}
{Douna}, V.~M., {Pellizza}, L.~J., {Mirabel}, I.~F., \& {Pedrosa}, S.~E. 2015, \aap, 579, A44, \dodoi{10.1051/0004-6361/201525617}

\bibitem[{{Draine} \& {Li}(2007)}]{Draine2007}
{Draine}, B.~T., \& {Li}, A. 2007, \apj, 657, 810, \dodoi{10.1086/511055}

\bibitem[{{Drout} {et~al.}(2023){Drout}, {G{\"o}tberg}, {Ludwig}, {Groh}, {de Mink}, {O'Grady}, \& {Smith}}]{Drout2023}
{Drout}, M.~R., {G{\"o}tberg}, Y., {Ludwig}, B.~A., {et~al.} 2023, Science, 382, 1287, \dodoi{10.1126/science.ade4970}

\bibitem[{{Eldridge} \& {Stanway}(2022)}]{Eldridge2022}
{Eldridge}, J.~J., \& {Stanway}, E.~R. 2022, \araa, 60, 455, \dodoi{10.1146/annurev-astro-052920-100646}

\bibitem[{{Eldridge} {et~al.}(2017){Eldridge}, {Stanway}, {Xiao}, {McClelland}, {Taylor}, {Ng}, {Greis}, \& {Bray}}]{Eldridge17}
{Eldridge}, J.~J., {Stanway}, E.~R., {Xiao}, L., {et~al.} 2017, \pasa, 34, e058, \dodoi{10.1017/pasa.2017.51}

\bibitem[{{Elmegreen} {et~al.}(2016){Elmegreen}, {Elmegreen}, {S{\'a}nchez Almeida}, {Mu{\~n}oz-Tu{\~n}{\'o}n}, {Mendez-Abreu}, {Gallagher}, {Rafelski}, {Filho}, \& {Ceverino}}]{Elmegreen2016}
{Elmegreen}, D.~M., {Elmegreen}, B.~G., {S{\'a}nchez Almeida}, J., {et~al.} 2016, \apj, 825, 145, \dodoi{10.3847/0004-637X/825/2/145}

\bibitem[{{Emami} {et~al.}(2019){Emami}, {Siana}, {Weisz}, {Johnson}, {Ma}, \& {El-Badry}}]{Emami2019}
{Emami}, N., {Siana}, B., {Weisz}, D.~R., {et~al.} 2019, \apj, 881, 71, \dodoi{10.3847/1538-4357/ab211a}

\bibitem[{{Estrada-Carpenter} {et~al.}(2020){Estrada-Carpenter}, {Papovich}, {Momcheva}, {Brammer}, {Simons}, {Bridge}, {Cleri}, {Ferguson}, {Finkelstein}, {Giavalisco}, {Jung}, {Matharu}, {Trump}, \& {Weiner}}]{Estrada2020}
{Estrada-Carpenter}, V., {Papovich}, C., {Momcheva}, I., {et~al.} 2020, \apj, 898, 171, \dodoi{10.3847/1538-4357/aba004}

\bibitem[{{Fabian}(2012)}]{Fabian12}
{Fabian}, A.~C. 2012, \araa, 50, 455, \dodoi{10.1146/annurev-astro-081811-125521}

\bibitem[{{Falc{\'o}n-Barroso} {et~al.}(2011){Falc{\'o}n-Barroso}, {S{\'a}nchez-Bl{\'a}zquez}, {Vazdekis}, {Ricciardelli}, {Cardiel}, {Cenarro}, {Gorgas}, \& {Peletier}}]{Falcon2011}
{Falc{\'o}n-Barroso}, J., {S{\'a}nchez-Bl{\'a}zquez}, P., {Vazdekis}, A., {et~al.} 2011, \aap, 532, A95, \dodoi{10.1051/0004-6361/201116842}

\bibitem[{{Fisher} {et~al.}(2014){Fisher}, {Bolatto}, {Herrera-Camus}, {Draine}, {Donaldson}, {Walter}, {Sandstrom}, {Leroy}, {Cannon}, \& {Gordon}}]{Fisher2014}
{Fisher}, D.~B., {Bolatto}, A.~D., {Herrera-Camus}, R., {et~al.} 2014, \nat, 505, 186, \dodoi{10.1038/nature12765}

\bibitem[{{Fragos} {et~al.}(2013){Fragos}, {Lehmer}, {Tremmel}, {Tzanavaris}, {Basu-Zych}, {Belczynski}, {Hornschemeier}, {Jenkins}, {Kalogera}, {Ptak}, \& {Zezas}}]{Fragos2013}
{Fragos}, T., {Lehmer}, B., {Tremmel}, M., {et~al.} 2013, \apj, 764, 41, \dodoi{10.1088/0004-637X/764/1/41}

\bibitem[{{Fruscione} {et~al.}(2006){Fruscione}, {McDowell}, {Allen}, {Brickhouse}, {Burke}, {Davis}, {Durham}, {Elvis}, {Galle}, {Harris}, {Huenemoerder}, {Houck}, {Ishibashi}, {Karovska}, {Nicastro}, {Noble}, {Nowak}, {Primini}, {Siemiginowska}, {Smith}, \& {Wise}}]{Fruscione06}
{Fruscione}, A., {McDowell}, J.~C., {Allen}, G.~E., {et~al.} 2006, in Society of Photo-Optical Instrumentation Engineers (SPIE) Conference Series, Vol. 6270, Society of Photo-Optical Instrumentation Engineers (SPIE) Conference Series, ed. D.~R. {Silva} \& R.~E. {Doxsey}, 62701V, \dodoi{10.1117/12.671760}

\bibitem[{Garmire {et~al.}(2003)Garmire, Bautz, Ford, Nousek, \& Jr.}]{Garmire03}
Garmire, G.~P., Bautz, M.~W., Ford, P.~G., Nousek, J.~A., \& Jr., G. R.~R. 2003, in X-Ray and Gamma-Ray Telescopes and Instruments for Astronomy, ed. J.~E. Truemper \& H.~D. Tananbaum, Vol. 4851, International Society for Optics and Photonics (SPIE), 28 -- 44, \dodoi{10.1117/12.461599}

\bibitem[{{Garnett} {et~al.}(1991){Garnett}, {Kennicutt}, {Chu}, \& {Skillman}}]{Garnett91}
{Garnett}, D.~R., {Kennicutt}, Robert~C., J., {Chu}, Y.-H., \& {Skillman}, E.~D. 1991, \apj, 373, 458, \dodoi{10.1086/170065}

\bibitem[{Garofali {et~al.}(2024)Garofali, Basu-Zych, Johnson, Tzanavaris, Jaskot, Richardson, Lehmer, Yukita, Hodges-Kluck, Hornschemeier, Ptak, \& Vulic}]{Garofali2024}
Garofali, K., Basu-Zych, A.~R., Johnson, B.~D., {et~al.} 2024, The Astrophysical Journal, 960, \dodoi{10.3847/1538-4357/ad0a6a}

\bibitem[{{Gelbord} {et~al.}(2009){Gelbord}, {Mullaney}, \& {Ward}}]{Gelbord2009}
{Gelbord}, J.~M., {Mullaney}, J.~R., \& {Ward}, M.~J. 2009, \mnras, 397, 172, \dodoi{10.1111/j.1365-2966.2009.14961.x}

\bibitem[{{G{\"o}tberg} {et~al.}(2023){G{\"o}tberg}, {Drout}, {Ji}, {Groh}, {Ludwig}, {Crowther}, {Smith}, {de Koter}, \& {de Mink}}]{Gotberg2023}
{G{\"o}tberg}, Y., {Drout}, M.~R., {Ji}, A.~P., {et~al.} 2023, \apj, 959, 125, \dodoi{10.3847/1538-4357/ace5a3}

\bibitem[{Greene {et~al.}(2020)Greene, Strader, \& Ho}]{Greene20}
Greene, J.~E., Strader, J., \& Ho, L.~C. 2020, Annual Review of Astronomy and Astrophysics, 58, 257, \dodoi{10.1146/annurev-astro-032620-021835}

\bibitem[{{Guseva} {et~al.}(2000){Guseva}, {Izotov}, \& {Thuan}}]{Guseva00}
{Guseva}, N.~G., {Izotov}, Y.~I., \& {Thuan}, T.~X. 2000, \apj, 531, 776, \dodoi{10.1086/308489}

\bibitem[{{G{\"u}ver} \& {{\"O}zel}(2009)}]{Guver2009}
{G{\"u}ver}, T., \& {{\"O}zel}, F. 2009, \mnras, 400, 2050, \dodoi{10.1111/j.1365-2966.2009.15598.x}

\bibitem[{{Houck} \& {DeNicola}(2000)}]{Houck-DeNicola2000}
{Houck}, J.~C., \& {DeNicola}, L.~A. 2000, Astronomical Data Analysis Software and Systems IX, ASP Conference Proceedings, 216

\bibitem[{Izotov {et~al.}(2019)Izotov, Guseva, Fricke, \& Henkel}]{Izotov2019}
Izotov, Y.~I., Guseva, N.~G., Fricke, K.~J., \& Henkel, C. 2019, Astronomy and Astrophysics, 623, \dodoi{10.1051/0004-6361/201834768}

\bibitem[{Johnson {et~al.}(2021)Johnson, Leja, Conroy, \& Speagle}]{Johnson2021}
Johnson, B.~D., Leja, J., Conroy, C., \& Speagle, J.~S. 2021, The Astrophysical Journal Supplement Series, 254, \dodoi{10.3847/1538-4365/abef67}

\bibitem[{{Kaaret} {et~al.}(2004){Kaaret}, {Ward}, \& {Zezas}}]{Kaaret04}
{Kaaret}, P., {Ward}, M.~J., \& {Zezas}, A. 2004, \mnras, 351, L83, \dodoi{10.1111/j.1365-2966.2004.08020.x}

\bibitem[{{Katz} {et~al.}(2024){Katz}, {Ji}, {Telford}, \& {Senchyna}}]{Katz2024}
{Katz}, H., {Ji}, A.~P., {Telford}, G., \& {Senchyna}, P. 2024, The Open Journal of Astrophysics, 7, 106, \dodoi{10.33232/001c.126253}

\bibitem[{Kauffmann {et~al.}(2003)Kauffmann, Heckman, Tremonti, Brinchmann, Charlot, White, Ridgway, Brinkmann, Fukugita, Hall, \u~Zeljko~Ivezi\'c, Richards, \& Schneider}]{Kauffmann2003}
Kauffmann, G., Heckman, T.~M., Tremonti, C., {et~al.} 2003, Monthly Notices of the Royal Astronomical Society, 346

\bibitem[{{Kehrig} {et~al.}(2018){Kehrig}, {V{\'\i}lchez}, {Guerrero}, {Iglesias-P{\'a}ramo}, {Hunt}, {Duarte-Puertas}, \& {Ramos-Larios}}]{Kehrig18}
{Kehrig}, C., {V{\'\i}lchez}, J.~M., {Guerrero}, M.~A., {et~al.} 2018, \mnras, 480, 1081, \dodoi{10.1093/mnras/sty1920}

\bibitem[{{Kehrig} {et~al.}(2015){Kehrig}, {V{\'\i}lchez}, {P{\'e}rez-Montero}, {Iglesias-P{\'a}ramo}, {Brinchmann}, {Kunth}, {Durret}, \& {Bayo}}]{Kehrig15}
{Kehrig}, C., {V{\'\i}lchez}, J.~M., {P{\'e}rez-Montero}, E., {et~al.} 2015, \apjl, 801, L28, \dodoi{10.1088/2041-8205/801/2/L28}

\bibitem[{{Kennicutt}(1998)}]{Kennicutt1998}
{Kennicutt}, Jr., R.~C. 1998, \araa, 36, 189, \dodoi{10.1146/annurev.astro.36.1.189}

\bibitem[{Kewley {et~al.}(2001)Kewley, Dopita, Sutherland, Heisler, \& Trevena}]{Kewley2001}
Kewley, L.~J., Dopita, M.~A., Sutherland, R.~S., Heisler, C.~A., \& Trevena, J. 2001, The Astrophysical Journal, 556, \dodoi{10.1086/321545}

\bibitem[{{Kobulnicky} {et~al.}(1999){Kobulnicky}, {Kennicutt}, \& {Pizagno}}]{Kobulnicky99}
{Kobulnicky}, H.~A., {Kennicutt}, Jr., R.~C., \& {Pizagno}, J.~L. 1999, \apj, 514, 544, \dodoi{10.1086/306987}

\bibitem[{{Kollmeier} {et~al.}(2019){Kollmeier}, {Anderson}, {Blanc}, {Blanton}, {Covey}, {Crane}, {Drory}, {Frinchaboy}, {Froning}, {Johnson}, {Kneib}, {Kreckel}, {Merloni}, {Pellegrini}, {Pogge}, {Ramirez}, {Rix}, {Sayres}, {S{\'a}nchez-Gallego}, {Shen}, {Tkachenko}, {Trump}, {Tuttle}, {Weijmans}, {Zasowski}, {Barbuy}, {Beaton}, {Bergemann}, {Bochanski}, {Brandt}, {Casey}, {Cherinka}, {Eracleous}, {Fan}, {Garc{\'\i}a}, {Green}, {Hekker}, {Lane}, {Longa-Pe{\~n}a}, {Mathur}, {Meza}, {Minchev}, {Myers}, {Nidever}, {Nitschelm}, {O'Connell}, {Price-Whelan}, {Raddick}, {Rossi}, {Sankrit}, {Simon}, {Stutz}, {Ting}, {Trakhtenbrot}, {Weaver}, {Willmer}, \& {Weinberg}}]{Kollmeier19}
{Kollmeier}, J., {Anderson}, S.~F., {Blanc}, G.~A., {et~al.} 2019, in Bulletin of the American Astronomical Society, Vol.~51, 274

\bibitem[{{Kroupa}(2001)}]{Kroupa2001}
{Kroupa}, P. 2001, \mnras, 322, 231, \dodoi{10.1046/j.1365-8711.2001.04022.x}

\bibitem[{{Lan{\c{c}}on} \& {Wood}(2000)}]{Lanccon2000}
{Lan{\c{c}}on}, A., \& {Wood}, P.~R. 2000, \aaps, 146, 217, \dodoi{10.1051/aas:2000269}

\bibitem[{{Lecroq} {et~al.}(2024){Lecroq}, {Charlot}, {Bressan}, {Bruzual}, {Costa}, {Iorio}, {Spera}, {Mapelli}, {Chen}, {Chevallard}, \& {Dall'Amico}}]{Lecroq2024}
{Lecroq}, M., {Charlot}, S., {Bressan}, A., {et~al.} 2024, \mnras, 527, 9480, \dodoi{10.1093/mnras/stad3838}

\bibitem[{{Lehmer} {et~al.}(2010){Lehmer}, {Alexander}, {Bauer}, {Brandt}, {Goulding}, {Jenkins}, {Ptak}, \& {Roberts}}]{Lehmer2010}
{Lehmer}, B.~D., {Alexander}, D.~M., {Bauer}, F.~E., {et~al.} 2010, \apj, 724, 559, \dodoi{10.1088/0004-637X/724/1/559}

\bibitem[{{Lehmer} {et~al.}(2015){Lehmer}, {Tyler}, {Hornschemeier}, {Wik}, {Yukita}, {Antoniou}, {Boggs}, {Christensen}, {Craig}, {Hailey}, {Harrison}, {Maccarone}, {Ptak}, {Stern}, {Zezas}, \& {Zhang}}]{Lehmer2015}
{Lehmer}, B.~D., {Tyler}, J.~B., {Hornschemeier}, A.~E., {et~al.} 2015, \apj, 806, 126, \dodoi{10.1088/0004-637X/806/1/126}

\bibitem[{Lehmer {et~al.}(2021)Lehmer, Eufrasio, Basu-Zych, Doore, Fragos, Garofali, Kovlakas, Williams, Zezas, \& Santana-Silva}]{Lehmer21}
Lehmer, B.~D., Eufrasio, R.~T., Basu-Zych, A., {et~al.} 2021, The Astrophysical Journal, 907, 17, \dodoi{10.3847/1538-4357/abcec1}

\bibitem[{{Leja} {et~al.}(2019{\natexlab{a}}){Leja}, {Carnall}, {Johnson}, {Conroy}, \& {Speagle}}]{Leja2019}
{Leja}, J., {Carnall}, A.~C., {Johnson}, B.~D., {Conroy}, C., \& {Speagle}, J.~S. 2019{\natexlab{a}}, \apj, 876, 3, \dodoi{10.3847/1538-4357/ab133c}

\bibitem[{{Leja} {et~al.}(2017){Leja}, {Johnson}, {Conroy}, {van Dokkum}, \& {Byler}}]{Leja2017}
{Leja}, J., {Johnson}, B.~D., {Conroy}, C., {van Dokkum}, P.~G., \& {Byler}, N. 2017, The Astrophysical Journal, 837, \dodoi{10.3847/1538-4357/aa5ffe}

\bibitem[{{Leja} {et~al.}(2019{\natexlab{b}}){Leja}, {Johnson}, {Conroy}, {van Dokkum}, {Speagle}, {Brammer}, {Momcheva}, {Skelton}, {Whitaker}, {Franx}, \& {Nelson}}]{Leja20192}
{Leja}, J., {Johnson}, B.~D., {Conroy}, C., {et~al.} 2019{\natexlab{b}}, \apj, 877, 140, \dodoi{10.3847/1538-4357/ab1d5a}

\bibitem[{{Liu}(2011)}]{Liu11}
{Liu}, J. 2011, \apjs, 192, 10, \dodoi{10.1088/0067-0049/192/1/10}

\bibitem[{{Lower} {et~al.}(2020){Lower}, {Narayanan}, {Leja}, {Johnson}, {Conroy}, \& {Dav{\'e}}}]{Lower2020}
{Lower}, S., {Narayanan}, D., {Leja}, J., {et~al.} 2020, \apj, 904, 33, \dodoi{10.3847/1538-4357/abbfa7}

\bibitem[{{Luridiana} {et~al.}(2002){Luridiana}, {Esteban}, {Peimbert}, \& {Peimbert}}]{Luridiana2002}
{Luridiana}, V., {Esteban}, C., {Peimbert}, M., \& {Peimbert}, A. 2002, \rmxaa, 38, 97, \dodoi{10.48550/arXiv.astro-ph/0205021}

\bibitem[{{Manzano-King} {et~al.}(2019){Manzano-King}, {Canalizo}, \& {Sales}}]{Manzano-King19}
{Manzano-King}, C.~M., {Canalizo}, G., \& {Sales}, L.~V. 2019, \apj, 884, 54, \dodoi{10.3847/1538-4357/ab4197}

\bibitem[{{Mezcua}(2019)}]{Mezcua19}
{Mezcua}, M. 2019, Nature Astronomy, 3, 6, \dodoi{10.1038/s41550-018-0662-2}

\bibitem[{{Mondal} {et~al.}(2025){Mondal}, {Saha}, {Borgohain}, {Smith}, {Windhorst}, {Reddy}, {Chen}, {Umetsu}, \& {Jansen}}]{Mondal2025}
{Mondal}, C., {Saha}, K., {Borgohain}, A., {et~al.} 2025, \apj, 988, 171, \dodoi{10.3847/1538-4357/ade2cd}

\bibitem[{Moon {et~al.}(2011)Moon, Harrison, Cenko, \& Shariff}]{Moon11}
Moon, D.-S., Harrison, F.~A., Cenko, S.~B., \& Shariff, J.~A. 2011, The Astrophysical Journal Letters, 731, L32, \dodoi{10.1088/2041-8205/731/2/L32}

\bibitem[{{Morrissey} {et~al.}(2007){Morrissey}, {Conrow}, {Barlow}, {Small}, {Seibert}, {Wyder}, {Budav{\'a}ri}, {Arnouts}, {Friedman}, {Forster}, {Martin}, {Neff}, {Schiminovich}, {Bianchi}, {Donas}, {Heckman}, {Lee}, {Madore}, {Milliard}, {Rich}, {Szalay}, {Welsh}, \& {Yi}}]{Morrissey2007}
{Morrissey}, P., {Conrow}, T., {Barlow}, T.~A., {et~al.} 2007, \apjs, 173, 682, \dodoi{10.1086/520512}

\bibitem[{{NASA/IPAC Extragalactic Database (NED)}(2019)}]{NED}
{NASA/IPAC Extragalactic Database (NED)}. 2019, NASA/IPAC Extragalactic Database (NED),  IPAC, \dodoi{10.26132/NED1}

\bibitem[{{Nersesian} {et~al.}(2024){Nersesian}, {van der Wel}, {Gallazzi}, {Leja}, {Bezanson}, {Bell}, {D'Eugenio}, {de Graaff}, {Kaushal}, {Martorano}, {Maseda}, \& {Zibetti}}]{Nersesian2024}
{Nersesian}, A., {van der Wel}, A., {Gallazzi}, A., {et~al.} 2024, \aap, 681, A94, \dodoi{10.1051/0004-6361/202346769}

\bibitem[{{Nersesian} {et~al.}(2025){Nersesian}, {van der Wel}, {Gallazzi}, {Kaushal}, {Bezanson}, {Zibetti}, {Bell}, {D'Eugenio}, {Leja}, {Martorano}, \& {Wu}}]{Nersesian2025}
{Nersesian}, A., {van der Wel}, A., {Gallazzi}, A.~R., {et~al.} 2025, \aap, 695, A86, \dodoi{10.1051/0004-6361/202452662}

\bibitem[{{Noll} {et~al.}(2009){Noll}, {Burgarella}, {Giovannoli}, {Buat}, {Marcillac}, \& {Mu{\~n}oz-Mateos}}]{Noll2009}
{Noll}, S., {Burgarella}, D., {Giovannoli}, E., {et~al.} 2009, \aap, 507, 1793, \dodoi{10.1051/0004-6361/200912497}

\bibitem[{{Oskinova} \& {Schaerer}(2022)}]{Oskinova2022}
{Oskinova}, L.~M., \& {Schaerer}, D. 2022, \aap, 661, A67, \dodoi{10.1051/0004-6361/202142520}

\bibitem[{{Pacifici} {et~al.}(2023){Pacifici}, {Iyer}, {Mobasher}, {da Cunha}, {Acquaviva}, {Burgarella}, {Calistro Rivera}, {Carnall}, {Chang}, {Chartab}, {Cooke}, {Fairhurst}, {Kartaltepe}, {Leja}, {Ma{\l}ek}, {Salmon}, {Torelli}, {Vidal-Garc{\'\i}a}, {Boquien}, {Brammer}, {Brown}, {Capak}, {Chevallard}, {Circosta}, {Croton}, {Davidzon}, {Dickinson}, {Duncan}, {Faber}, {Ferguson}, {Fontana}, {Guo}, {Haeussler}, {Hemmati}, {Jafariyazani}, {Kassin}, {Larson}, {Lee}, {Mantha}, {Marchi}, {Nayyeri}, {Newman}, {Pandya}, {Pforr}, {Reddy}, {Sanders}, {Shah}, {Shahidi}, {Stevans}, {Triani}, {Tyler}, {Vanderhoof}, {de la Vega}, {Wang}, \& {Weston}}]{Pacifici2023}
{Pacifici}, C., {Iyer}, K.~G., {Mobasher}, B., {et~al.} 2023, \apj, 944, 141, \dodoi{10.3847/1538-4357/acacff}

\bibitem[{{Pakull} \& {Angebault}(1986)}]{Pakull86}
{Pakull}, M.~W., \& {Angebault}, L.~P. 1986, \nat, 322, 511, \dodoi{10.1038/322511a0}

\bibitem[{{Pakull} \& {Mirioni}(2002)}]{Pakull02}
{Pakull}, M.~W., \& {Mirioni}, L. 2002, arXiv e-prints, astro, \dodoi{10.48550/arXiv.astro-ph/0202488}

\bibitem[{{Pakull} \& {Motch}(1989)}]{Pakull89}
{Pakull}, M.~W., \& {Motch}, C. 1989, in European Southern Observatory Conference and Workshop Proceedings, Vol.~32, European Southern Observatory Conference and Workshop Proceedings, 285

\bibitem[{{Peletier}(2013)}]{Peletier2013}
{Peletier}, R.~F. 2013, in Secular Evolution of Galaxies, ed. J.~{Falc{\'o}n-Barroso} \& J.~H. {Knapen} (Secular Evolution of Galaxies), 353, \dodoi{10.48550/arXiv.1210.2127}

\bibitem[{{Plat} {et~al.}(2019){Plat}, {Charlot}, {Bruzual}, {Feltre}, {Vidal-Garc{\'\i}a}, {Morisset}, {Chevallard}, \& {Todt}}]{Plat19}
{Plat}, A., {Charlot}, S., {Bruzual}, G., {et~al.} 2019, \mnras, 490, 978, \dodoi{10.1093/mnras/stz2616}

\bibitem[{{Ponnada} {et~al.}(2020){Ponnada}, {Brorby}, \& {Kaaret}}]{Ponnada20}
{Ponnada}, S., {Brorby}, M., \& {Kaaret}, P. 2020, \mnras, 491, 3606, \dodoi{10.1093/mnras/stz2929}

\bibitem[{{Prestwich} {et~al.}(2015){Prestwich}, {Jackson}, {Kaaret}, {Brorby}, {Roberts}, {Saar}, \& {Yukita}}]{Prestwich2015}
{Prestwich}, A.~H., {Jackson}, F., {Kaaret}, P., {et~al.} 2015, \apj, 812, 166, \dodoi{10.1088/0004-637X/812/2/166}

\bibitem[{Prestwich {et~al.}(2013)Prestwich, Tsantaki, Zezas, Jackson, Roberts, Foltz, Linden, \& Kalogera}]{Prestwich13}
Prestwich, A.~H., Tsantaki, M., Zezas, A., {et~al.} 2013, The Astrophysical Journal, 769, 92, \dodoi{10.1088/0004-637x/769/2/92}

\bibitem[{{Quintero} \& {Eenens}(2024)}]{Quintero2024}
{Quintero}, E.~A., \& {Eenens}, P. 2024, \mnras, 532, 2604, \dodoi{10.1093/mnras/stae1670}

\bibitem[{{Rauch}(2003)}]{Rauch2003}
{Rauch}, T. 2003, \aap, 403, 709, \dodoi{10.1051/0004-6361:20030412}

\bibitem[{{Roy} {et~al.}(2025){Roy}, {Krumholz}, {Salvadori}, {Meynet}, {Ekstr{\"o}m}, {Vink}, {Sander}, {Sutherland}, {Paul}, {Pallottini}, \& {Sk{\'u}lad{\'o}ttir}}]{Roy2025}
{Roy}, A., {Krumholz}, M.~R., {Salvadori}, S., {et~al.} 2025, \aap, 696, A29, \dodoi{10.1051/0004-6361/202553697}

\bibitem[{{S{\'a}nchez-Bl{\'a}zquez} {et~al.}(2006){S{\'a}nchez-Bl{\'a}zquez}, {Peletier}, {Jim{\'e}nez-Vicente}, {Cardiel}, {Cenarro}, {Falc{\'o}n-Barroso}, {Gorgas}, {Selam}, \& {Vazdekis}}]{Sanchez2006}
{S{\'a}nchez-Bl{\'a}zquez}, P., {Peletier}, R.~F., {Jim{\'e}nez-Vicente}, J., {et~al.} 2006, \mnras, 371, 703, \dodoi{10.1111/j.1365-2966.2006.10699.x}

\bibitem[{{Sartorio} {et~al.}(2023){Sartorio}, {Fialkov}, {Hartwig}, {Mirouh}, {Izzard}, {Magg}, {Klessen}, {Glover}, {Chen}, {Tarumi}, \& {Hendriks}}]{Sartorio2023}
{Sartorio}, N.~S., {Fialkov}, A., {Hartwig}, T., {et~al.} 2023, \mnras, 521, 4039, \dodoi{10.1093/mnras/stad697}

\bibitem[{{Schaerer}(1996)}]{Schaerer96}
{Schaerer}, D. 1996, \apjl, 467, L17, \dodoi{10.1086/310193}

\bibitem[{{Schaerer} {et~al.}(2019){Schaerer}, {Fragos}, \& {Izotov}}]{Schaerer19}
{Schaerer}, D., {Fragos}, T., \& {Izotov}, Y.~I. 2019, \aap, 622, L10, \dodoi{10.1051/0004-6361/201935005}

\bibitem[{{Schaerer} {et~al.}(2022){Schaerer}, {Marques-Chaves}, {Barrufet}, {Oesch}, {Izotov}, {Naidu}, {Guseva}, \& {Brammer}}]{Schaerer2022}
{Schaerer}, D., {Marques-Chaves}, R., {Barrufet}, L., {et~al.} 2022, \aap, 665, L4, \dodoi{10.1051/0004-6361/202244556}

\bibitem[{Senchyna \& Stark(2019)}]{Senchyna19}
Senchyna, P., \& Stark, D.~P. 2019, Monthly Notices of the Royal Astronomical Society, 484, 1270, \dodoi{10.1093/mnras/stz058}

\bibitem[{Senchyna {et~al.}(2020)Senchyna, Stark, Mirocha, Reines, Charlot, Jones, \& Mulchaey}]{Senchyna20}
Senchyna, P., Stark, D.~P., Mirocha, J., {et~al.} 2020, Monthly Notices of the Royal Astronomical Society, 494, 941, \dodoi{10.1093/mnras/staa586}

\bibitem[{{Shenar} {et~al.}(2020){Shenar}, {Gilkis}, {Vink}, {Sana}, \& {Sander}}]{Shenar2020}
{Shenar}, T., {Gilkis}, A., {Vink}, J.~S., {Sana}, H., \& {Sander}, A.~A.~C. 2020, \aap, 634, A79, \dodoi{10.1051/0004-6361/201936948}

\bibitem[{Shirazi \& Brinchmann(2012)}]{Shirazi12}
Shirazi, M., \& Brinchmann, J. 2012, Monthly Notices of the Royal Astronomical Society, 421, 1043, \dodoi{10.1111/j.1365-2966.2012.20439.x}

\bibitem[{{Simmonds} {et~al.}(2021){Simmonds}, {Schaerer}, \& {Verhamme}}]{Simmonds21}
{Simmonds}, C., {Schaerer}, D., \& {Verhamme}, A. 2021, \aap, 656, A127, \dodoi{10.1051/0004-6361/202141856}

\bibitem[{{Sparre} {et~al.}(2017){Sparre}, {Hayward}, {Feldmann}, {Faucher-Gigu{\`e}re}, {Muratov}, {Kere{\v{s}}}, \& {Hopkins}}]{Sparre2017}
{Sparre}, M., {Hayward}, C.~C., {Feldmann}, R., {et~al.} 2017, \mnras, 466, 88, \dodoi{10.1093/mnras/stw3011}

\bibitem[{{Stanway} {et~al.}(2014){Stanway}, {Eldridge}, {Greis}, {Davies}, {Wilkins}, \& {Bremer}}]{Stanway2014}
{Stanway}, E.~R., {Eldridge}, J.~J., {Greis}, S. M.~L., {et~al.} 2014, \mnras, 444, 3466, \dodoi{10.1093/mnras/stu1682}

\bibitem[{{Storey} \& {Hummer}(1995)}]{Storey1995}
{Storey}, P.~J., \& {Hummer}, D.~G. 1995, \mnras, 272, 41, \dodoi{10.1093/mnras/272.1.41}

\bibitem[{{Strickland} \& {Heckman}(2009)}]{Strickland2009}
{Strickland}, D.~K., \& {Heckman}, T.~M. 2009, \apj, 697, 2030, \dodoi{10.1088/0004-637X/697/2/2030}

\bibitem[{{Sun} {et~al.}(2012){Sun}, {Chen}, {Feng}, {Chu}, {Chen}, {Wang}, \& {Li}}]{Sun2012}
{Sun}, W., {Chen}, Y., {Feng}, L., {et~al.} 2012, \apj, 760, 61, \dodoi{10.1088/0004-637X/760/1/61}

\bibitem[{{Sz{\'e}csi} {et~al.}(2015){Sz{\'e}csi}, {Langer}, {Yoon}, {Sanyal}, {de Mink}, {Evans}, \& {Dermine}}]{Szcesi15}
{Sz{\'e}csi}, D., {Langer}, N., {Yoon}, S.-C., {et~al.} 2015, \aap, 581, A15, \dodoi{10.1051/0004-6361/201526617}

\bibitem[{{Sz{\'e}csi} {et~al.}(2025){Sz{\'e}csi}, {Tramper}, {Kub{\'a}tov{\'a}}, {Kehrig}, {Kub{\'a}t}, {Krti{\v{c}}ka}, {Sander}, \& {Garcia}}]{Dorottya2025}
{Sz{\'e}csi}, D., {Tramper}, F., {Kub{\'a}tov{\'a}}, B., {et~al.} 2025, \aap, 703, A131, \dodoi{10.1051/0004-6361/202452483}

\bibitem[{{Tacchella} {et~al.}(2023){Tacchella}, {Johnson}, {Robertson}, {Carniani}, {D'Eugenio}, {Kumari}, {Maiolino}, {Nelson}, {Suess}, {{\"U}bler}, {Williams}, {Adebusola}, {Alberts}, {Arribas}, {Bhatawdekar}, {Bonaventura}, {Bowler}, {Bunker}, {Cameron}, {Curti}, {Egami}, {Eisenstein}, {Frye}, {Hainline}, {Helton}, {Ji}, {Looser}, {Lyu}, {Perna}, {Rawle}, {Rieke}, {Rieke}, {Saxena}, {Sandles}, {Shivaei}, {Simmonds}, {Sun}, {Willmer}, {Willott}, \& {Witstok}}]{Sandro2023}
{Tacchella}, S., {Johnson}, B.~D., {Robertson}, B.~E., {et~al.} 2023, \mnras, 522, 6236, \dodoi{10.1093/mnras/stad1408}

\bibitem[{{Thuan} \& {Izotov}(2005)}]{Thuan05}
{Thuan}, T.~X., \& {Izotov}, Y.~I. 2005, \apjs, 161, 240, \dodoi{10.1086/491657}

\bibitem[{Thygesen {et~al.}(2023)Thygesen, Plotkin, Soria, Reines, Greene, Anderson, Baldassare, Owens, Urquhart, Gallo, Miller-Jones, Paul, \& Rollings}]{Thygesen23}
Thygesen, E., Plotkin, R.~M., Soria, R., {et~al.} 2023, Monthly Notices of the Royal Astronomical Society, 519, 5848, \dodoi{10.1093/mnras/stad002}

\bibitem[{{Topping} {et~al.}(2020){Topping}, {Shapley}, {Reddy}, {Sanders}, {Coil}, {Kriek}, {Mobasher}, \& {Siana}}]{Topping2020}
{Topping}, M.~W., {Shapley}, A.~E., {Reddy}, N.~A., {et~al.} 2020, \mnras, 495, 4430, \dodoi{10.1093/mnras/staa1410}

\bibitem[{{Triani} {et~al.}(2024){Triani}, {Di Stefano}, \& {Kewley}}]{Triani2024}
{Triani}, D.~P., {Di Stefano}, R., \& {Kewley}, L.~J. 2024, arXiv e-prints, arXiv:2405.08121, \dodoi{10.48550/arXiv.2405.08121}

\bibitem[{{Van Bever} {et~al.}(1999){Van Bever}, {Belkus}, {Vanbeveren}, \& {Van Rensbergen}}]{VanBever1999}
{Van Bever}, J., {Belkus}, H., {Vanbeveren}, D., \& {Van Rensbergen}, W. 1999, \na, 4, 173, \dodoi{10.1016/S1384-1076(99)00011-1}

\bibitem[{{Verner} {et~al.}(1996){Verner}, {Ferland}, {Korista}, \& {Yakovlev}}]{Verner1996}
{Verner}, D.~A., {Ferland}, G.~J., {Korista}, K.~T., \& {Yakovlev}, D.~G. 1996, \apj, 465, 487, \dodoi{10.1086/177435}

\bibitem[{Virtanen {et~al.}(2020)Virtanen, Gommers, Oliphant, Haberland, Reddy, Cournapeau, Burovski, Peterson, Weckesser, Bright, {van der Walt}, Brett, Wilson, Millman, Mayorov, Nelson, Jones, Kern, Larson, Carey, Polat, Feng, Moore, {VanderPlas}, Laxalde, Perktold, Cimrman, Henriksen, Quintero, Harris, Archibald, Ribeiro, Pedregosa, {van Mulbregt}, \& {SciPy 1.0 Contributors}}]{Virtanen2020}
Virtanen, P., Gommers, R., Oliphant, T.~E., {et~al.} 2020, Nature Methods, 17, \dodoi{10.1038/s41592-019-0686-2}

\bibitem[{{Wan} {et~al.}(2024){Wan}, {Tacchella}, {Johnson}, {Iyer}, {Speagle}, \& {Maiolino}}]{Wan2024}
{Wan}, J.~T., {Tacchella}, S., {Johnson}, B.~D., {et~al.} 2024, \mnras, 532, 4002, \dodoi{10.1093/mnras/stae1734}

\bibitem[{{Wang} {et~al.}(2023){Wang}, {Leja}, {Bezanson}, {Johnson}, {Khullar}, {Labb{\'e}}, {Price}, {Weaver}, \& {Whitaker}}]{Wang2023}
{Wang}, B., {Leja}, J., {Bezanson}, R., {et~al.} 2023, \apjl, 944, L58, \dodoi{10.3847/2041-8213/acba99}

\bibitem[{{Webb} {et~al.}(2014){Webb}, {Cseh}, \& {Kirsten}}]{Webb2014}
{Webb}, N.~A., {Cseh}, D., \& {Kirsten}, F. 2014, \pasa, 31, e009, \dodoi{10.1017/pasa.2014.1}

\bibitem[{{Wenger} {et~al.}(2000){Wenger}, {Ochsenbein}, {Egret}, {Dubois}, {Bonnarel}, {Borde}, {Genova}, {Jasniewicz}, {Lalo{\"e}}, {Lesteven}, \& {Monier}}]{Wenger2000}
{Wenger}, M., {Ochsenbein}, F., {Egret}, D., {et~al.} 2000, \aaps, 143, 9, \dodoi{10.1051/aas:2000332}

\bibitem[{{Wilms} {et~al.}(2000){Wilms}, {Allen}, \& {McCray}}]{Wilms2000}
{Wilms}, J., {Allen}, A., \& {McCray}, R. 2000, \apj, 542, 914, \dodoi{10.1086/317016}

\bibitem[{{Xiao} {et~al.}(2018){Xiao}, {Stanway}, \& {Eldridge}}]{Xiao2018}
{Xiao}, L., {Stanway}, E.~R., \& {Eldridge}, J.~J. 2018, \mnras, 477, 904, \dodoi{10.1093/mnras/sty646}

\bibitem[{Xiao {et~al.}(2024)Xiao, Oesch, Elbaz, Bing, erica J.~Nelson, Weibel, Illingworth, {van Dokkum}, Naidu, Daddi, Bouwens, Matthee, Wuyts, Chisholm, Brammer, Dickinson, Magnelli, Leroy, Schaerer, Herard-Demanche, Lim, Barrufet, Endsley, Fudamoto, G\'omez-Guijarro, Gottumukkala, Labb\'e, Magee, Marchesini, Maseda, Qin, Reddy, Shapley, Shivaei, Shuntov, Stefanon, Whitaker, \& Wyithe}]{Xiao24}
Xiao, M., Oesch, P.~A., Elbaz, D., {et~al.} 2024, Nature, 635, \dodoi{10.1038/s41586-024-08094-5}

\bibitem[{{York} {et~al.}(2000){York}, {Adelman}, {Anderson}, {Anderson}, {Annis}, {Bahcall}, {Bakken}, {Barkhouser}, {Bastian}, {Berman}, {Boroski}, {Bracker}, {Briegel}, {Briggs}, {Brinkmann}, {Brunner}, {Burles}, {Carey}, {Carr}, {Castander}, {Chen}, {Colestock}, {Connolly}, {Crocker}, {Csabai}, {Czarapata}, {Davis}, {Doi}, {Dombeck}, {Eisenstein}, {Ellman}, {Elms}, {Evans}, {Fan}, {Federwitz}, {Fiscelli}, {Friedman}, {Frieman}, {Fukugita}, {Gillespie}, {Gunn}, {Gurbani}, {de Haas}, {Haldeman}, {Harris}, {Hayes}, {Heckman}, {Hennessy}, {Hindsley}, {Holm}, {Holmgren}, {Huang}, {Hull}, {Husby}, {Ichikawa}, {Ichikawa}, {Ivezi{\'c}}, {Kent}, {Kim}, {Kinney}, {Klaene}, {Kleinman}, {Kleinman}, {Knapp}, {Korienek}, {Kron}, {Kunszt}, {Lamb}, {Lee}, {Leger}, {Limmongkol}, {Lindenmeyer}, {Long}, {Loomis}, {Loveday}, {Lucinio}, {Lupton}, {MacKinnon}, {Mannery}, {Mantsch}, {Margon}, {McGehee}, {McKay}, {Meiksin}, {Merelli}, {Monet}, {Munn}, {Narayanan}, {Nash}, {Neilsen}, {Neswold}, {Newberg}, {Nichol}, {Nicinski},
  {Nonino}, {Okada}, {Okamura}, {Ostriker}, {Owen}, {Pauls}, {Peoples}, {Peterson}, {Petravick}, {Pier}, {Pope}, {Pordes}, {Prosapio}, {Rechenmacher}, {Quinn}, {Richards}, {Richmond}, {Rivetta}, {Rockosi}, {Ruthmansdorfer}, {Sandford}, {Schlegel}, {Schneider}, {Sekiguchi}, {Sergey}, {Shimasaku}, {Siegmund}, {Smee}, {Smith}, {Snedden}, {Stone}, {Stoughton}, {Strauss}, {Stubbs}, {SubbaRao}, {Szalay}, {Szapudi}, {Szokoly}, {Thakar}, {Tremonti}, {Tucker}, {Uomoto}, {Vanden Berk}, {Vogeley}, {Waddell}, {Wang}, {Watanabe}, {Weinberg}, {Yanny}, {Yasuda}, \& {SDSS Collaboration}}]{York00}
{York}, D.~G., {Adelman}, J., {Anderson}, John~E., J., {et~al.} 2000, \aj, 120, 1579, \dodoi{10.1086/301513}

\end{thebibliography}
\bibliographystyle{aasjournal}

\end{document}